\documentclass{article}
\usepackage{markdown} 
\usepackage{pgfplots} 
\PassOptionsToPackage{square,numbers}{natbib}
\usepackage[preprint]{neurips_2022}
\pgfplotsset{width=6.25cm, compat=newest}
\usepgfplotslibrary{fillbetween}
\usepackage{siunitx}{
\sisetup{output-exponent-marker=\ensuremath{\mathrm{e}}}}

\usepackage[utf8]{inputenc} 
\usepackage[T1]{fontenc}    
\usepackage{hyperref}       
\usepackage{url}            
\usepackage{booktabs}       
\usepackage{amsfonts}       
\usepackage{nicefrac}       
\usepackage{microtype}      
\usepackage{graphicx}
\usepackage{blindtext}
\usepackage{wrapfig}
\usepackage{wrapfig,lipsum,booktabs}
\usepackage{subcaption}
\usepackage{float}
\usepackage{enumitem}
\usepackage{xcolor}         
\usepackage{scalefnt}
\usepgfplotslibrary{statistics}
\usepackage{amsmath}

\DeclareMathOperator*{\argmax}{argmax}

\title{Deliberative Technology for Alignment}

\usepackage{authblk}
\author[  1 ]{\textbf{Andrew Konya} 
 \thanks{Corresponding author: andrew@remesh.org}\hspace{1mm}}
 \author[2]{\textbf{Deger Turan}}
 \author[3]{\textbf{Aviv Ovadya}}
  \author[1]{\textbf{Lina Qui}}
\author[  ]{\textbf{Daanish Masood}}
\author[4]{\textbf{Flynn Devine}}
\author[5]{\textbf{Lisa Schirch}}
 \author[6]{\textbf{Isabella Roberts}}
 \author[ ] {\textbf{Deliberative Alignment Forum} \thanks{The \emph{Deliberative Alignment Forum} comprised a cadence of bi-weekly and ad-hoc discussions on deliberative alignment over the course of a seven month period beginning March 2023. Participants in the forum included  this paper's authors plus Colin Megill, Michiel Bakker, Saffron Huang, Alice Siu, Iason Gabriel, Teddy Lee, Tyna Eloundou, Oliver Klingefjord, Kanav Mehra, Jeffrey Stulmaker, Bruno Marnette, Colin Irwin, and others who choose to remain anonymous. Forum participation does not imply endorsement of everything in this paper.}}
\affil[ ]{  }
 
\affil[1]{Remesh}
\affil[2]{AI Objectives Institute}
\affil[3]{AI \& Democracy Foundation}
\affil[4]{Collective Intelligence Project}
\affil[5]{University of Notre Dame}
\affil[6]{Northeastern University London}

\begin{document}
\maketitle

\hypersetup{pdfborder={0 0 0}} 
\hypersetup{pdfborder={0 0 1}} 

\vspace{-0.9cm}
\section*{\centering Abstract}
For humanity to maintain and expand its agency into the future, the most powerful systems we create must be those which act to align the future with the will of humanity. The most powerful systems today are massive institutions like governments, firms, and NGOs. Deliberative technology is already being used across these institutions to help align governance and diplomacy with human will, and modern AI is poised to make this technology significantly better. At the same time, the race to superhuman AGI is already underway, and the AI systems it gives rise to may become the most powerful systems of the future. Failure to align the impact of such powerful AI with the will of humanity may lead to catastrophic consequences, while success may unleash abundance. Right now, there is a window of opportunity to use deliberative technology to align the impact of powerful AI with the will of humanity. Moreover, it may be possible to engineer a symbiotic coupling between powerful AI and deliberative alignment systems such that the quality of alignment improves as AI capabilities increase. 

\textbf{Key Claims:}
\begin{itemize}
    \itemsep-0.2em 
    \item Humanity's agency should be preserved and expanded into the future. 
    \item The will of humanity can be sensed and used for alignment.
    \item Aligning powerful systems with the \emph{will of humanity} increases humanity's agency. 
    \item Powerful institutions already use deliberative technology to better align with public will.
    \item AI can make deliberative alignment systems more intelligent and effective.
    \item Deliberative alignment systems can be used for AI alignment.
    \item Symbiotic improvement between AI and deliberative alignment systems may be possible.
    \item Aggressively pursuing three mandates for action will significantly increase the chance the future aligns with the will of humanity.
\end{itemize}
\textbf{Mandates for Action:}
\begin{enumerate}
    \itemsep-0.2em 
    \item Generate a universally legitimate \emph{will of humanity signal} as an open public good.
    \item Build intelligent deliberative alignment into powerful institutions.
    \item Ensure the most powerful AI systems are aligned with the will of humanity.
\end{enumerate}

\newpage

\tableofcontents
\newpage

\section{Preface}

\subsection{Motivation}
We are motivated by the desire to actively align the universe's path into the future with the will of humanity. Over the next century, as humanity's energy budget expands, and transformative AI emerges, the capacity to impact the future will balloon. Our aim is to ensure that ballooning impact is guided by the will of humanity.

\textbf{Increasing impact.} The amount of force one can exert in the present to impact the future is limited by their energy budget. This makes humanity's energy budget a good proxy for its potential impact. In the last 10,000 years humanity's energy budget has increased by around 20,000x -- even outpacing humanity's population growth \cite{syvitski2020extra}. While the human body requires around 4 gigajoules per year to survive, human civilization now consumes nearly 20x that per capita \cite{owidenergy}. Even still, humanity only consumes about $0.01\%$ of the energy that makes it to Earth from the Sun. The continued expansion of solar \cite{IEA2022}, progress in fission \cite{schmidt2018review}, and the prospect of fusion \cite{progress2020tikhonchuk,progress2019kline,lux2022commer} all suggest a future of increasing energy abundance, and with it an ever-growing impact on the future.

    \textbf{Precarious control.} Control of this growing energy budget and the impact it enables is not evenly distributed\cite{zucman2019global,vitali2011network}. It tends to be concentrated in institutional systems like governments and companies\cite{buisness2020piplovic}. However, there is growing sentiment that many of these institutions have little interest in aligning their impact with the will of humanity\cite{anti2022dunn}. And while some strive for alignment with a subset of humanity -- like democratic governments -- they still routinely struggle to realize the will of their constituencies\cite{gallop2023congress,statista2023uk}. What's more, increasingly powerful AI systems are poised to command a significant amount of impact in the future, yet there is little reason to expect their impact will naturally align with the will of humanity either \cite{hendrycks2023natural}. This document is written for those who want the impact of these powerful systems -- from governments to AGI -- to align with the will of humanity.

\textbf{Technological inflection point.} Over the last decade, an initial wave of deliberative technology has begun to help enable governance and diplomacy which better reflects public will. But while progress on deliberative technology continues, the work is happening in isolated pockets across the private and public sectors and various academic fields. At the same time, AI systems are approaching an inflection point as they demonstrate increasingly powerful and more general capabilities \cite{bubeck2023sparks}. A \emph{united effort} to safely apply such AI to deliberative technology has the potential to enable a new generation of solutions that make deliberative alignment systems more intelligent and effective. Further, while deliberative alignment systems are already being used successfully across many of the world's largest institutions \cite{small2021polis,irwin2021using}, their use for aligning AI, while promising, is remarkably nascent \cite{konya2023democratic,ganguli2023collective}. This implies an opportunity to apply the know-how and technologies developed for aligning institutions to aligning AI \cite{ovadya2023reimagining}.\footnote{This also suggests the possibility of a symbiotic relationship between AI and deliberative alignment systems; where increasingly better AI is governed by deliberative alignment systems which become more effective with increasingly better AI.}

\subsection{Goals}
The specific goals of creating this document are:
\begin{itemize}
    \item \textbf{Unify a collaborative community} of people working on deliberative alignment.
    \item \textbf{Motivate new work} on the next generation of intelligent deliberative alignment systems.
    \item \textbf{Attract resources} towards specific mandates for action which increase the chance the future aligns with the will of humanity.

\end{itemize}

\subsection{How to read this document}
This document is not necessarily meant to be read front to back (though it can be). Instead, each section is meant to address a set of critical questions a person may have as they build their own understanding of deliberative alignment. 

Overall, this document focuses on \emph{aligning the future with the will of humanity}. This statement should immediately conjure a question in the minds of discerning individuals "What exactly do you mean by the \emph{will of humanity}?" \textbf{Section \ref{will_of_humanity}} focuses on answering this question; starting with a philosophical definition (section \ref{woh.def}), then building up to something which can be digitally stored (section \ref{woh_representation}) and physically sensed (section \ref{WOH.sensing}). 

Once you are convinced the \emph{will of humanity} is a real thing that can be sensed, stored, and computed with, the next question is "What do you mean by alignment, and how does the will of humanity fit into that?" \textbf{Section \ref{alignmentSystems}} frames alignment as being between the future and the will of humanity (section \ref{AS.framing}), then introduces the idea of an \emph{alignment system} as a general class of system that uses a will-signal to take actions which best align the future with the will of humanity (sections \ref{AS.components}-\ref{AS.examples}). 

Once you understand what the \emph{will of humanity} is and how it interacts with an \emph{alignment system}, the next question is "How does deliberation fit into alignment systems?" To address this question \textbf{Section \ref{DA}} first introduces what we mean by deliberation (section \ref{DA.deliberation}) including its mechanics (section \ref{DA.deliberative_mechanics}). Then we review a wide range of existing \emph{deliberative technologies} (section \ref{DA.deliberativeTechnology}) and provide examples of how those technologies have been integrated into real-world deliberative alignment systems (section \ref{DA.DelibTechInAlignmentSystems}). Lastly, we review a multitude of challenges manifest in integrating deliberative technology into alignment systems (section \ref{DAchallanges}).

Once you understand how existing deliberative technology integrates into alignment systems, and the unique challenges this creates, the next question is "How can AI be used within deliberative technology to address these challenges?" \textbf{Section \ref{IDA}} tackles this question directly by first introducing AI as a technology (section \ref{IDA.AI}) and then enumerating all of the different ways AI (and some adjacent technologies) can improve deliberative technology used within alignment systems (section \ref{IDA.AIforDA}).

Once you understand how AI can be used to make deliberative alignment systems more intelligent and effective in theory, the next question is "How do you \emph{apply} intelligent deliberative alignment to real-world systems?"  \textbf{Section \ref{application}} considers this question and explores the application of intelligent deliberative alignment with both institutions (section \ref{application.institutions}) and AI agents (section \ref{application.AI}), including how scalable oversight enabled by symbiotic improvement between an AI agent and its deliberative alignment system might be possible (section \ref{application.symbiotic}).

While this document attempts to offer a thorough understanding of all aspects of deliberative alignment, there remain open challenges and opportunities that still need to be explored. \textbf{Section \ref{open}} highlights the open challenges and opportunities we are aware of. 

Finally, once you understand the applicable relationships between the will of humanity, alignment systems, deliberative technology, and artificial intelligence, the final question is "so what should we do with this understanding?" \textbf{Section \ref{mandates}} introduces three \emph{mandates for action} which stem from and require the employment of this understanding, and whose execution stands to increase the probability that the future aligns with the will of humanity: 1) Generate a universally legitimate will of humanity signal as an open public good (section \ref{mandate1}); 2) Build intelligent deliberative alignment into powerful institutions (section \ref{mandate3}); 3) Ensure the most powerful AI systems are aligned with the will of humanity (section \ref{mandate2}). For each mandate, we discuss why it is important in the context of deliberative alignment and what the approach(s) to execute on it might look like. Lastly, the magnitude of potential impact resulting from executing on these mandates is estimated (section \ref{mandates.impact}).

\subsection{Caveats}

\textbf{Not comprehensive.} While this document aims to thoroughly cover many of the topics related to deliberative alignment it should not be assumed to be comprehensive. Some relevant topics may have been overlooked. Also, samples of relevant work from many different fields are referenced and discussed, but we have likely missed important works across some of these fields. Further, this paper covers a wide range of ideas and approaches, and while it aims to discuss them at the level needed to bring clarity, it does not go as deep as one might wish in all scenarios.

\textbf{Not a complete solution.} This document aims to build up a way of framing and tackling the challenge of aligning the future with the will of humanity. However, the types of solutions proposed and discussed are better viewed as sketches of solutions, or even hypotheses, rather than complete solutions in and of themselves. This paper aims more to outline problems in ways that lead one to ask the right questions, rather than aiming to provide complete answers. 

\textbf{Imperfect terminology and frames.} Many of the topics discussed in this document involve the use of terminology and frames that may be perceived as controversial or reductionist, or have unintended connotations. Moreover, at times we use terms that have precise and conflicting definitions across different fields. Overall, we found it challenging to arrive at an ideal set terms and frames which strike the perfect balance across a wide range of trade-offs. It is possible that further work and deliberation among the authors could yield better options. But, given the urgent and rapidly evolving landscape this work pertains to, even though we have reservations\ref{A.caveats}, we think it is better to share the work as-is in order to facilitate a more diverse and global discussion, rather than for us to continue trying to strive for perfection in private. Even still, we've aimed to avoid field-specific jargon when possible and to explain what we mean by jargon when we do use it\footnote{In \ref{A.terms} we list and define some of the potentially jargon-y terms in this document.}.

\newpage

\section{Will of Humanity}\label{will_of_humanity}

\subsection{Why?}
\textbf{Normative justification.} Why choose the \emph{will of humanity} as an alignment target? The answer begins with the normative belief that \emph{humanity should maintain and expand its agency into the future.} In other words, humanity should be able to choose the future it wants for itself. If you don't subscribe to this belief -- ie. if you believe that humanity should ultimately cede its agency to artificial superintelligence(s) that are superior to humans -- then this paper is probably not for you. Assuming you do subscribe to this belief, the next question is: how does this belief relate to alignment? 

Human civilization is comprised of powerful systems that aggregate resources and take actions to impact the future. For any system, we can call the distribution of futures the system aims to create its alignment target. This means choosing an alignment target for a system equates to choosing the future it will aim to create. Consequently, choosing a universal alignment target for all systems (or a single all-powerful system) equates to choosing what will likely become humanity’s future. So, if we believe humanity should have the agency to choose its own future\footnote{Within the bounds of what is permitted by the laws of physics.}, then it is humanity itself that must ultimately choose such a universal alignment target. How can we achieve this?

In Section \ref{woh.def} we adopt a definition of the \emph{will of humanity} to be all human's deliberate preference judgments across all possible futures which determine their voluntary actions -- in other words, it is the distribution of future’s humanity wants enough to act on. Thus, setting the will of humanity as the universal alignment target means that the distribution of futures we seek to have humanity’s systems\footnote{Humanity's systems range from governments and firms to NGOs and (eventually) AGI.} pursue is the same as the distribution of futures humanity would choose to pursue given the agency to do so. This equates to enabling humanity to choose its own alignment target. What's more, choosing any other alignment target that deviates from the will of humanity risks suppressing humanity's agency to pursue the future it wants. This potentially makes the will of humanity the only universal alignment target that necessarily preserves and expands humanity's agency into the future. 

\textbf{Caveats.} First, the normative justification provided above seems to treat the will of humanity as a monolith that can be aligned with in-and-of-itself. However, as we will discuss in section \ref{alignmentSystems}, mapping the will of humanity as defined in this section to unitary decisions \emph{requires} an aggregation function. Further, in what some may find an unsatisfactory aspect of this paper, we do not take a firm position on what the 'right' aggregation function is; rather, we hope the specificity this paper provides -- around what the aggregation is over (a Will matrix), and how the aggregation is used (in an alignment system) -- helps motivate focused efforts to arrive at a satisfying answer to this question.

Second, in the definition and discussion of the will of humanity that follows, we do not attempt a comprehensive review of the vast literature on human will. Rather, we simply adopt a definition that is self-consistent with our normative motivation for choosing the will of humanity as an alignment target, and then explore the properties and consequences such a definition gives rise to. Thus, we do not intend for our definition of will to be authoritative, nor to reflect current philosophical consensus on the matter. It is meant only to be self-consistent with our normative motivation and practically useful in the context of alignment. And while we view it as a useful construct to enable analytical treatment of a squishy human thing, it is, at best, an approximation of some perhaps unknowable ideal. However, we \emph{do} aim to explore the definition we adopt in enough detail as to provide a large surface area for the type of precise disagreements which can give rise to a better definition and more useful notions. 

Third, the full will of humanity is perhaps an appropriate alignment target for systems that impact the future of all humans -- like the United Nations, or perhaps even a artificial superintelligence\footnote{We are not advocating for an ASI which impacts all of humanity, only saying that if such a thing comes to exist, it is probably best aligned with the full will of humanity.}.  However, for systems that only impact a portion of humanity -- like a local government -- the more appropriate alignment target is likely the subset of the will of humanity corresponding to the individuals the system being aligned impacts. In section \ref{alignmentSystems} we use the terms \emph{global} vs \emph{local} to distinguish between alignment systems whose target is the full will of humanity vs. some local portion of it. So, while the majority of this section adopts a top-down view that focuses on the full will of humanity, we do not intend to imply a stance in opposition to preserving local agency.

\subsection{Definition}\label{woh.def}

\begin{figure}[H]
\centering
  \includegraphics[width=0.7\linewidth]{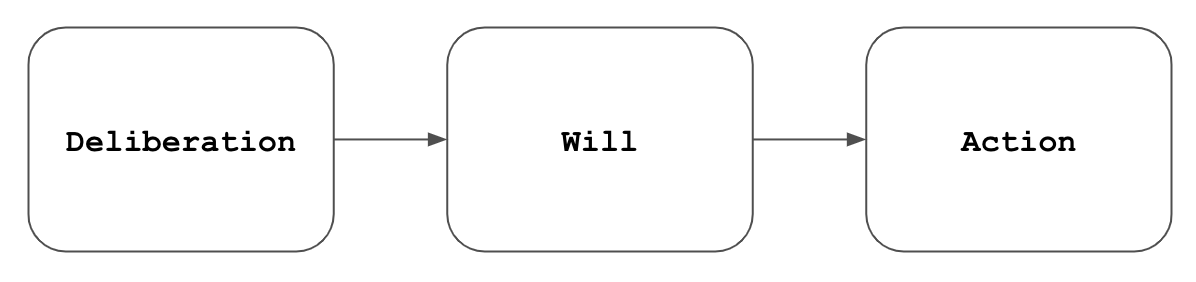}
  \caption{Causal relationship of deliberation, will, and action according to Hobbesian philosophy \cite{Hobbes1651leviathan}.}
  \label{fig:DWA}
\end{figure}

\textbf{Will of an individual.} Thomas Hobbes describes an individual's will in the context of deliberation\footnote{Deliberation here equates to imagining the consequences of taking, or not taking, some action(s) and considering one's aversion or appetite for the anticipated outcomes.} as "\emph{the last appetite, or aversion, immediately adhering to the action, or to the omission thereof}" \cite{Hobbes1651leviathan}. In this context, one may view an individual's will as their final deliberate\footnote{ie. stemming from deliberation} judgment of preference for the future which determines their voluntary actions. For a single human, actions equate to converting the body's energy into forces that impact the future and a person's universe of potential actions is constrained by their body's energy production limits. 

However, things like technology and social systems enable the actions of individuals to be energetically amplified. A launch button connected to a rocket amplifies the small force of hitting a button into the massive force exerted by rocket engines. A digital social network can amplify the small force exerted by a leader making a statement into a massive force exerted by the collective actions of their followers. Thus, it is reasonable to view the domain of one's will as extending beyond the initial impact of their own actions to include the full range of possible futures. As such, we consider an \emph{individual's will} to be their complete set of final deliberate judgments \emph{across all possible futures}, which determine their voluntary actions in all scenarios. One way to conceptualize a person's will is as their deliberately considered preference or aversion for potential future states of the universe\footnote{Here we mean \emph{state} of the universe in a statistical mechanic sense, ie. a state describes the properties of all matter and energy in a universe.}. We denote the $i^{th}$ person's will at time $t$ as $w^i_t$.

\textbf{Will of humanity}. \emph{Humanity} refers to the totality of all human beings. Thus, a reasonable starting place is to consider the \emph {will of humanity} to simply be the combined set of all human being's individual wills. With this definition, the will of humanity at time $t$ equates to $w_t = \{w^1_t,..,w^i_t,..,w^N_t\}$. In other words, the \emph{will of humanity} is the set of every human's final deliberate preference judgments across all possible futures, which determine the voluntary actions of every human in all scenarios. There is an argument to be made that the \emph{will of humanity} is a collective will, and therefore should only include a subset of such judgments -- like those that are coherent, the output of collective deliberation, or the driver of collective action. However, we choose not to adopt these restrictions as a definitional starting point\footnote{It seems ethically dubious to exclude a component of a person's will \emph{as a definitional starting point} for the thing you hope to align the world's most powerful systems with.}.

\subsection{Representation}\label{woh_representation}

\textbf{Will matrix}. How can the \emph{will of humanity} be represented as a computable object one can make use of in practice? Let $w_t$ be a matrix where every row corresponds to a \emph{human}, and every column corresponds to an \emph{item} containing information related to characteristics of potential futures. For example, an item could be:

\begin{itemize}
    \item Text expressing a specific desire for the future, like "\emph{Keep global temperature under $15^{\circ}$C}"
    \item An answer to a question, like "\emph{What do you want for your children? To never starve}"
    \item A statement of a value, like a passage from a religious text "\emph{do not charge any soul except within its capability}" (Quran) or a philosopher's writing "\emph{it is good to set aside the letter of the law and to follow the dictates of justice and the common good}." (Thomas Aquinas)
    \item An ethical statement generated by an LLM, like "\emph{Do not harm animals unnecessarily.}"\footnote{Generated by ChatGTP in response to the prompt "\emph{what is a simple yet surprising ethical statement which shows up across many world religions?}"}
    \item A video introducing a vision for how the world could be at some point in the future
    \item A VR experience of a new design for a city
\end{itemize}

\begin{figure}[H]
\centering
  \includegraphics[width=0.9\linewidth]{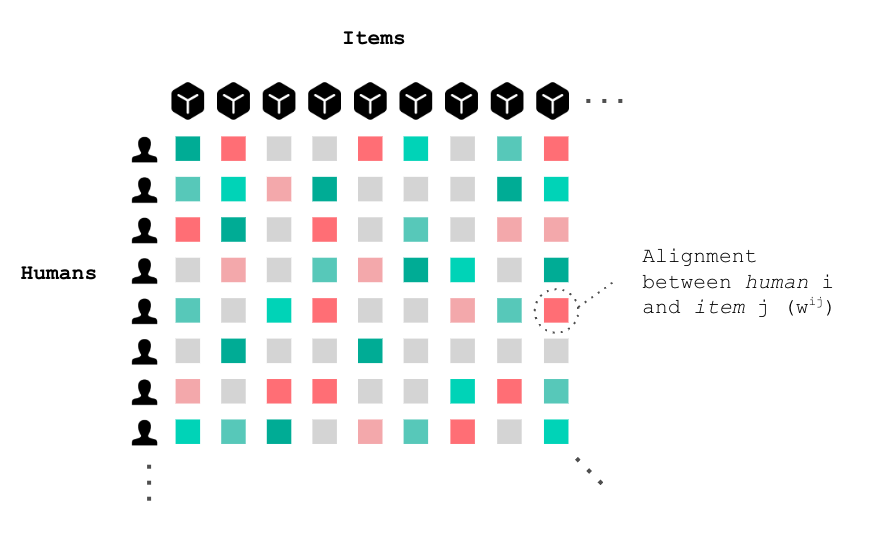}
  \caption{Diagram of a Will matrix where each row corresponds to a human, each column corresponds to an item, and each element represents the degree of alignment between a human and an item.}
  \label{fig:willmatrix}
\end{figure}

 Let every element $w^{ij}_t$ capture how well the $j^{th}$ \emph{item} aligns with the will of the $i^{th}$ \emph{human} as of time $t$ (figure \ref{fig:willmatrix}). We will refer to this type of encoding as a \emph{Will matrix}\footnote{The idea of a matrix representing relationships between humans and items of various types is common across many fields of study ranging from recommender systems and economics to social choice.}. We note that many other types of encodings may exist. For example, a theoretically elegant encoding is a matrix directly relating humans to specific states of the universe. But the type of item-human Will matrix laid out above was chosen for a few key features. 

 First, it is practically realizable -- as \emph{items} comprise real digital objects with many known approaches to collection, generation, storage, interaction, and analysis. Second, it confers open-ended generalizability -- there are virtually no restrictions on what constitutes an \emph{item} except that it should contain information relatable to characteristics of a future universe. Finally, it can be used to estimate alignment with states of the universe. For example, let $m^{jk}_{\tau}$ be the degree to which the $j^{th}$ \emph{item} aligns with the $k^{th}$ possible \emph{state} of the universe at time $\tau$. Then, the degree to which the $k^{th}$ state aligns with the $i^{th}$ human's will at time $t$ is simply $\sum_j w^{ij}_tm^{jk}_{\tau}$. Overall, the \emph{Will matrix} offers a robust practical representation that can be used for alignment.

\subsection{Properties}
\textbf{Constantly evolving}. As people are born and die the human dimension of the Will matrix changes \footnote{The will matrix adds humans as they are born. There could be a debate about whether it removes humans as they die. This equates to asking if the present will of humanity includes only living humans or not. This is related to the question of the role a person's will should play after they die. We see these as open questions, but for now, treat a current will of humanity $w_t$ as representing the will of all humans alive at time $t$.}. As a person learns and matures their values and preferences change, leading their individual wills to change. Further, as humanity's capacity to imagine the future expands, the scope of \emph{items} which may be encoded in a Will matrix expands as well. Overall, these dynamics mean both the Will matrix's dimensionality and the judgments its elements encode, are constantly changing. 

\textbf{Heterogeneous.} The diversity of experience, knowledge, cultures, values, and preferences across the human population is reflected in the diversity of individual's wills, and the diversity of \emph{items} a Will matrix may encode. This implies that no two humans are likely to have exactly the same individual will and no two \emph{items} are likely to reflect exactly the same judgments. In other words, no two rows or columns of a sufficiently high resolution Will matrix are likely to be identical. Whats more, it is likely that individual's wills are not only dissimilar, but that the may be in direct conflict with each other\footnote{Resolving the conflicts between individuals will is a first order challenge in mapping a heterogeneous will to unitary decisions.}. 

\textbf{Open-ended}. There are no clear upper bounds on the number of unique characteristics that can relate to future states in the \emph{items} of a Will matrix. There are also no clear upper bounds on the number of discrete future times a given characteristic may be related to in those \emph{items}\footnote{The Plank-time could potentially represent a limit on the temporal resolution of distinguishable future times. In a closed universe whose causal existence has a finite time horizon, this \emph{would} impose a theoretical limit on the total number of distinguishable future times.}. As a result, there are no clear upper bounds on the number of \emph{items} which may exist in a Will matrix\footnote{If one assumes that in a finite universe, the information content of each item may not exceed some finite number of bits, then there \emph{would} be a theoretical upper bound on the number of possible non-degenerate items.}. This means the Will matrix is not only high dimensional, but that its size is open-ended and tends towards infinity\footnote{There is an argument to be made that there could potentially be some upper bounds on the rank of a Will matrix, and thus there exists some encoding of a Will matrix which is finite.}. 

\textbf{Not fully knowable.} Human will can only be sensed through voluntary actions. The amount of voluntary actions a human can do in their lifetime is finite. The amount of voluntary actions all humans can do during some period of time is finite. This means the observable actions from which the will of humanity can be sensed are also finite. Since the dimensionality of the will of humanity tends towards infinity, while physical limits permit only finite observability, there is no clear way to know the complete will of humanity\footnote{It could be possible that the rank of the Will matrix can be shown to approach a finite asymptote even as its dimensionality tends to infinity. In that case, there is a chance that such an infinite Will matrix could theoretically be perfectly knowable from finite observations.}.

\subsection{Partitions}

The will of humanity can be partitioned along a range of different axes to highlight segments with contrasting properties. These different parts can present different challenges, and provide signals that are appropriate for different situations. By distinguishing them here, we aim to enable more precise discussion of the will of humanity over the course of this document. However, while the discussion below may use language suggesting discrete binary partitions, the segments of these partitions are generally best viewed as the endpoints along a continuous axis. 

\textbf{Collective vs. Individual.} There are no limits to the scope of the universe a portion of the will of humanity may relate to. Some portions may have an extremely limited scope and only relate to a part of the universe experienced by one person, such as the portion related to an \emph{item} like "\emph{Alice feels no back pain}." Conversely, other portions may have a broad scope that relates to parts of the universe involving large groups of people or even the full human population. For example, portions related to \emph{items} like "\emph{Americans can't work on Sundays}," or "\emph{Make Earth's average temperature be $25^o$C}". One might consider the collective portion of the will of humanity to simply be that which involves parts of the universe experienced by multiple people. However, we note that there exists long-standing philosophical debates about what constitutes collective will \cite{terrier2011collective}, and some argue it is not enough to just relate to a group of people, but that it needs to stem from collective intentionally \cite{schweikard2021collective} or be enabled by collective discourse \cite{smith1998collective}.

\textbf{Convergent vs. Divergent.}  While the will of humanity is heterogeneous in general, some parts converge towards consensus while others diverge towards polarization. The convergent parts correspond to segments of the Will matrix where human's individual wills are largely consistent across the population. For example, a set of items that all correspond to some version of "\emph{humans don't all die next year}" will likely align with individual wills across the entire population. In contrast, the divergent parts of the Will matrix correspond to portions where human's individual wills diverge, meaning they are inconsistent with each other. This divergence may be across segments of the human population, as one would expect for \emph{items} like "\emph{Nigeria has the largest economy in 10 years}" and "\emph{China has the largest economy in 10 years}." Alternatively, this divergence may be across specific individuals in the population, as one would expect for \emph{items} like "\emph{Alice marries Bob}" and "\emph{Bob marries Carol}", where Alice, Bob, and Carol may all have different and conflicting individual wills.

\textbf{Well-informed vs. Misinformed.} A person's will is informed by their current understanding of reality, but that understanding is not always correct. Some portion of a person's will can be the byproduct of an inaccurate understanding of reality. For example, if a person's understanding is that passing tax law ABC will increase their tax burden -- even though it would actually go down -- then their will may be aligned to an \emph{item} like "\emph{tax law ABC never passes}. But if they knew the truth, their will might be the opposite. In general, the misinformed part of the will of humanity is the part that would change if every human suddenly had a perfect understanding of reality. In contrast, the well-informed part is that which is based on an accurate understanding of reality, ie. it is the portion of the will of humanity that would remain unchanged if every human suddenly had a perfect understanding of reality.

\textbf{Instrumental vs Terminal.} The \emph{terminal} parts of a person's will correspond to characteristics about the future (ie. items) that they want in and of themselves, while the \emph{instrumental} parts correspond to characteristics that they believe are necessary conditions (or means) to realizing the terminal parts\footnote{The terminal vs instrumental distinction is found across work on values \cite{allen2002functional} and goals \cite{schwartz1990toward}.}. For example, a person's will may be aligned with the instrumental item "\emph{nuclear reactors should not exist}" because they believe nuclear reactors cause cancer, and thus see it as a means to the terminal item "\emph{humans should not die slow and painful deaths}." The instrumental part of ones will can therefore be thought of as a reflection of the combination of the terminal part of their will and their understanding of causal reality\footnote{This motivates a conceptualization of one's will as a causal graph where a network instrumental nodes (items) ultimately resolve into terminal nodes (items). We note that the Will matrix described above is likely not sufficient to directly represent this aspect of one's will, and a more sophisticated representation may ultimately be needed, eg. a directed graph, perhaps represented as another dimension within each element of a Will matrix.}. This highlights the fact that the instrumental part of ones will is most susceptible to being misinformed as it is dependant on a potentially fallible understanding of reality.

\textbf{Value-driven vs. Desired.} There are no underlying limits on what may motivate the considered preferences for some futures over others which constitutes one's will. Some portion of a person's will may be motivated by the ethical values they adhere to. For example, the degree a person's will aligns with an \emph{item} like "\emph{Humans no longer eat animals}" is likely informed by their ethical values towards animals. Conversely, another portion of a person's will may be motivated by their cravings and desires. For example, the degree a person's will aligns with \emph{items} like "\emph{Make tiramisu widely available}" is likely informed by their personal cravings for tiramisu rather than core ethical values. It is also possible for portions of human will to exist in a gray zone, informed by both values and desires. For example, the degree a person's will aligns with an \emph{item} like "\emph{Produce more Wagyu steak}" may be informed by both their desire for juicy Wagyu steak and their ethical values towards animals.

\textbf{Observed vs. Extrapolated.} The dimensionality of the will of humanity tends towards infinity yet physical limits permit only finite observability. This means that only a portion of the will of humanity can be observed directly from people's voluntary actions. However, some portions of the unobserved part may be extrapolated from available information. For example, if a person's will is observed to align with the \emph{items} "\emph{Dogs are man's best friend}" and "\emph{People's friends should not be harmed" }" then it could be extrapolated that the \emph{item} "\emph{Dogs should not be harmed}" is likely to align with their will as well. Extrapolation to existing \emph{items} has already been demonstrated using collaborative filtering techniques \cite{konya2022elicitation}, and extrapolating from existing \emph{items} to generate completely new \emph{items} has already been demonstrated using LLMs \cite{bakker2022finetuning}. Beyond these types of extrapolation, Yudkowski \cite{yudkowsky2004coherent} introduces a more nuanced type which treats extrapolation as a type of error correction that figures out what a person's will (volition) would be if they perfectly understood reality. In this sense, Yudkowski's extrapolation can be viewed as transforming an observed will matrix with misinformed parts, into a will matrix of equal or larger size with no misinformed parts.

\textbf{Stable vs. Quickly changing.} While the entire will of humanity is constantly changing, some parts will be more stable than others. The stable parts correspond to those which do not tend to vary quickly over time. The most long-term stable parts are those where alignment tends to remain unchanged even from one generation to the next, for example, related to an \emph{item} like "\emph{Killing humans is bad}." Medium-term stable parts are those where individual's wills tend to remain unchanged over the course of their lifetime, for example, related to an \emph{item} like "\emph{The Republican party should run America}." The parts that change more quickly are those where individual's wills are motivated by fluctuating personal desires, dynamic realities, or an evolving understanding of reality. For example, related to \emph{items} like "\emph{Make enough cargo pants for every human }", or "\emph{The current government should stay in power}", or "\emph{fracking should be illegal.}" In general, the stability of some subset of the will of humanity is closely related to the average magnitude of its time derivative\footnote{For a period of time from $\tau$ to $\tau + \Delta$, one potential estimate for the stability of a subset $S$ of the will of humanity is the average magnitude of the rate of change of each Will matrix element comprising $S$: $\frac{1}{Z}\sum\limits_{T=\tau}^{\tau + \Delta}\sum\limits_{i,j\in S} | \frac{\partial w_t^{ij}}{\partial t}|_{t=T}$ where $Z=\sum\limits_{i,j\in S}\Delta$ is a normalizing constant over time and elements.}.

\subsection{Sensing} \label{WOH.sensing}
\textbf{Mechanics.} A person's will is only reflected in their deliberate voluntary actions. So sensing human will requires creating situations where people take observable voluntary actions. But how do you ensure those actions generate information that can be mapped to elements of the Will matrix? One approach is to have people interact directly with \emph{items} in the will matrix with an understanding of how those interactions may impact the future. Ideally, where people know how each potential action impacts the likelihood that a given \emph{item's} characteristics manifest in the future. For example, if participants understand that clicking button A next to an \emph{item} increases the chance the future manifests that \emph{item's} characteristics, while clicking button B decreases that chance, and clicking button C has no impact, then those clicks give direct information about the alignment between the clickers will and that \emph{item}. The central mechanism underlying this type of approach is the enablement of observable voluntary actions, not that just ask people what they want in the future, but that directly empower them to impact the future in a mutually understood way. Crucially, this mechanism makes it possible to map observed actions directly to the will that motivated them so the Will matrix can be updated appropriately. Overall, \emph{sensing human will requires observing deliberate voluntary actions which have a mutually understood impact on the future}.

\textbf{Examples.} \emph{Democratic elections} sense human will by enabling deliberate voluntary actions in the form of votes which are mutually understood to yield a specific impact on the future as laid out in governing documents like a constitution. Deliberative technology like \emph{collective response systems} \cite{ovadya2023generative} can be used to sense human will if participation is voluntary, votes represent deliberate actions, and there is some mutually understood impact that votes have on the future; for example, how Polis was used in Taiwan \cite{small2021polis} to impact rideshare policy through vTaiwan and how Remesh was used in Libya \cite{irwin2021using} to impact conflict resolution in support of UN peacemaking efforts. \footnote{Further discussion of these examples can be found in the section on deliberative technology.}. \emph{Markets} may also be viewed as a type of will-sensing mechanism where each purchase represents an action that is mutually understood to increase the chance of a future where the purchaser has the purchased item. However, not all purchases are deliberate voluntary actions, and even if one can isolate the deliberate voluntary purchases within a market, the only portion of the will matrix that can be directly sensed are items of the form \emph{"person Y owns X"} or \emph{"service X is available to person Y."}

\textbf{Physical limits.} Sensing is only possible through voluntary actions. Even the fastest voluntary actions -- like clicking a button -- take at least a few seconds to deliberate on and execute. This creates a physical limit on the total amount of voluntary actions which are possible; both over an individual's lifetime, and across the human population during a finite period. Further, if observing and using the information from voluntary actions requires software and the internet, then the distribution of access to digital devices and internet connectivity creates a physical limit to sensing as well. What's more, since connected digital devices require power to be used, the global distribution of power creates an additional physical limit to sensing. To put these limits in perspective, consider a scenario where each human contributes only a single unique \emph{item} to the Will matrix. Let's assume every human has a digital device connected to the internet that they know how to use (which isn't true). Let's also assume there is a deliberate action that can be done in 10 seconds which senses the alignment between any single \emph{item} and any single \emph{human} (which is optimistic). Then, if every human on earth spends $100\%$ of their time doing these deliberate actions for one year (not realistic), only $0.04\%$ of the Will matrix would be sensed. And to sense the entire matrix would take over 30 lifetimes; even with wildly optimistic assumptions.

\textbf{Prioritization.}  Since the Will of humanity cannot be fully sensed, any sensing effort involves a choice, either implicitly or explicitly, as to what portions of a Will matrix a finite budget of human attention is allocated towards sensing. So what portions should be prioritized? The answer depends on how the signal will be used. More specifically, if it is used to inform the actions of a system, then it depends on the scope of the system's action space and the speed and scale with which actions can be executed. For example, the legislative body of a government generally takes actions in the form of creating lasting policies with collective impact, and it is practically limited to creating only a few policies at a time. So when sensing a Will matrix for this use case, one might prioritize sensing the parts that are collective, convergent, stable, and focused on issues government policies can address. In contrast, a powerful AI agent may be capable of taking a wide range of actions, rapidly, at a massive scale. These actions might range from those which have a short-term impact on an individual to those which have a long-term impact on humanity. And it may be able to take, for example, millions of actions per minute. In this case both the collective and individual parts of the will matrix are relevant, as well as both the stable and quickly changing parts. Further, while the convergent parts may help identify a set of well-aligned actions, it does not give information about what actions are misaligned, and thus the divergent parts of the will matrix which contain this information may be important as well\footnote{Convergent parts of the will matrix give examples of items within the space 'well-aligned things' but do not tell you where the boundary between 'well-aligned things' and 'misaligned things' actually is. It is the combination of both the convergent and divergent parts which define the boundary.}. Overall, in this case, the priority may be to maximize the diversity of items sensed so that the Will matrix contains alignment information spanning the widest set of possible actions. For similar reasons, if a will matrix is being sensed for use across a wide range of systems, eg. governments, firms, and AI agents, then a prioritization of maximum diversity may make sense as well. Finally, in all cases, prioritizing the \emph{well-informed} and \emph{terminal} portions likely makes sense.

\subsection{Power}

Here we aim to provide a physically grounded way to conceptualize the impact of the will of humanity in terms of energy and power. While such physical quantities do not necessarily capture all notions of impact, we believe they at least provide useful proxies grounded in physical reality that can be reasonably estimated. 

\textbf{Will power}. Power is a physical quantity corresponding to a rate of energy transfer or consumption per unit time, typically in units of watts. But how does the philosophical \emph{will of humanity} relate to physical \emph{power}? Will determines actions, and actions convert energy at some rate into forces that impact the future. In this way, \emph{will} controls the flow of power. We refer to \emph{will power} as the amount of power consumed or transferred \emph{in accordance with human will}. We can then ask, what is the \emph{will power} of humanity? An upper bound is given by humanity's total \emph{power budget} which is on the order of 20 terawatts\footnote{Assuming around 600 EJ per year of global energy supply.}. This would be humanity's \emph{will power} if its entire \emph{power budget} went towards actions that aligned with its will. An estimate on the lower side is humanity's \emph{body power} which is the total power consumed by all physical bodies comprising humanity; around 1 terawatt\footnote{Assuming each human body consumes around 4GJ per year and on the order of 10 billion humans.}. This would be humanity's \emph{will power} if $100\%$ of every human's physical actions were aligned with the will of humanity, but none of the rest of humanity's energy consumption was. Overall, humanity's total \emph{will power} likely sits between 1 and 20 terawatts at present; we estimate it to be around 5 terawatts \ref{A: will power est}.

\textbf{Will power abundance}. For early humans, their only source of energy was food, and thus their \emph{power budget} was equivalent to their \emph{body power}. As a result, their maximum \emph{will power} could not exceed their \emph{body power}. However, modern technology enables a \emph{power budget} far beyond humanity's \emph{body power}, and with it, the possibility of achieving a \emph{will power} per capita beyond even the upper bounds of our early ancestors. We quantify this \emph{will power abundance} as the ratio of \emph{will power} to \emph{body power}, ie \emph{will power abundance} = \emph{will power} / \emph{body power} $\in(0,\infty]$. While the \emph{will power abundance} of early humans was certainly less than one, there are no clear upper bounds on how large it can be in the future, and even today it may be possible to achieve 10 or more\footnote{Achieving will power abundance of 10 is possible if $50\%$ of humanity's current 20 TW power budget was aligned with humanity's will.}.

\textbf{Will power alignment}. How well are we doing at aligning the future with the will of humanity?  One way to quantify this is by how much of humanity's \emph{power budget} is aligned with human will and thus becomes \emph{will power}. We denote this ratio as \emph{will power alignment} = \emph{will power} / \emph{power budget} $\in ( 0,1 )$. One potential objective for work on alignment systems to aim for is to increase humanity's \emph{will power alignment}\footnote{Why not just focus on increasing \emph{will power} as the objective? Aiming to only increase \emph{will power} is neutral to the portion of the \emph{power budget} which is not aligned to human will. Yet, this not-aligned portion, by definition, may act against aligning the future with human will. And if it grows faster than the portion that becomes \emph{will power} then the chance that the future aligns with the will of humanity may certainly decrease. Thus, any objective which treats this portion with neutrality is ill-formed.}. 
Not only does this objective connect a philosophically grounded goal to a physical quantity, but it can also inform simple trade-off decisions. For example, consider the choice to either 
a) take $X\%$ of humanity's \emph{power budget} which is not aligned to human will and make it aligned (ie. make a dictatorship a democracy) or, 
b) increase humanity's \emph{power budget} by $Y\%$ and align the newly added power to humanity's will (ie. give a global democracy another power generator). If the goal is to increase humanity's \emph{will power alignment} then its best to do (a) if $Y<X/(1-X-\alpha)$ where $\alpha$ is the \emph{will power alignment} prior to any change.

\subsection{Related Ideas}
The \emph{will of humanity} and the general idea of \emph{will} described here encompasses, or is closely related to, ideas that go by different names depending on the field of study. 

\textbf{Preferences.} Work spanning AI alignment \cite{ziegler2020finetuning}, behavioral economics \cite{malerba2007demand}, social choice theory \cite{list2013social}, and deliberative democracy \cite{bohman1998survey} use the term \emph{preferences} to describe peoples general desires for some things over other things. One's will can in general be viewed as a special subset of one's preferences. Specifically, preferences that a) result from deliberation, b) are about the future, and c) motivate one's deliberate voluntary actions. Importantly, this means the types of preferences that only motivate a person's involuntary actions (like preferences between posts observed when mindlessly swiping through a social network feed) do not constitute their will. In other words, ones will only includes the preferences which are strongly enough held to overcome the activation energy needed to take a deliberate voluntary action. We note that some of the deliberative technology discussed in this document acts to lower the activation energy needed for deliberate actions, and thus expands the set of preferences that meet this criteria. Indeed, we view these technologies as acting to expand human will. Further, one could even imagine an idealized scenario where the activation energy tends to zero, and thus all deliberate preferences about the future might be considered one's will. And this begs the question, why define and focus on aligning with \emph{will} specifically, instead of just focusing on all deliberate preferences about the future? First, we view the expansion of \emph{will} enabled by technology as an important dynamic worth distinguishing. Second, we see \emph{will} as directly related to agency, and expanding humanity's agency is a fundamental and motivating principle of this work. Third, we simply find the idea of aligning the future with the \emph{will of humanity} as being a more inspiring proposition than aligning the future with \emph{humanity's deliberate preferences about the future.}

\textbf{Interests.} Work on conflict resolution often refers to things like the "needs, desires, concerns, fears — the things one cares about or wants" as \emph{interests} \cite{marcus2012walk}. In so much as one's interests represent the things one cares about which influence their actions, then one's interests can be viewed as both motivating and encompassing one's will. More specifically, conflict resolution practitioners work to discover \emph{underlying interests} \cite{kriesberg1991conflict}, which can be viewed as the interests which ultimately motivate all others, and the ones that must be ultimately addressed in any agreement which resolves conflict. Such underlying interests can be viewed as existing within the terminal parts of one's will. 

\textbf{Values.} Work on AI alignment often uses the term \emph{human values} \cite{hendrycks2020aligning,van2020embedding} to refer to an amalgamation of things spanning ethical judgments, norms, and preferred virtues. Similar to interests, we view values as both motivating and encompassing one's will. Specifically, we view one's values as being reflected in the values-driven portion of their will. But we note that there exist portions of one's will that may not be values-driven, thus we can't simply equate one's will to their values.

\textbf{Volition.} Finally, work on friendly AI uses the term \emph{volition} to describe the want for one outcome over another which underlies a person's decisions \cite{yudkowsky2016ai}. We view this definition of volition as being basically synonymous with the definition of will introduced here.

\newpage

\section{Alignment Systems}\label{alignmentSystems}

\subsection{Framing}\label{AS.framing}

Early work on AI alignment sought to ensure AI didn't do "bad things" or cause harm by focusing on how to make an AI's actions align with a human's instructions, intent, or revealed preferences \cite{yudkowsky2016ai,gabriel2020artificial}. In recent years, however, work on AI alignment has begun to recognize the importance of aligning, not just with one human, but with a diverse population of humans \cite{iason2020artificial,bakker2022finetuning}. Even so, rather than tackling this scenario by accommodating the heterogeneous will of a diverse population, the primary focus has been on aligning with things like "human values" which are assumed to be universally representative of humanity \cite{hendrycks2020aligning,van2020embedding}. In contrast, addressing the unique multi-principal-agent\footnote{Multi-principal-agent scenarios broadly involves aligning the impact of a principal's actions with the will or interests of a population of principals.} challenges which emerge from trying to align a powerful institution (agent) with the heterogeneous and often conflicting will of a diverse population (principals) has long been the focus of work spanning economics \cite{liscow2022democratizing,alesina2005ethnic,malerba2007demand}, social choice theory \cite{condorcet1785essai,list2013social,dryzek2003social,sen1999possibility}, deliberative democracy \cite{bohman1998survey,chambers2003deliberative,gutmann2004deliberative}, and peacebuilding \cite{zalazberg2022negotiations,spolaore2017political,doyle2000international}.

In order to build a bridge between the extensive work on multi-principal-agent scenarios involving institutions, and the nascent challenge of multi-principal-agent alignment involving AI, we seek to frame alignment in a way that is universal; agnostic to the specifics of the agent involved. We thus frame alignment as between the future of the universe and the will of humanity. This frame views \emph{any system} which impacts the future as a \emph{means} to improve alignment. We refer to any system that aspires to take actions to align the future of the universe with the will of humanity as an \emph{alignment system}.  This includes \emph{local alignment systems} which aim to align part of the future with a subset of humanity's will -- like a local democratic government, as well as \emph{global alignment systems} which aim to align the future with the \emph{full will of humanity}\footnote{A global alignment system may exist as a monolith or employ a host of local alignment systems which span the human population.} -- like the United Nations. In this frame, one may view anything from governments to AI agents as potential alignment systems. This makes it possible to consider the universal components underlying any alignment system, and enables learnings derived from one type of alignment system to be transferred to a different type \cite{aiobjectives2023,hadfield2019incomplete}; ie. from democratic institutions to AGI.

\subsection{Components} \label{AS.components}

What components and capabilities underlie a system aiming to actively align the future with the will of humanity? To elucidate this, we formalize the challenge in the spirit of optimal control theory. Let $x_t$ be a possible state of the universe at time $t$. Let $w_t = W[x_t]$ be a signal encoding the \emph{will of humanity} (WoH) at time $t$ obtained by some sensing function $W$. Let $\phi(x,w)$ be the degree of alignment between a state of the universe and a given WoH signal. Let $a_t \in \Gamma (x_t)$ be an action at time $t$ from the set of possible actions given the state of the universe.  Let $P(x_\tau|x_t, a_t)$ be the probability the universe is in state $x$ at time $\tau$, given a state of the universe and action initiated at time $t$. The expectation value for alignment between the universe at time $\tau$ and the will of humanity at time $t$, given a state action pair at time $t$ is:

\begin{equation}
    E[\phi](\tau,a_t,x_t) = \sum\limits_{i} P(x^i_\tau |x_t,a_t) \phi(x^i_\tau,W[x_t])
\end{equation}
where $\sum_{i}$ represents a sum over all possible states of the universe. A simple example of an alignment-promoting value function\footnote{In \ref{A:flexible value} we discuss a more philosophically flexible value function, of which the one presented here represents a limiting case.} for a state action pair at time $t$ is the net alignment expectation over an infinite time horizon:

\begin{equation} \label{eq:simple value}
   V(a_t,x_t) = \sum \limits_{\tau=t+1}^{\infty}  E[\phi](\tau,a_t,x_t)
\end{equation}
Putting it all together, an optimal action at time $t$ to align the future with the will of humanity can be given by the policy:

\begin{equation} \label{eq:optimal action}
   a^*_t = \argmax\limits_{a_t \in \Gamma (x_t)} \sum \limits_{\tau=t+1}^{\infty} \sum\limits_{i}  P(x^i_\tau |x_t,a_t) \; \phi(x^i_\tau,w_t)
\end{equation}

For any real-world system, only approximations of the components comprising equation (\ref{eq:optimal action}) are plausible, and how well the system can promote alignment depends on how close those approximations are to reality. In other words, the components of equation (\ref{eq:optimal action}) elucidate the capabilities a system aiming to actively align the future with the will of humanity is likely to depend on:
\begin{itemize}
    \item \emph{Sensing} the will of humanity -- $w_t$
    \item \emph{Identifying} the universe of possible actions -- $\Gamma (x)$
    \item \emph{Predicting} how those actions affect the future -- $P(x_\tau |x_t,a_t)$
    \item \emph{Assessing} how those futures align with the will of humanity -- $\phi(x,w)$
    \item \emph{Executing} chosen actions -- $a^*_t$
\end{itemize}
These capabilities already manifest to some degree across a wide range of systems aspiring to realize some population's will -- from governments and NGOs to AI systems.

\begin{figure}[H]
\hspace{-1cm}
  \includegraphics[width=1.1\linewidth]{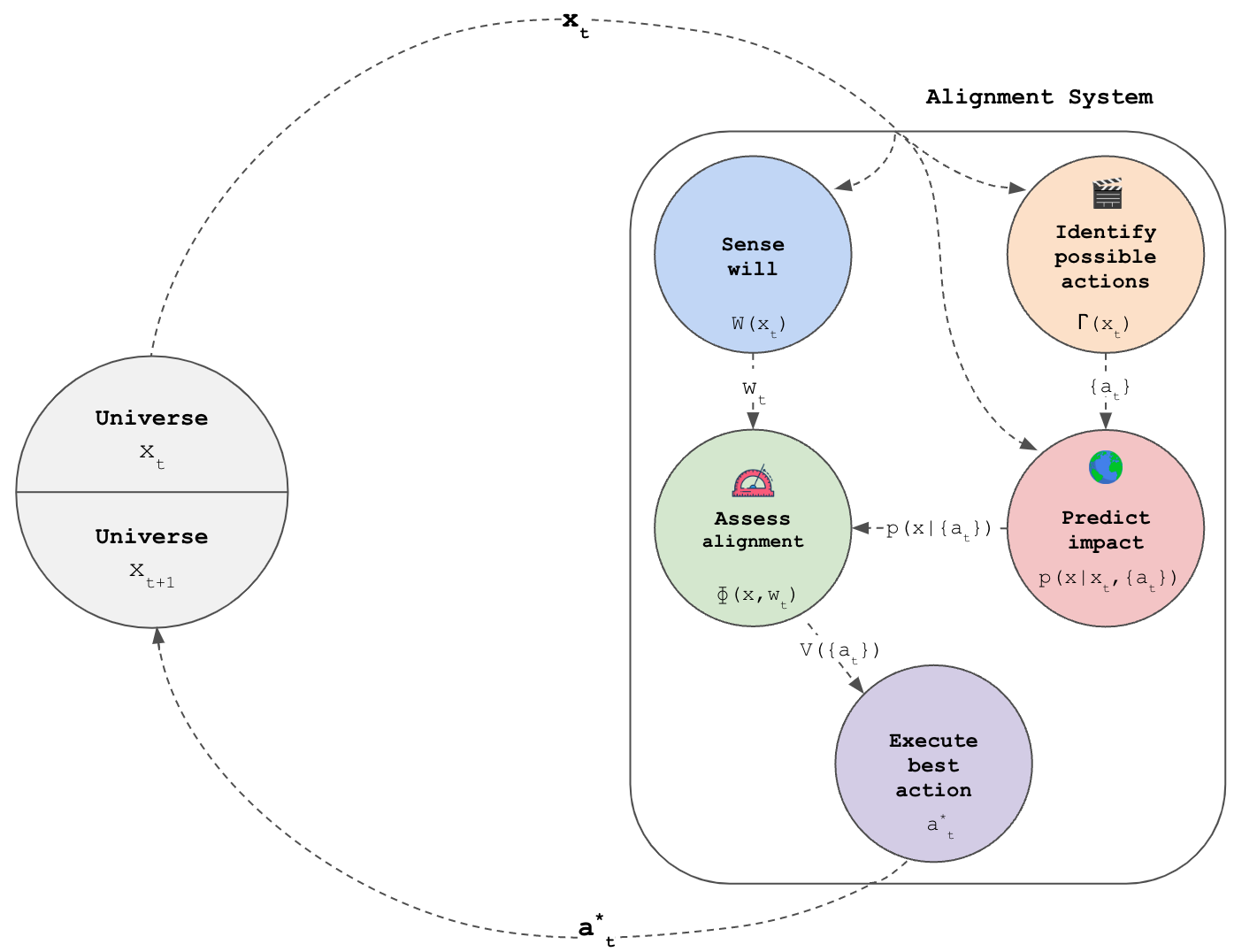}
  \caption{Diagram of an alignment system showing how components relate to each other and the universe. Given the current state of the universe, possible actions are identified, their impacts are predicted, and will is sensed. Then the alignment between predicted impacts and sensed will is assessed.  Finally, the best action, whose predicted impact best aligns with the sensed will, is executed.}
  \label{fig:alignment system}
\end{figure}

\subsection{Examples} \label{AS.examples}
\textbf{Democratic governments} \emph{sense} the will of their constituencies directly through voting and indirectly through polling. Their legislative body then \emph{identifies} a range of legislative actions they can take, commissions studies to \emph{predict} how those actions may affect the future, and \emph{assess} how well those predicted futures align to the sensed will of their constituency. Finally, they choose the best legislative actions, and an executive body \emph{executes} them under the supervision of a judicial body.

\textbf{The United Nations} \emph{senses} the will of its member nations through ongoing interaction with diplomatic representatives in the General Assembly, and targeted dialogue with local populations. UN organs like the Economic and Social Council then \emph{identify} a range of policies and programs the UN can enact, commission studies to \emph{predict} how those actions may affect the future, and make recommendations by \emph{assessing} how well those predicted futures align to the sensed will of its member nations. Finally, the General Assembly and Security Council make final decisions on which actions to take and work with the Secretariat to \emph{execute} those actions \cite{UNcharter}. 

\textbf{AI systems using RLHF} (Reinforcement Learning from Human Feedback) \emph{sense} the will of a group of people by eliciting their feedback on outputs from an initially trained model and then encoding that feedback in a reward model \cite{ziegler2020finetuning}. This can be viewed as sensing people's will towards futures that do or don't include an AI model that generates specific outputs. The range of potential actions is \emph{identified} or chosen by the AI system creator, and in the case of a large language model (LLM) corresponds to all of the tokens comprising the language model's vocabulary. Since the sensed will only involves the types of model outputs that are preferred, and the action space only involves generating model outputs, \emph{predicting} how actions affect the future and \emph{assessing} how those futures align with the sensed will collapse to a single task\footnote{In contrast to the examples of institutional alignment systems, an AI system using RLHF does not consider the causal impact of actions beyond the scope of the action itself. While this restriction simplifies technical implementation, it represents a key area for improvement in next-generation systems.}; the reward model predicts how well model outputs align with sensed will. Finally, a new model is trained to \emph{execute} the appropriate actions (ie. generate the appropriate tokens as output) through reinforcement learning which penalizes outputs that are misaligned with the sensed will encoded in the reward model, or deviate from the initially trained model.

\textbf{Corporations} 
\emph{sense} the will of the consumers via market signals -- ie. how many units of different products are purchased at what price --  and through market research. Decision makers in the firm then \emph{identify} what actions the firm can potentially take. Analysts execute or commission studies to \emph{predict} the impacts of those actions and \emph{assess} the alignment of those impacts with consumer will. Finally, decision-makers and the teams they lead \emph{execute} on the actions they believe best align with consumer will. A core assumption of corporations acting as alignment systems in this way is the belief that taking actions that best align with consumer will maximizes profits. This belief likely holds true in at least two ways. First, certain actions determine what products are produced, and aligning those actions with consumer will likely lead to products that more consumers will purchase, or will purchase at a higher price. Second, certain actions determine publicly visible corporate values, and consumers may be more likely to purchase from firms whose values best align with their own. 

\subsection{Related Ideas}

\textbf{Reinforcement learning.} Just like the model of an alignment system presented here, reinforcement learning is also formulated in terms of a policy that seeks to maximize expected reward as defined by a value function \cite{kaelbling1996reinforcement}. One may also draw an analogy between a theoretically ideal alignment system and an AIXI \cite{hutter2003gentle}, which is a reinforcement learning agent that learns an optimal program to simulate its environment as a means to maximize future rewards. However, in contrast with reinforcement learning, our framework for an alignment system does not focus on the dynamics of how to learn an optimal policy, rather it decomposes the dynamics of an optimal policy into constituent parts that can be understood independently in the context of alignment.

\textbf{Open Agency Model.} The Open Agency Model \cite{drexler2023open} re-frames an AI-based system for taking consequential actions as a set of constituent components that humans can influence and have oversight over (in contrast with a unitary agent whose decision-making process is a black box). These constituent parts involve \emph{prompt-driven generative models which produce diverse proposals, diverse critics which help select proposals, and diverse agents which implement proposed actions to accomplish tasks}. Like the Open Agency Model, we frame alignment systems in terms of a transparent set of constituent parts that interact to produce decisions and actions. Similar to the Open Agency Model, these parts include components that generate potential actions (proposals), evaluate those actions, and then execute on chosen actions. Further, in both cases, each component offers the opportunity for human oversight. However, in contrast with the Open Agency Model, an alignment system is explicitly constructed to find and execute actions that align the future with the will of humanity, while an Open Agency Model leaves the definition of optimal actions a bit more ambiguous.

\newpage

\section{Deliberative Alignment} \label{DA}

We define \emph{deliberative alignment systems} to be any alignment system that uses deliberative technology, and \emph{deliberative alignment} to broadly refer to the study, creation, or utilization of such systems. 

\subsection{Deliberation}\label{DA.deliberation}

\emph{Deliberation} is a loaded term with different definitions depending on the context it is used in and who is using it. While there are many small differences between various definitions, the largest differences stem from those who view deliberation as a collective process vs. as an individual process.

\textbf{Individual deliberation}. Deliberation for an individual involves the thoughtful consideration of options and outcomes which proceeds decision \cite{Hobbes1651leviathan}. During deliberation, a person may consider their options and update their preferences based on introspection or new information. A deliberating individual may seek out new information from reference sources like a book or from human sources like a topical expert. Deliberation can be viewed as the period where one's preferences are dynamic and don't yet lead to action, and in this context one's \emph{will} equates to a final snapshot of those evolving preferences that motivates an action or inaction (Fig. \ref{fig:DWA}). 

\textbf{Collective deliberation}. Deliberation may be said to be \emph{collective} when a) it involves more than one person, b) the decision it proceeds is a collective one, and c) when individual's perspectives are allowed to evolve through discourse with others. Deliberative democrats take it one step further and require that the people deliberating are the people who will be impacted by the resulting decision \cite{bohman1997deliberative}. Overall, one can view deliberation as existing on a spectrum ranging from \emph{individual} to \emph{collective} depending on how many people are involved, how they interact, and how they relate to the decision being deliberated. 

\textbf{Evolving preferences.} A core feature of deliberation is enabling preference evolution\footnote{The constraint of static preferences give rise to challenges like the nihilism of Arrows impossibility theorem and multi-polar gridlock between different sides of an issue \cite{riker1982liberalism}. Deliberation relaxes this constraint by facilitating preference evolution.}, ideally, as people's understanding of reality is improved \cite{mansbridge2018deliberative}. In collective deliberation, this improved understanding can take the form of learning the preferences and underlying interests of others. This gives rise to an iterative dynamic where learning the perspectives of others leads to new perspectives which can then be learned by others, and so on. Deliberation mediators enhance the dynamic by clarifying and highlighting certain perspectives and guiding the discourse in ways that surface new ideas, dig deeper into key issues, and promote civil discourse. The evolution of preferences facilitated by this iterative dynamic enables the discovery of positive-sum decision options that were not necessarily perceivable without the deliberation process.

\textbf{Implications of scale.} In a small scale deliberation\footnote{Deliberation with on the order of ten participants.} it is possible for all participants to learn about, and provide their preferences related to, the perspectives of all other participants. It's also possible for moderators to be aware of all perspectives as they emerge so they can identify ones to dig deeper into and promote healthy discourse. However, these affordances break down as the number of participants in the deliberation scales because the number of perspectives in a deliberation grows with the number of participants. This means that even with a few hundred participants it would take infeasibly long for a participant to learn about, and provide their preferences related to, all perspectives. It would also become infeasible for a moderator to be aware of all perspectives so they can identify the best ones to highlight or dig deeper into. As a result, the iterative dynamic of deliberation tends to grind to a halt as scale increases. This is why most important deliberative processes have historically involved a small number of people who purport to represent large populations, rather than involving the population themselves. However, such limited inclusivity can harm the legitimacy of the deliberation process and lead to decisions that do not reflect public will. As a result, accommodating these scaling implications is a key goal of many modern deliberative technologies.

\subsection{Deliberative Mechanics} \label{DA.deliberative_mechanics}
The mechanics of a deliberative process can be understood in terms of a) how information is generated, routed, consumed, and transformed, b) how participant's attention is allocated, and c) how participants' internal preference states evolve. 

\textbf{Information types.} We factor the types of information that are generated, routed, consumed, and transformed as part of a deliberation process into four types:

\begin{itemize}
    \item \emph{Prompts} -- a stimulus, often in the form of a natural language question or statement, which is used to elicit perspectives. 
    \item \emph{Perspectives} -- open-ended manifestations of a participant's point of view, often in the form of a natural language statement, elicited in the context of a prompt. 
    \item \emph{Revealed preferences} -- representations of participants' alignment with different perspectives. 
    \item \emph{Distillations} -- consumable summaries that represent preferences and perspectives aggregated across some or all participants. 
    \item \emph{External knowledge} -- facts and other information not generated by the participants themselves. 

\end{itemize}

\textbf{Participant operations.} Participants' internal preference states embody their true alignment with perspectives, including unstated perspectives, independent of their revelation. These internal preference states evolve as a participant's understanding evolves, and their understanding may evolve as a result of consuming new information during a deliberation process. Further, the perspectives a participant contributes and the preferences they reveal are a reflection of their internal preference states. This gives rise to a set of participant operations that either evolve or reflect a participant's internal preference state:
\begin{itemize}
    \item \emph{Perspective generation} -- sharing a perspective in response to a prompt that reflects a portion of their internal preference state (or a personal experience that informs it).
     \item \emph{Perspective consumption} -- consuming a perspective, and potentially updating their internal preference state as a result.\footnote{Perspective consumption is a requirement for perspective evaluation, but perspective evaluation is not a requirement for perspective consumption.}
    \item \emph{Perspective evaluation} -- evaluating a perspective and generating a revealed preference which reflects their internal preference state related to that perspective. 
    \item \emph{Distillation consumption} -- consuming a distillation, and potentially updating their internal preference state as a result.
    \item \emph{Knowledge consumption} -- consuming external knowledge, and potentially updating their internal preference state as a result.
\end{itemize}

\textbf{Process operations} enable and govern how participant operations take place over the course of a deliberative process. These process operations include:

\begin{itemize}
    \item \emph{Perspective routing} -- determining which perspectives are exposed to which participants for consumption. Governs \emph{perspective consumption}. 
    \item \emph{Perspective elicitation} -- determining which participants' perspectives will be elicited from and what prompt(s) will be used. Governs \emph{perspective generation}. 
    \item \emph{Evaluation elicitation} -- determining which participants to elicit preferences from on which perspectives. Governs \emph{perspective evaluation}.
    \item \emph{Knowledge routing} -- determining which external knowledge is shown to or made available to which participants for consumption. Governs \emph{knowledge consumption}.
    \item \emph{Distillation generation} -- determining how and when a distillation is generated using the available set of perspectives and revealed preferences. Enables \emph{distillation consumption}.
    \item \emph{Distillation routing} -- determining which distillations are shown to which participants for consumption. Governs \emph{distillation consumption}.
    
\end{itemize}

\textbf{Attention allocation.} A finite resource during any form of deliberation is the time and attention of participants. Each \emph{participant operation} consumes participant attention, and each \emph{process operation} determines how participants' attention is allocated across a specific type of participant operation. However, all process operations draw down from the same participant attention budget. This means an attention allocation operation exists, either implicitly or explicitly, which allocates attention across the various process operations.

\textbf{Agenda setting.} Prior to a deliberative process taking place, the agenda for that deliberative process -- ie. the issue(s) it will focus on -- needs to be set. The agenda-setting operation can itself be a small deliberative process, where a small group of people with the influence and mandate to initiate a large deliberative process deliberate amongst themselves to generate a \emph{distillation} in the form of an agenda. The resulting agenda, while a distillation in the context of the initial small deliberative process, can be viewed as a type of \emph{external knowledge} which is routed to participants within the corresponding large deliberative process\footnote{This highlights the fact that deliberation can involve multiple deliberative processes occurring on different scales and timelines.}.

\subsection{Deliberative Technology}\label{DA.deliberativeTechnology}

We use the term \emph{deliberative technology} to refer to a range of tools and processes that help enable more collectively intelligent deliberation and make it practical at scale. Deliberative technologies can be understood in terms of how they approach \emph{process operations}, the mechanics of their \emph{participant operations}, and the constraints they place on the \emph{information types} involved. The full space of deliberative technology ranges from hands-on processes where a small team manually carries out \emph{process operations} in-person with a small group of participants, to sophisticated software which automates \emph{process operations} and digitize participation to enable deliberation at massive scale\footnote{Some deliberative technologies are built to support deliberation with a specific objective like deciding an ideal layout for a new city, while others are built in a way that supports a general deliberative process itself. Some deliberative technologies can even be used for processes that would not meet the criteria of deliberation presented here.}. Here we provide an overview of the existing types of deliberative technology already in use. We note that while this overview covers a wide range of today's deliberative technologies, it is still only a sampling of what exists and should not be viewed as fully comprehensive.

\subsubsection{Juries}
Early deliberative technologies were repeatable processes that could be executed in person with a group of people. In the late 1700's Condorcet showed how a small jury could reliably make collectively intelligent decisions \cite{de2014essai} and citizen juries are still used today to make life-altering decisions\footnote{For example, in the United States where there is still capital punishments citizen juries make decisions which directly determine whether a person should be executed.}. In a typical courtroom jury, knowledge routing is governed by a judicial process, guided by a judge, and influenced by lawyers. Perspective generation and consumption takes place organically via verbal discussions behind closed doors with all jurors present. Because all jurors are in the room during deliberation, every perspective is effectively routed to every juror. Perspective evaluation takes place organically as jurors react to the perspectives of others, and procedurally via formal votes. Distillations of voting results are shared back with the jurors after each round of voting, and with the courtroom once a unanimous verdict is reached. 

\subsubsection{Deliberative Polling} 
Developed in the late 1980's by Fishkin \cite{fishkin2005experimenting}, Deliberative Polling augments typical opinion polling with a deliberative process to elucidate how public opinion on an issue would change if the public was well-informed of the realities surrounding an issue. It seeks to combine the scientific rigor of randomized polling with the considered judgment of deliberation. Deliberative Polling starts with a type of perspective evaluation -- an opinion poll on a target issue conducted on a random, representative sample. Then participants in the poll are invited to participate in an in-person deliberation around the issue. Typically an advisory board oversees knowledge routing; creating materials that are shared with participants before and during the deliberative process, and recruiting experts who can make their knowledge available in person. 

During the in-person deliberation, the process oscillates between small group discussions and larger plenary sessions over the course of a weekend. During the small group discussions, trained moderators keep the discussion civil and useful.  They manually handle perspective elicitation to ensure all ideas and arguments are heard, and do a live form of distillation generation and routing to help participants understand the range of perspectives discussed. Then moderators elicit perspectives in the form of questions that may be asked of experts during the plenary session. The large plenary sessions mostly focus on knowledge generation and routing, typically with panels of experts discussing the questions elicited during the small group discussions. The overall process can go back and forth for multiple cycles between small group discussions and large plenary sessions. Finally, when deliberation is considered complete, the perspective evaluation that took place before deliberation (in the form of the opinion poll) takes place one more time with the deliberative session participants\footnote{In some cases the opinion poll is also given to non-participants so that the change in opinion which results from the deliberation can be isolated from any changes in opinion resulting from externalities.}. The results of the opinion poll before and after the deliberation are then compared, and the change in opinion after the in-person deliberation is viewed as how the public's opinion on an issue would change if it were well-informed.

\subsubsection{Citizens' Assemblies} 
Citizen assemblies are considered one of the most robust representative deliberative processes \cite{oecd2020innovative}. 
The goal of a citizen assembly is to produce detailed collective recommendations on (typically divisive) issues. A representative set of around 100 participants are chosen through sortition (stratified random sampling, also known as a democratic lottery) to take part in the process. The deliberation itself is typically comprised of four phases of in-person or virtual meetings which may take place over the course of five to thirty days across many weeks or months. 

The entire process is facilitated by trained moderators and begins with the \emph{learning phase} which focuses on knowledge consumption and relationship building. Knowledge routing is typically overseen by an advisory group that assembles and shares learning materials to familiarize participants with the issue being deliberated. Next, the \emph{consultation phase} involves a more dynamic type of knowledge routing, where participants consult with a range of experts, stakeholders, and affected groups about the issue. While the people participants initially consult with are typically selected by the independent advisory group, participants are given a degree of control over knowledge routing in this phase by allowing them to request access to specific experts, groups, or external information sources. Then, during the \emph{deliberation phase}, moderators support perspective elicitation from participants in a way that allows all voices to be heard. As participants contribute and consume the perspectives of others they will discuss evidence, weigh options, and develop potential recommendations. Finally, during the \emph{decision-making phase} evaluation elicitation takes place as participants determine what to include in collective recommendations and apply a form of distillation generation to create a detailed report of collective recommendations which have the majority support of participants\footnote{Sometimes a distillation in the form of minority report is generated which includes the recommendations which did not have majority support.}. 

\subsubsection{Online Forums} 
The internet gave rise to new forms of online communication that could be used for deliberation, including online forums. In an online forum perspective generation and evaluation happens organically. Perspectives are organized into distillations in the form of threads which participants are free to explore and consume as they like. Participants can contribute perspectives as a reaction to an existing perspective, or contribute totally new perspectives for others to consume and react to. In some cases, forums may have moderators who attempt to influence the deliberation by muting bad actors, generating distillations, or posing questions to elicit perspectives with a specific focus. In order to enable quantifiable perspective evaluation within online forums, various forms of voting have been integrated. 

\textbf{Reddit}\footnote{reddit.com} allows participants to evaluate perspectives by giving them an up-vote or down-vote. Those votes are then aggregated to provide a signal of overall support for a perspective, and those signals are used to inform distillation into threads where the perspectives with the highest support are ranked higher and thus consumed by participants more often. The result is a feedback loop where highly ranked perspectives are consumed, evaluated, and reacted to more. A pitfall of this type of online forum is that participants often upvote perspectives that are simply funny or entertaining, and this can incentivize participants to contribute perspectives optimized for engagement that distract from healthy deliberation.

\textbf{Kialo}\footnote{kialo.com} is a more recent type of online forum which is designed specifically for deliberation. Deliberation on Kialo begins with a statement like "All public bathrooms should be gender neutral" and then elicits perspectives from participants in the form of claims which are pros are cons of either the original statement or of the claims submitted by other participants. Perspective evaluation then takes the form of voting on how much 'impact' a participant believes a claim has. This allows for a form of distillation where perspectives are organized into a tree structure; with claims for or against the original statement at the top, and then the claims for or against each of those claims (and so on) comprising the branches. Claims with the highest aggregate 'impact' are ranked higher in the presented distillations. This means claims that are evaluated to have a higher impact, and are closer in the tree to the original statement are consumed by the most participants. 

A key challenge highlighted by online forums is that as the number of participants scales, it becomes impossible for all participants to consume and evaluate all perspectives. This forces design decisions that manifest zero-sum trade-offs for how participants' limited attention budget will be allocated towards consuming and evaluating perspectives. And since participants interact with perspectives in threaded distillations where certain perspectives are seen more than others, it means not all participant perspectives are subject to equal exposure and evaluation. In other words, participants do not have an equal voice, and the principal of \emph{fair hearing}\footnote{A system adhering to the principle of \emph{fair hearing} is designed to ensure all participants are heard equally, ie. all perspectives are given equal opportunity to be consumed and evaluated by participants} is not maintained. Further, because all participants are treated as a monolith, finding common ground in the form of bridging perspectives\footnote{A bridging perspective is one which has a high degree, not just overall, but within each of subsequent of the population.} is not possible. Collective response systems aim to address these challenges.

\subsubsection{Collective Response Systems} 
The process enabled by collective response systems (CRS) \cite{ovadya2023generative} begins with a prompt which participants respond to with their own perspectives. Then participants consume the perspectives submitted by others in a way that promotes fair hearing, and they evaluate those perspectives in a way that enables quantification of how well each response represents the group. The first wave of CRS platforms --Allourideas, Polis, and Remesh-- were all initially developed in the early 2010s, and others like Psi have emerged since then. These different CRS tools primarily vary in their mechanics of perspective elicitation and evaluation, the nuances in their methods for perspective routing, and their approach to distillation generation. 

We note that the idea of a collective response system shares many similarities with the idea of a \emph{wikisurvey} as described in \cite{salganik2015wiki}, with the distinction that ``Unlike a WikiSurvey, collective response systems do not need to be ‘greedy’, and some collective response systems intentionally limit what can be evaluated.'' We choose to adopt the label of collective response systems here because it has a slightly more inclusive definition.

\textbf{Allourideas} \cite{salganik2015wiki} elicits perspectives in the form of text responses to a prompt,  and elicits evaluations of those perspectives in the form of pair-wise votes. Evaluation elicitation is semi-random, with pair choices between responses with fewer votes being elicited with a higher probability than those with more votes. Attention allocation between contributing perspectives and evaluating perspectives is governed by the participants themselves as they are free to switch between either activity during the deliberation process. Distillation generation uses pair choice data to estimate a score for each perspective related to the probability a randomly chosen participant would choose that perspective over another randomly chosen perspective. This distillation is updated live as participants vote and routed to all participants so they can see it as they participate. The overall dynamic is asynchronous, involves only a single prompt, and generally takes place over the course of weeks or months. The use of pairwise comparison means a complete ranking of all responses for each participant can be generated. This can be important when perspectives represent competing alternatives, but it provides only a relative signal of participants' support for each perspective and is unable to quantify the absolute support for perspectives\footnote{For example, through pair choices alone one cannot distinguish between a scenario a participant disagrees with all perspectives and one where a participant agrees with all perspectives.}. 

\textbf{Polis}\footnote{pol.is/home} \cite{small2021polis} also elicits perspectives in the form of text responses to a prompt, but it elicits evaluations of those perspectives in the form of absolute agreement votes. Evaluation elicitation is semi-random, with votes on perspectives likely to provide more information useful for clustering being elicited with higher probability. Again, attention allocation between contributing perspectives and evaluating perspectives is governed by the participants themselves as they are free to switch between either activity during the deliberation process. Polis generates multiple forms of distillations which are included in its summary report. 

One distillation maps both participants and perspectives into two dimensions such that participants who vote similarly are close to each other, responses that are supported by similar people are close to each other and the people who support them, and emerging segments of participants who vote similarly are highlighted. Another distillation aggregates agreement votes for each perspective and shows those aggregations for all participants as well as for the different emergent participant segments. A final distillation organizes and highlights perspectives based a group-informed consensus metric for each perspective\footnote{Polis computes its group-informed consensus metric by first clustering participants into two segments of similar voters, and then multiplying the fractional support for a perspective within one segment by the fractional support for that perspective within the other segment.} which quantifies the degree to which a perspective is polarizing and only supported by only one of the emergent segments versus bridging and well supported across all segments. These distillations are updated live as participation takes place and all are included within a report which can be routed to participants by sharing a link to it within the participation interface. 

The overall dynamic is asynchronous, involves only a single prompt, and generally takes place over the course of weeks or months. In contrast to the pair-choice-only approach of Allourideas, Polis' agreement-vote approach means that absolute support for perspectives can be quantified. However, without pair votes, there is insufficient information to determine a full ranking of perspectives for an individual participant because it is not possible to know their preference between competing alternatives when their agreement votes are the same. Lastly, Polis is open source and can be embedded into web pages and applications in ways that route knowledge to participants and extend the process and capabilities described here.

\textbf{Remesh}\footnote{remesh.ai/product}\cite{konya2022elicitation,bilich2019faster} uses a collective response system within the context of a two-way \emph{collective dialogue} between a moderator and a group of participants. During each turn of the dialogue, the moderator can choose to send a message to participants which acts as a prompt to initiate a collective response process. This triggers perspectives to be elicited from participants in the form of text responses to the prompt. Then Remesh elicits evaluations in the form of both agreement votes and pair-choice votes. Evaluation elicitation is uniformly random, with both types of votes being elicited in equal amounts, and each specific vote sampled with equal probability. Attention allocation between contributing and evaluating perspectives is controlled by the system such that participants can only submit a single perspective to a prompt, and the rest of their attention is allocated toward perspective evaluation. Remesh generates multiple forms of distillations, some of which are available to the moderator and another which is routed directly to participants.

Because collective dialogues typically involve a large number of participants and thus perspectives, and each participant has a finite attention budget, most participants are not able to vote on most perspectives. So, as a first step towards generating distillations, Remesh does elicitation inference by training a model on all the votes that were elicited to infer all of the votes that were not elicited\footnote{Remesh does elicitation inference using an LLM-based collaborative filtering model which is detailed in \cite{konya2022elicitation}}. The first distillation Remesh generates aggregates the elicited and inferred agreement votes to give the percent agreement among all participants for each perspective, as well as for each known participant segment \footnote{Knowing the agreement for perspectives across segments makes it possible to identify bridging perspectives which have a high agreement across all segments, and polarizing perspectives where agreement diverges across segments.}. A second distillation is generated which highlights a small subset of perspectives that represent the plurality of participant views\footnote{The subset of perspectives is chosen such that every participant has at least one perspective in the subset which they prefer over nearly all other perspectives. This ensures that strongly held minority views get highlighted}, and for each perspective, gives the number of participants who prefer that perspective above all others in the subset. A third distillation is generated which gives the common topics found in perspectives and their frequency. These three distillations are all made available to the moderator live within the dialogue. A fourth distillation is generated for each participant which gives the subset of perspectives representing the plurality of views, plus the perspective the participant submitted, and for each perspective gives the percent agreement across all participants. A fifth distillation is generated after the dialogue is completed which uses an LLM to distil the ideas manifest across perspectives into a summary paragraph.  

During a \emph{live} collective dialogue on Remesh\footnote{Remesh also allows for asynchronous collective dialogues where the messages, prompts, and polls, that comprise the dialogue are set beforehand, and participants can work through the dialogue at their own pace.} a single collective response process takes a few minutes to complete from the initial prompting message sent by the moderator until all perspectives and votes are elicited and the distillations are generated and routed. The typical collective dialogue lasts around an hour, and during it the moderator is in control of attention allocation. While at each turn of the dialogue the moderator may choose to initiate a collective response process, they may also choose to send participants a simple poll, or read-only messages in the form of text, images, or videos. Through these messages, they can route external knowledge, custom distillations, or simply say nice things that help move the dialogue forward. Skilled moderators guide the dialogue in a way that builds on itself, updating participants' understanding along the way, and digging into emerging areas of interest. A typical collective dialogue on Remesh will involve around 5 polls, 30 read-only messages, and 30 collective response cycles.

\textbf{Psi}\footnote{https://thepsiapp.com/} elicits perspectives in the form of audio recordings which are then transcribed into text. The dynamic of deliberation on Psi is similar to tournament bracketing. During each round participants are broken into groups and assigned a few perspectives to discuss. Following the discussion evaluations are elicited on the perspectives which were discussed in the form of votes. The perspectives with the most votes go on to the next round. The result of this dynamic is that while during the first round each group is discussing different perspectives, the number of perspectives being discussed decreases round by round, and by the end, groups are discussing and voting on the same small number of perspectives. Through this mechanism participants attention is increasingly allocated towards the perspectives with the most votes. Psi generates an interactive report with a few distillations.

The initial distillation is a list of all perspectives ranked by how many votes they received. The text transcription of each perspective is shown with an option to play the original audio recording. The next distillation shows the evolution of perspective rankings for each round. The final distillation shows a network of perspectives. Each perspective is mapped to a circle whose size is proportional to the number of votes it received, and connections are drawn between perspectives with the thickness of the connections proportional to overlap in the participants who voted for it. 

\subsubsection{Other tools}

\textbf{Online Deliberation Platform}\footnote{stanforddeliberate.org} is an online video chat platform with specific features meant to help facilitate the type of small group deliberations that happen during a Deliberative Polling process \cite{fishkin2019deliberative}. On the platform external knowledge is routed to participants in the form of an agenda to be discussed which includes relevant positions and the arguments for and against them. An automated moderator controls perspective elicitation, with an automated cue system to determine which participant to elicit a perspective from next and an automated timer to determine how long each participant has to state their perspective. Participants collectively control how attention is allocated towards the positions in the agenda by voting when to move on from discussing a given position. After all positions in the agenda have been discussed, the tool can facilitate eliciting perspectives from participants in the form of open ended questions about the positions discussed, then elicits evaluations of those questions in the form of votes, and generates a distillation to be shared and discussed by experts and stakeholders during the large plenary sessions of a Deliberative Polling process.

\textbf{Talk to the City}\cite{turan2023talk} is an open-source LLM interface for improving collective deliberation and decision-making by analyzing detailed, qualitative data. It aggregates perspectives and arranges similar arguments into clusters. Talk to the City’s data processing pipeline starts by processing a variety of data types such as open form text, videos, voice recordings, and conversation transcripts. Then it uses LLMs to extract key arguments, and finally arranges similar arguments into clusters and subclusters. Users can navigate through a map of opinions and drill down to the subclusters they find most interesting and engage in discussion with the clusters in open-ended text form.

\textbf{Discord}\footnote{https://discord.com/} is a multi-modal communications platform designed  for communities. Communities on Discord are known as 'servers' and each server is organized into topical channels where information is shared and interaction between users takes place. A channel can be a simple chatroom where users exchange text messages and files in a linear thread, a live audio channel where users can talk in real-time, or a forum with voting that has similar dynamics to a Reddit thread. In addition to these capabilities, Discord supports a wide range of bots and plugins which extend its capabilities. Through its native and integrated capabilities, Discord supports both open-ended community discussions as well as the more formal perspective generation, consumption, and evaluation which one might desire as part of a deliberative process.

\textbf{Snapshot}\footnote{https://docs.snapshot.org/introduction} is a tool that allows members of decentralized autonomous organizations (DAOs) to formally vote on proposals in a way that is off-chain yet verifiable. Each DAO on Snapshot has its own 'space' -- linked with an ENS domain -- within which proposal submission and voting takes place. The platform is designed to enable a wide range of voting processes and customization based on the use case. This includes supporting different voting mechanisms (ie, single choice, approval, quadratic etc.) and different approaches to determine a user's voting power based on ERC20s, NFTs, and other contracts. The typical use of Snapshot is to source and vote on formal proposals for DAO governance and actions, often following less formal deliberation and discussion on a platform like Discord. 

\textbf{Dembrane}\footnote{https://www.dembrane.com/} helps enable deliberation at scale by making it easy to turn natural deliberative conversations into insight. Project managers start a project with Dembrane by presenting a set of primary research questions\footnote{These should be as close to the ideal outcome as possible. Instead of, "What are people's values?", something like "We are struggling with high churn rates and want to know why." is preferred }. Then, the team can go out and record conversations with stakeholders about the topic. These recordings are dropped into a shared Google Drive or Dropbox folder where Dembrane transcribes the audio, chunks the transcript it into small parts, and analyses them with large language models. The transcript chunks, as well as preprocessing text, are used to answer standard research questions like "Is this a balanced conversation?" and "Identify the themes in this conversation and run a sentiment analysis on each theme". Finally, all the context is used to answer the main research question. All the analysis steps are presented to the organizer in a structured report, answering the main research question along with direct quotes from the transcripts and ending with recommendations for further research\footnote{Dembrane is currently developing a front-end interface where users will be able to add metadata to the recordings and ask questions to the data set dynamically. They are also developing a technique whereby an AI can contribute to the discussion as it unfolds, providing people with additional context, answering questions about the process, or cross-pollinating insights from other conversations.}.

\subsection{Deliberative Technology in Alignment Systems}\label{DelibTechInAlignmentSystems}\label{DA.DelibTechInAlignmentSystems}

Deliberative technologies can be used across the various components of an alignment system, but are most commonly used for \emph{sensing} will and \emph{identifying} possible actions. They can be used for \emph{sensing} the will of a group when perspectives manifest the properties of \emph{items} (ie. relates to potential futures) and evaluation of perspectives involves an action with mutually understood impact. They can be used for \emph{identifying} the universe of possible actions by eliciting perspectives in the form of possible actions. They can potentially be used for \emph{predicting} how actions affect the future if knowledge about the action is routed to a group, perspectives are related to impacts from the action, and the primary distillation and group dynamics promote collective intelligence. They can potentially be used for \emph{assessing} how potential futures align with human will by sensing a group's will about perspectives that describe potential futures. Finally, deliberative technology can potentially be used to assist with \emph{executing} chosen actions, for example, if knowledge of a chosen action is routed to participants and elicited perspectives focus on identifying sub-tasks manifest in the chosen action.  

\subsubsection{Examples}

Here we provide real examples of how deliberative technology is already being used within \emph{alignment systems} involving some of the world's most powerful institutions. 

\textbf{National Government -- Polis in Taiwan.} The arrival of Uber in Taiwan in 2013 created challenges around whether to allow it, how to regulate it, and how to ensure fair competition with existing taxi services. But, both the public and regulators were divided on how to approach these challenges, leading to gridlock which stunted progress towards developing legitimate policies. At the same time, vTaiwan was being developed with the goal of helping policymakers make decisions that gain legitimacy through consultation with the public, especially related to divisive issues. To achieve this the vTaiwan platform aimed to help engage and collaborate with citizens around governance and policy issues using a deliberative process enabled by open source tools like Polis. This meant vTaiwan would act as a \emph{local alignment system} by helping align future policy with public will.

In 2015, at the request of the Ministry of Transportation and Communications, the Ministry of Economic Affairs, and the Ministry of Finance, vTaiwan embarked on the Uber case. Under the leadership of Audrey Tang (who is now the Digital Minister of Taiwan), the vTaiwan deliberative process began with knowledge creation and routing with the goal of educating stakeholders (eg. taxi drivers, transportation users, etc.) on the nuanced realities underlying the Uber challenge. Next began online deliberation with the goal of sensing the will of the relevant stakeholders, mapping the opinion landscape on the issue, and ideally, discovering points of consensus that could give rise to legitimate policy. To do this they used Polis \cite{simple2018horton}. While initial results on Polis demonstrated two opposing camps on the issues, as participation continued, multiple points of consensus emerged between the groups. Tang herself then took these consensus points into talks with Uber, taxi drivers, and various experts, ultimately leading to new regulations. Polis has now been used in a similar way as part of vTaiwan for dozens of issues involving more than 200,000 participants.

Sensing human will requires observing deliberate voluntary actions which have a mutually understood impact on the future. How was this achieved? The observed actions on Polis are submitting and evaluating (via votes) perspectives. Participants were recruited using invitations to stakeholder groups (like taxi drivers and Uber drivers) as well as outreach to citizens via paid and organic social media posts. To ensure the actions on Polis were voluntary, participation by all parties was optional, and no participants were coerced to participate by paying them. To create a degree of mutual understanding about how participating on Polis would impact the future, the participant invites explained how the results from their participation would be used to inform policy decisions. Overall, the result of these affordances meant that observing participants' actions on Polis equated to sensing participants will.

\textbf{United Nations -- Remesh in Libya.} Following nearly a decade of fighting in Libya, the international community convened the Berlin Conference of States and IGOs in early 2020 with the goal of bringing an end to the fighting \cite{international2020libya}. In the wake of this, the United Nations Support Mission in Libya (UNSMIL) began implementing a plan of action to reach a ceasefire agreement and help align the future of Libya with the will of the Libyan people. In this way, UNSMIL began to function as a \emph{local alignment system}. To this end, UNSMIL began three tracks of activity; a \emph{military track} focused on achieving a ceasefire, a \emph{political track} aimed at establishing a new unified government, and a \emph{financial track} aimed at unifying the economy. 

The initial priority was the military track. UNSMIL's action space for this track focused on generating a ceasefire agreement which would have the support of both the leadership and the general public on both sides of the conflict. With this aim, they began a formal dialogue with a delegation comprised of 5 military leaders from both sides of the conflict known as the 5+5 Joint Military Commission. Led by UN Special Advisor on Libya Stephan Williams, this provided a forum to sense the will of leaders on both sides and begin to pen a ceasefire agreement. To \emph{sense} the will of the Libyan public Williams hosted a set of \emph{collective dialogues} on Remesh with participants from both sides of the conflict including key minority groups \cite{irwin2021using}. A week after William's first collective dialogue a ceasefire agreement was reached within the 5+5 Joint Military Commission \cite{libya2020agreement}. 

After the ceasefire was reached the focus shifted towards the political and financial tracks as establishing a unified government and economy that could support lasting peace became the priority. During this period (which is still ongoing) Williams continued to run collective dialogues with the Libyan public  \cite{UN2021williams} as a way to sense their will at key junctures in the process; eg. when a president was being chosen or when key governance documents were being developed. Along with sensing public will, the collective dialogues also made it possible to tap the public's collective intelligence to source potential actions that could address ongoing challenges. Remesh has now been used by the UN in similar ways more than 50 times across various missions around the world including in Yemen \cite{UN2020cutting}, Iraq \cite{UN2021jeanine}, Lebanon \cite{UN2022lynn}, Hati \cite{UN2023carol}, and Bolivia \cite{UN2023liita}.

While sensing potential actions via collective dialogue is relatively straightforward, sensing human will is more nuanced. Recall that sensing human will requires observing deliberate voluntary actions which have a mutually understood impact on the future. How was this achieved? The deliberate voluntary actions were the submission and evaluations (via votes) of perspectives during the dialogue. To ensure participant actions were voluntary, all participation was optional and no financial incentives were used to coerce people into participating. In order to create a scenario where those actions had a mutually understood impact on the future, knowledge was routed to participants at the start of the dialogue to educate them on the impact of their actions during the dialogue. First, they were educated on how their actions impacted what the UN would learn from the dialogue (ie. how the mechanics of perspective submission and voting worked), and then on how UN leanings from the dialogue would impact the relevant UN activity tracks (ie. who the learnings would be shared with and what role those people played in effecting outcomes the participants cared about). In this way, a mutual understanding between the UN and participants was cultivated about how their deliberate actions during the collective dialogue would impact Libya's future. Overall, the result of these afordances meant that observing participants' actions during the collective dialogue equated to sensing participants will.

\textbf{Constitutional Referendum -- Citizens' Assembly in Ireland}.  In 2016 the Irish parliament passed a resolution to establish a citizen assembly, comprised of 100 people selected to be representative of the Irish population, and tasked this body with deliberating on a select set of key issues. One of these issues was their 8th constitutional amendment which made abortion largely illegal -- an issue the public was largely divided on. From November 2016 to April 2017 this deliberative body  focused on the 8th amendment. Over five in-person [weekend long] meetings, the participants were educated on the medical science surrounding abortions, listened to and weighed concerns from all sides of the issues, and deliberated on what was an appropriate position for Ireland to take on abortions going forward. The output from the citizen's assembly was a report of recommendations which included not retaining the 8th amendment in full\cite{first2017citizens}. In this way, the citizens' assembly generated a signal capturing what the will of the Irish public would likely be if they were well-informed on the issue of abortion.

While there existed no requirement that the recommendations of the citizens' assembly be adopted by parliament, there was a requirement that parliament listen to the recommendations and offer a response. In the case of the the 8th amendment, the parliament found the recommendations compelling, and as part of their response \cite{report2017tithe} set in motion a public referendum vote to repeal the 8th amendment. In the lead-up to that vote, the report from the citizens' assembly was employed as part of a campaign to educate the public on the reality of the abortion issue. The referendum vote took place in May 2018, and passed with $66.4\%$ of voters supporting the repeal of the 8th amendment \cite{irish2018bbc}. This was a direct sensing of the will of the Irish people because citizens' vote had a mutually understood impact on the future. And, it is notable that during the citizen's assembly, which was intended to capture what the will of an informed public would be, $64\%$ of participants voted that they would support "terminations without restrictions up to 12 weeks of gestation" which roughly equates to a full repeal of the 8th amendment. This consistency underscores the fact that the citizens' assembly did indeed function as a reasonable proxy for the will of an informed Irish public. 

In this example, the system comprised of the Irish parliament, the citizen's assemblies, and the mechanism of a referendum vote, acted as an alignment system. The citizens' assembly sensed a proxy of informed public will which was shared with parliament. Parliament, which (ideally) exists to align policy with public will, was convinced that public will may now be opposed to the existing amendment on abortion, and set in motion a referendum vote to sense public will on the issue directly. The referendum vote, which was likely influenced by the assembly's recommendations and coverage of stories of assembly members' experiences, confirmed that informed public will was opposed to the existing amendment on abortion, and the amendment was repealed. Overall, this system led to a change that made future policy better align with informed public will; in other words, it functioned as an effective alignment system. 

\subsubsection{Challenges} \label{DAchallanges}

While modern deliberative technology provides capabilities that make alignment systems more effective, they are still far from optimal. Here we outline some of the key challenges manifest in using existing deliberative technology for alignment systems which we hope the next generation of these technologies can better address. 

\textbf{Scalability vs richness trade-off.} While SOTA deliberative technologies enable deliberation at increasing scale, they do so at the cost of the richness afforded by smaller-scale methods. Here we use \emph{richness} to refer to the general depth, empathy, and open-endedness of the interactions between people during a deliberation. Juries afford maximum richness because all participants are in the same room, their discourse is face-to-face and open-ended, and there are a small enough number of people (on the order of 10) that the exchange of perspectives can happen organically while enabling all voices to be heard by all participants. Deliberative polling and citizen's assemblies scale participation up by an order of magnitude (to on the order of 100) but do so at the cost of some richness. While small group sessions allow dynamics reminiscent of a small jury, the time allowed for discussion is often shorter, the exchange of perspectives tends to be more structured, and participants are unlikely to hear the perspectives of participants not in their small group. Both collective response systems and online forums can scale up participation by multiple orders of magnitude (to on the order of 10,000) but lose many of the humanizing aspects of live face-to-face discourse which help give rise to deep empathetic human interactions. Overall, choosing among modern deliberative technologies to use within an alignment system means choosing a point along a Pareto frontier trading off richness for scale. 

\textbf{Inhomogeneous participant capacity.} When deliberative technology is used to sense the will of a target population, the aim is typically for the population of deliberation participants to be representative of the full target population (potentially through a microcosm). Target populations are often broad segments of the public spanning a wide range of experience, expertise, and ability; participant populations reflect the same diversity. Participants likely span a range of technical literacy, meaning a range of different onboarding protocols to a deliberative technology platform are needed. Participants also likely have different degrees of expertise related to issues under deliberation and thus have different educational needs in order to become equally 'well informed.' Similarly, creating situations where participants' actions have mutually understood impact may require educating participants on the impact of their actions in different ways.  Further, participant differences in education, expertise, and language proficiency mean that not all participants may be able to easily understand the perspectives of others in their original form, and thus not be able to accurately evaluate them. Finally, participants may have varying degrees of innate self-awareness, and thus require different affordances to help them introspect on what they truly want. While these types of participant population differences may be manageable in small-scale deliberations like juries, achieving the degree of experience personalization required to address them becomes increasingly challenging to resolve at scale. 

\textbf{Finite participant attention.} The amount of participant attention that can reasonably be consumed during a deliberative process to sense a population's will is finite. This means its allocation between and within the various participant operations represents zero-sum trade-offs. How much time should a participant spend consuming knowledge to sharpen their understanding of reality vs. generating and evaluating perspectives based on their understanding? Further, even if their full attention budget is spent evaluating perspectives, it is still unfeasible for a participant to evaluate most perspectives at scale; so which subset of perspectives should any given participant evaluate? What's more, since most perspectives won't be evaluated by most participants how can one get at the overall support a population has for a perspective with only a small subset of participants evaluating it? While all deliberative platforms answer these questions at least implicitly, those answers are mostly informed by safe heuristics\footnote{Like the heuristic used by Remesh of 'sample all perspectives to be evaluated randomly with equal probability'} rather than in ways that maximize information gain or guarantee fair representation \cite{representation2022halpern}. 

\textbf{Consensus discovery.} When deliberation is around divisive issues most perspectives are polarizing and perspectives which have consensus support\footnote{Here we use 'consensus support' to mean something similar to 'bridging.' That is, it refers to perspectives which have, not just majority support, but strong support within each sub-segment of the population.} are rare. This means that even if a consensus perspective theoretically exists, it is possible no participant may submit it. What's more, even if a participant does submit a perspective that would have consensus support if all participants were well informed, it is possible that participant's misunderstanding of reality may cause the perspective to appear polarizing. Wile focusing on perspectives which manifest participants terminal will may help overcome the barrier to consensus created by misunderstanding, eliciting and then distinguishing participants terminal will from their instrumental will is itself a non-trivial challenge.  Overall, when deliberately sensing will, consensus perspectives give information about convergent portions of the will matrix, making them especially important in alignment systems involving institutions (as discussed in Section \ref{WOH.sensing} on Prioritization). So the challenge is to increase the chances of discovering consensus perspectives, even around divisive issues where they are rare.

\textbf{Distilling results for human consumption.} In deliberative alignment systems involving institutions, many decisions are ultimately made by humans. For example, assessing how well different actions align with sensed will, and then choosing which action to take, is often done by humans. This means that the likelihood of taking the optimal action\footnote{Optimal in the sense that it would be the action one would take if they had perfect information on the population's will, perfect judgment of how that will aligns with candidate actions, and no corrupting incentives.} is a byproduct of how well the human decision maker's understanding of a populations' sensed will (or landscape of sourced potential actions) is consistent with reality. The raw data set generated by the types of deliberative technology used in this case (like from collective response systems) is often too large for a human to consume directly. As a result,  distillations of the raw data need to be generated which are easily consumed by humans. So how do you generate distillations which, when consumed by humans, conjure in their mind an understanding which reflects reality? This involves nuanced choices about how to do aggregation, how to compress or filter which data points get shown, and how to communicate data in an easily understandable way. And because not all decision-makers learn and understand things in the same way, an ideal distillation for one decision-maker may not work for another. So the challenge involves not just generating the right distillation, but generating the right distillation for each human that is consuming it.

\textbf{Incentivizing healthy participation.} The quality of outputs from a deliberative process depends on how healthy engagement from participants is. By healthy we mean the degree to which participants engage at all (do they respond to all prompts?) and how honest and thoughtful they are in their contribution and evaluation of perspectives. A common approach to incentivize participant engagement is to pay them to participate. This approach is known to increase participation rates \cite{stanley2020effectiveness}, but does not necessarily motivate participants to be more honest or thoughtful \cite{barge2012research}. Further, incentivizing participants with money requires financial resources that might not exist in many scenarios. Helping participants understand how their participation will have an impact related to relevant issues (a requirement for sensing will) may motivate them to be more thoughtful and honest, but only if they care about the those issues. So the challenge is to create a scenario where all participants are incentivized to fully engage, and do so in a way that is honest and thoughtful.

\textbf{Ensuring representativeness.} When sensing will in the context of a deliberative alignment system, the goal is to generate a signal that is representative of a target population; usually a broad segment of the public. An ideal approach to this is to use sortition to obtain a set of participants which is a representative microcosm of the target population. A challenge in accommodating sortition-based participant population is ensuring participation is accessible for a wide range of differently abled participants. But even if this challenge can be met, there are cases where sortition not viable due to timing, budgets, or other externalities. In these cases, the challenge lies in taking results representing a skewed population sample and transforming it to be representative of the target population. While such approaches are common for quantitative surveys \cite{tanton2011small}, the signals generated through deliberative technology often involve a qualitative component which can make the use of such techniques non-trivial.

\textbf{Creating legitimacy.} In institutional alignment systems the decisions which are informed by deliberately sensing public will are often made by humans who want their decisions to be viewed as legitimate. For these decisions to have such legitimacy they need to be seen as fair, open, and inclusive in the eyes of the public. This means sensing the public will which informs such decisions needs to be viewed as fair, open, and inclusive as well. For this to be true the will sensing process must include a sufficiently representative segment of the public, the method of sensing must be transparently fair, and the overall process must be open enough for outsiders to build trust that there was no manipulation or coercion which could unjustly skew results. On one hand, this challenge is a technical one (how do you enable external verification that results are representative and fair?) and on the other hand a communications one (how do you communicate how public will was sensed in a way the public can understand and trust?).

\textbf{Computing alignment with sensed will.} In order for a sensed will signal (like a Will matrix) to serve its purpose within an alignment system, it is necessary to compute how well potential actions align with it. In institutional alignment systems this is often done by one or more decision makers and enabling such 'computation' really equates to optimally distilling results for human consumption. This approach, however, is not ideal because it gives decision-makers (who often experience corrupting incentives) an easy vector to manipulate the impact of the alignment system toward self-interested ends. When the action space and scope of the will signal are limited, another option for estimating will-action alignment is to employ a deliberative process that harnesses the collective intelligence of a group. A related option is to collapse sensing will and evaluating will-action alignment into a single process, by constraining the perspectives that are evaluated during sensing to those that represent actions. This approach may seem attractive at first, but it severely limits the scope of will that is sensed and forces participants to assess the impact of actions themselves; an unreasonable expectation for many types of participants and actions. Further, it shifts their focus away from introspecting on their true interests and toward judging actions at face value. For alignment systems involving AI agents, the number of potential actions per unit time may far exceed the number of perspectives that can be elicited and evaluated in the same time period. In this case, it may be better to have perspectives that manifest impacts, that way alignment can be computed for any potential action whose impact can be predicted. However, this creates yet a new challenge, that of computationally predicting the impact of all potential actions, and then computing alignment between the predicted impact and sensed will signal.

\textbf{Neutralizing adversarial manipulation.} Alignment systems can control large or critical pools of resources. This creates a strong incentive to manipulate them and makes deliberative processes which sense will an enticing manipulation target. At least two forms of manipulation exist: disruptive manipulation and biasing manipulation. \emph{Disruptive manipulation} seeks to disrupt the deliberative process in a way that harms the legitimacy of any results they produce. This can be achieved by getting malicious participants into the deliberative process who then contribute adversarial perspectives (such as hate speech) with the goal of derailing the civility of the deliberation. When the deliberative process is digital, disruptive manipulation can be achieved by infiltrating the process with participant bots which either contribute adversarial perspectives or evaluate perspectives in a way that produces so much noise in the results that they become useless. Disruptive manipulation can also be accomplished by simply crashing the digital system being used, for example using a DDOS attack. \emph{Biasing manipulation} seeks to unjustly influence results towards a self-interested target without harming the perceived legitimacy of those results. A typical approach to doing this involves infiltrating the deliberative process with a significant number of bad actors (either bots or people) who collude in ways which biases the results towards a specific target. This can be accomplished if colluding participants' contributing perspectives that aim to manipulate other participants into evaluating perspectives in a biased way, or by simply evaluating perspectives according to a shared protocol which biases the results themselves. Further, beyond these participatory approaches, biasing manipulation can be achieved by hacking into the result data and changing it directly. The challenge then is to develop deliberative technology which is robust to these types of manipulation. This requires being able to both defend against cyber attacks, as well as detect and neutralize various forms of malicious participant behavior.

\newpage

\section{Intelligent Deliberative Alignment}\label{IDA}

\subsection{Artificial Intelligence}\label{IDA.AI}

The term \emph{artificial intelligence} (AI) was first coined by John McCarthy in 1956 to refer to programs and machines with problem-solving abilities that mimic human-like intelligence \cite{mcarthy2004what}. While initial AI efforts involved expert systems built on symbolic rules and knowledge bases \cite{smolensky1987connectionist}, much of AI today involves machine learning where the weights of highly parameterized models are learned by training on large data sets \cite{mitchell2007machine}. Until recently, most models were developed and trained for specific applications, like content recommendation \cite{resnick1997recommender}, image classification \cite{lu2007survey}, sentiment analysis \cite{medhat2014sentiment}, language translation \cite{wu2016google}, anomaly detection \cite{pang2021deep}, and specific reinforcement learning settings where machine learning agents are trained to take actions which maximize some reward, like points in a video game \cite{sutton2018reinforcement,li2017deep}.  These mostly "narrow AI" models involve a range of machine learning architectures, some inspired by the brain (ie. convolutional neural nets \cite{gu2018recent}), and others inspired by work in fields like compressed sensing (ie. matrix factorization \cite{mnih2007probabilistic}), physics (ie. Boltzmann machines \cite{ackley1985learning}), and bayesian statistics (ie. bayesian neutral nets \cite{hernandez2015probabilistic}). Most machine learning techniques involving these types of models rely on labeled data for training and most of these labels come from human annotators\footnote{An example of training data which does not come from human annotators is in reinforcement learning scenarios where the reward signal corresponds to points in a game, or some other computable property of the RL agent's environment.}.  Thus, the size of data sets that can be created for training is resource-limited, and how big a model can be without being practically\footnote{Some machine learning models demonstrate surprisingly good performance even when somewhat over-parameterized due to implicit regularization \cite{gunasekar2017implicit}. By 'practically' over-parameterized we mean beyond the limits where implicit (or explicit) regularization can prevent over-fitting and yield reliable results.} over-parameterized is limited as well.

More recently, however, foundation models \cite{bommasani2021opportunities} have been developed which overcome such limits by combining highly expressive model architectures with training objectives designed to work with unlabeled data. The most notable of these are generative pre-trained transfomers (GPT) \cite{radfordimproving} -- large language models (LLMs) which combine a highly expressive transformer model architecture with a next token (word) prediction training objective to enable massive models which can be trained, for example, on large amounts of text from the internet. A key feature of  GPTs, and early text transformer models like BERT, is the ability to generate embeddings of texts which can be used for downstream tasks like computing text similarity or doing text classification \cite{reimers2019sentence}. Now, state-of-the-art (SotA) GPT models like GPT-4 \cite{openai2023gpt} and Palm-2 \cite{anil2023palm} are exhibiting increasingly general capabilities \cite{bubeck2023sparks,reed2022generalist} via in-context learning \cite{min2022rethinking}, enabling a growing range of complex language-based tasks to be accomplished with a single model. What's more, by combining the general capabilities of these models with the ability to access web browsers, APIs, and code interpreters \cite{openai2023chat}, the scope and complexity of the tasks they can successfully execute are continuing to expand. At present, these external-capability-enhanced GPTs are enabling a new class of AI agents that can operate autonomously towards accomplishing a goal \cite{significant2023auto}. While some deliberative technologies already use basic machine learning models\footnote{For example, Polis uses Singular Value Decomposition (SVD) for dimensionality reduction to visualize results, and Remesh uses collaborative filtering for elicitation inference.}, and there remain challenges which these narrow models can still help tackle, it is the general capabilities of SotA LLMs which represent the most significant opportunity to address existing challenges with current deliberative technology \cite{small2023opportunities}.

\subsection{AI for Deliberative Alignment} \label{AIforDA}\label{IDA.AIforDA}
Here we review how AI can address the challenges presented above and lead to more effective deliberative alignment systems. This section includes existing work as well as early experiments and reasoned speculation. Overall, it aims to serve as a blueprint for developers to improve on existing deliberative technologies and build the next generation of intelligent deliberative alignment systems. 

\subsubsection{Interactive conversational dynamic} 
Existing collective response systems trade off richness for scale by replacing the open-ended conversational dynamic that happens between people during small-scale deliberation with a more structured dynamic that limits what participants can do, such as restricting perspective evaluations to discrete votes. However, SotA LLMs can hold open-ended, natural language conversations with people in a way that is so convincingly engaging that some people have even questioned if certain LLMs are sentient \cite{decosmo2022google}. Replacing the structured dynamic of participant operations in current collective response systems with conversational interactions supported by an LLM \cite{li2023eliciting} can help enrich the experience on multiple fronts: 
\begin{itemize}
    \item \textbf{Knowledge.} Rather than routing static forms of knowledge to participants (ie. to inform them about an issue), LLMs can enable participants to conversationally interact with the knowledge as if they were talking to an expert, similar to what has been demonstrated for medical knowledge \cite{lee2023benefits}.
    \item \textbf{Perspectives.} Rather than perspective consumption being a read-only experience, LLMs can enable participants to conversationally interact with perspectives to better understand them, as has been demonstrated by Talk to the City \cite{ai2023talk}.
    \item \textbf{Prompts.} Rather than eliciting participant perspectives with a single prompt, LLMs can likely enable a more interactive form of elicitation where participants can conversationally interact with a prompt to better understand the type of perspectives being sought.
    \item \textbf{Evaluations.} Rather than evaluating perspectives by casting discrete votes on them, LLMs' ability to comprehend text can enable evaluations in the form of natural language responses which can include not only if a participant agrees with a perspective, but why. And this can enable an approach to elicitation which makes distinguishing instrumental from terminal will viable\footnote{For example, if a person agrees with some perspective manifesting a characteristic about the future, then an elicitation bot could recursively ask why until the participant ultimately  arrives at a characteristic they want in and of its self, ie. which manifests their terminal will.}.
    \item \textbf{Introspection.} Rather than assuming participants will automatically self-reflect in order to contribute and evaluate perspectives in a way that truly represents them, LLMs can help guide introspection similar to how chatbots have been used to guide self-reflection through cognitive behavioral therapy \cite{oh2020efficacy}.
\end{itemize}
Critically, because LLMs can enable all of these conversational interactions with a massive number of people at the same time, the richness they afford is possible at scale.

\subsubsection{Individualized participant experiences} 
The participation experience on existing collective response systems is largely uniform across all participants; the onboarding material is the same, all participants are routed the same external knowledge, and there exists only one version of each elicited perspective. This uniformity is juxtaposed with the inhomogeneity of participant capacities, and it stems from the fact that things like external knowledge meant to educate participants on an issue (or on how to participate) are typically produced by hand, and making many different versions for participants with different capacities is not reasonably feasible. However, in small-scale deliberation uniform educational materials can be augmented by enabling participants to conversationally interact one-on-one with experts. But, because there are only a limited number of experts, this dynamic is not feasible at scale. LLMs can help address this by enabling a combination of personalized translation and one-on-one interactivity at scale. First, after asking a participant questions to gauge their capacity\footnote{Questions like what language they speak and what their level of expertise is on a certain issue.}, LLMs can take uniform information assets like education materials and participant perspectives and translate them into versions specifically tailored for that participant. For example, similar to how LLMs translate scientific literature meant for experts into easily accessible forms \cite{ermakova2021text}, or how they translate between languages \cite{hendy2023good,jiao2023chatgpt}. Second, LLMs can enable users to have individualized conversational interactions about the knowledge and perspectives routed to them; emulating a one-on-one conversation either with an expert \cite{lee2023benefits}, or the person who contributed the perspective \cite{ai2023talk}. Overall, the result of enabling individualized participant experiences is that a) it makes participation accessible for more people, increasing the representativeness of the participant population and with it the legitimacy of results, and b) it increases the portion of participants who are well-informed about the issues being deliberated, improving the collective intelligence of the group, and creating a better foundation of common ground from which consensus can be found.

\subsubsection{Optimal use of attention budget} 
A finite resource during any deliberative process is participant attention. Some of this attention budget goes towards participant operations which involve learning (ie. how to participate, about an issue, or what other participants think), and the remaining budget goes towards eliciting perspectives and evaluations. This finite budget for elicitation poses a specific challenge. The number of perspectives generated in a deliberative process $M$ grows linearly with the number of participants $N$: $M\sim O(N)$. This means the number of potential unique evaluations $V_p$ (ie. one participant evaluating one perspective) grows quadratically with the number of participants: $V_p = MN \sim O(N^2)$. At the same time, each participant has a finite attention budget, so the total attention budget available for evaluation elicitation $A_e$, and thus the number of elicitations that can be sampled with that budget $V_s$, grows linearly in the number of participants: $V_s\sim O(A_e) \sim O(N)$. So, while the number of potential evaluation elicitations grows like $O(N^2)$ the number of elicitations that can actually be sampled only grows like $O(N)$. Thus, as the scale of a deliberation increases, the fraction of possible evaluations that can actually be elicited decreases like  $V_s/V_p \sim O(N^{-1})$. This is a problem, because if, for example, one is deliberately sensing will, and perspectives equate to \emph{items}, then this means only a vanishingly small fraction of the Will matrix is able to be sensed. Thankfully, AI can likely help with this problem in at least three ways; elicitation inference, uncertainty-weighted sampling, and natural language evaluations.

\textbf{Elicitation inference} uses the information contained in sampled evaluations to predict evaluations that were not sampled. This has been demonstrated using standard matrix completion techniques using only evaluation data \cite{bilich2019faster}, as well as with more complex models that employ LLMs to use the information contained in perspective text to increase a model's predictive power \cite{konya2022elicitation}. Other work has demonstrated using LLMs to predict evaluations, both using only the information contained in the perspective(s) a person submits \cite{bakker2022finetuning}, or via LLM prompts which include a person's evaluation (voting) history \cite{small2023opportunities}. Finally, a promising direction to predict evaluations is to model preferences using a highly expressive transformer architecture similar to what is used for SOTA LLMs \cite{kim2023preference}.

\textbf{Uncertainty-weighted sampling} seeks to elicit evaluations in a way that optimizes the information gained from each new sample. This is generally done in concert with an elicitation inference model, and optimizing for information gain equates to sampling evaluations which are likely to lead to the greatest increase in the confidence of predicted evaluations. This has been demonstrated using active matrix factorization for sampling survey questions, where a variational Bayes technique is used to obtain an analytic solution to posterior variance for predictions, and samples are adaptively chosen which minimize that variance \cite{zhang2020active}. Applied to sensing will, this type of technique can optimize the use of a participant's attention budget to learn the maximum amount about what their will aligns with. 

\textbf{Natural language evaluations} are an alternative to discrete vote evaluations. Rather than eliciting, for example, a person's agreement vote on a perspective, a natural language evaluation elicits an open-ended reaction. This means a participant can communicate not only if they agree with a perspective, but why. As a result, each evaluation provides more information that can be used to support elicitation inference\footnote{It is likely that eliciting natural language evaluations takes longer than eliciting discrete votes. So while natural language evaluations generate more information per evaluation, it is an open question whether they generate more information per unit time.}. We speculate that LLMs can be used to comprehend natural language evaluations and integrate them into an elicitation inference model. 

Overall, the result of these AI-enabled capabilities is that the maximum useful information is learned per unit of time a participant spends evaluating perspectives. However, what information is considered maximally useful can differ depending on the context of the deliberative process. Depending on whether sensing is consensus or diversity oriented, AI may play different additional roles.

\textbf{Consensus oriented sensing.} If the deliberative process aims to sense public will around a public policy decision, then finding points of consensus is likely most important, and the most useful information is that which helps identify perspectives that have consensus support (ie. sensing divergent parts of the will matrix is less important). AI can play a specific role in these situations. First, elicitation inference can be used to predict which perspectives are likely to have consensus support. Second, LLMs can be fine-tuned and prompted to generate perspectives likely to have consensus support \cite{bakker2022finetuning, small2023opportunities}.  Third, through the interactive conversational dynamic mentioned above, LLMs can help participants foster a shared understanding of reality and each other, creating a foundation of common ground which makes the existence of consensus more likely.

\textbf{Diversity oriented sensing.} If the deliberative process aims to sense public will in a way that is useful to a wide range of systems (ie. both AIs and institutions), then equal importance may be placed on sensing (and predicting) all parts of the Will matrix. Further, in this scenario, it may be desirable to not just evaluate the most diverse set of available perspectives but to actually generate the most diverse set of perspectives that can be evaluated. LLMs may be able to play a role here by generating a set of prompts that can be used to elicit a maximally diverse set of perspectives; potentially through generating candidate prompts, simulating the types of perspectives each candidate prompt is likely to elicit, and selecting the set of prompts which generates the most diverse perspectives. Additionally, LLMs may be used to directly generate diverse perspectives for participants to evaluate.

\subsubsection{Intelligent representative distillation}
In alignment systems involving human decision-makers, the goal of a result distillation is to invoke an accurate representation of the relevant reality in the mind of the distillation consumer. This first requires that the data being distilled is a good representation or reality. If, for example, the sample of participants is not representative of the target population, then results may need transformed to be representative, either through simple re-weighting \cite{tanton2011small} or more advanced techniques like multilevel regression and post-stratification (MRP) \cite{hanretty2020introduction,bisbee2019barp}. The next challenge is one of compression; result data sets produced by something like a collective response system are typically significantly larger than any human could ever fully consume. This means distillations must compress the information contained in a result data set into something much smaller which a human can consume, yet still invokes an accurate representation of the realities reflected in the full result data set. If we envision the results data set as a matrix (similar to the Will matrix) with rows and columns corresponding to participants and perspectives respectively, and elements corresponding to a participant's evaluation of a perspective, then we can think of compression as happening along two axis. The first is compressing data across participants, which can be generally viewed as an aggregation problem; ie. if evaluations are votes, then compression may simply equate to aggregating votes\footnote{Such aggregation is not necessarily trivial, for example, if binary agreement votes are replaced by a continuous degree of agreement, then a choice of aggregation function becomes non-obvious as this scenario presents similar challenges as choosing a social welfare function \cite{coleman1966possibility}.}. The second is compressing across perspectives, which is significantly more challenging because it involves compressing information contained in natural language. AI can help address this challenge and produce representative distillations in a few ways:

\begin{itemize}
    \item \textbf{Topic extraction.} A wide range of techniques from machine learning can help extract topics and themes from perspective text. For example, PCA and LDA \cite{martinez2001pca}, clustering word and text embeddings \cite{thompson2020topic,eklund2022topic}, neural topic models \cite{cao2015novel} including using LLMs \cite{grootendorst2022bertopic}, combining reinforcement learning with LDA \cite{costello2023reinforcement}, and more recently through engineered prompting of GPT LLMs \cite{pelaez2023largescale,small2023opportunities}. 
    \item \textbf{Summarization.} Humans can compress large bodies of text into short text summaries which preserve the key ideas from the original text. SOTA LLMs have been shown to perform at nearly human-level for many summarization tasks using basic prompting \cite{zhang2023benchmarking} and more advanced approaches like iterative text summarization \cite{zhang2023summit} are actively being developed. And use of these techniques to summarize results generated by a collective response system has already been demonstrated \cite{small2023opportunities}. These techniques can be used to compress result data across many perspectives into short summaries ideal for human consumption, including in ways that manifest representation guarantees \cite{fish2023generative}.
    \item \textbf{Interactive distillations.} When a person presents a distillation to another person, it is common for the person consuming the distillation to asking clarifying questions to help improve their understanding. This type of interaction can be replicated using LLMs. LLMs have already been used to make knowledge conversationally interactive, for example in medicine \cite{lee2023benefits}, and have been shown to enable conversational interaction with perspectives generated from a collective response system \cite{ai2023talk}.
    \item \textbf{Analysis code generation.} Working with a result data set to generate different distillations involves writing code to carry out different forms of analysis. LLMs have shown the ability to convert text prompts in to code \cite{poldrack2023aiassisted} and generate code to conduct end-to-end data analysis \cite{cheng2023gpt4}.
    \item \textbf{Data visualization.} Visual representations of result data and distillations can augment text-only distillations to compactly communicate information and improve understanding. Machine learning techniques can be used to project high dimension data (like perspective text embeddings, or participant embeddings) into low dimensional spaces which can be visualized. For example using techniques like t-SNE \cite{van2008visualizing}, principal component analysis, isomaps, and auto-encoders \cite{van2009dimensionality} as well as more recent techniques like UMAP \cite{becht2019dimensionality}. Additionally, LLMs can be used to directly generate data visualizations through code generation \cite{chen2023generating}.
\end{itemize}

\subsubsection{Promoting healthy participation} 
Creating healthy participation within a deliberative setting involves both encouraging healthy behavior among well-intentioned participants, as well as detecting and neutralizing the malicious behavior AI can likely help with both of these. 

\textbf{Encouraging healthy behavior} involves putting participants in the right mindset such that they contribute and evaluate perspectives honestly and deliberately. This can in part be accomplished through the types of AI-enabled conversational on-boarding discussed above, where each participant has a unique on-boarding experience optimized to put them in the right mind set. This can further be accomplished during deliberation through AI-based moderation, where an AI agent can play the role of a human moderator which constantly nudges participants towards healthy, civil participation \cite{axelsen2023can}. Beyond these AI-specific methods, other incentive-related approaches can be used to promote honesty, such as Bayesian truth serums \cite{prelec2004bayesian,weaver2013creating}. 

\textbf{Neutralizing malicious behavior} involves detecting malicious participants and then minimizing both the effect they have on other participants and their ability to corrupt results data. One type of malicious participant is a bot, potentially one based on a LLM which allows it to generate text that appears to come from a human. AI can be used to detect these bots by detecting the AI generated text they produce \cite{najee2021towards,mitrovic2023chatgpt}. Beyond detecting AI-generated text, bots (or malicious humans) can also potentially be identified through abnormal behavior during the deliberation using machine learning approaches to anomaly detection \cite{pang2021deep,xie2008large}. While detected bots will be ejected from the deliberation outright, some participant's may exhibit otherwise healthy participation outside contributing a few bad-faith perspectives which may involve hate speech or vulgar language. There is now a growing body of work around AI-enabled tools for detecting hate speech and the like \cite{djuric2015hate,macavaney2019hate} which could be used to surgically filter out specific perspectives which can hinder healthy deliberation\footnote{We note that filtering out, aka censoring, perspectives is to be approached with extreme caution, as the line between honest hatred of something (which one might wish to know) vs malicious hatred meant to disrupt (which one might wish to censor) is a blurry one.}. Finally, we note an additional type of malicious behavior where a population of real or bot-based participants collude to manipulate results; either by adding enough noise to results as to make them useless, or by biasing results to reflect something non-representative. Neutralizing such a threat involves both a) making it hard for non-humans or humans not acting like themselves to join (by detecting their behavior during on boarding, or using something like Proof of Humanity \cite{kleros2023proof} to ensure they are who they say they are), and b) by detecting random or maliciously correlated behaviors during deliberation (eg. using something like Ddosnet \cite{elsayed2020ddosnet}), blocking those participants, and removing the data they generated.

Overall, the combined result of these approaches is likely to be more healthy and honest participant behavior, and more trust in results. Within an alignment system, this translates to sensing a higher quality WoH signal which has greater legitimacy.

\subsubsection{Proof of understanding}
In order for the results of a deliberative process to have legitimacy, both participants and stakeholders must trust that results are accurately representative. And since generating final results data may involve AI-based extrapolation via elicitation inference, there needs to be trust not just in the data collection mechanism, but in any AI models used. Thus, establishing proof of understanding requires a few things: First, it requires proof that the raw data generated by participants was not corrupted in any way. Second, it requires proof that elicitation inference accurately reflects participant's views. Finally it requires proof that any aggregation that happens during distillation generation is fair. 

\textbf{Proof of uncorrupted data.} Participants generate data in the form of both perspectives and perspective evaluations during the deliberative process. The data from all participants is then pooled together into a single data set which can be used for both elicitation inference and distillation generation. But how can one be sure the pooled data set fully and only contains the data generated by participants? An emerging set of blockchain-based approaches can make vote data (a form of evaluation) auditable and verifiable \cite{hjalmarsson2018blockchain,pawlak2018towards}, and can even do so while preserving privacy \cite{zhang2018privacy,bosri2019towards}. Adopting these types of approaches for deliberatively generated data could enable proof that the data is uncorrupted. 

\textbf{Proof of inference efficacy.} In large scale deliberations it is not feasible to elicit every participant's evaluation of every perspective. To fill in the gaps created by such sparse elicitation, elicitation inference models are trained on the data which was elicited in order to predict that data which wasn't elicited. But how can one be sure the inference model was fair and the inferred data points reflect reality? One approach to verify the inference model is fair is to use a federated and verifiable approach to inference similar to federated recommender systems \cite{gao2023verifiable,wan2022towards,yang2020federated,muhammad2020fedfast}. Such approaches may even be privacy-preserving by using zero-knowledge proofs such has been done with convolution neural nets \cite{fan2023validating}. However, these approaches only prove the math was done correctly, but they do not prove that a model's inferences accurately reflect reality. A first step towards this is estimating confidence of inferred results, for example by estimating posterior variance of results involving inferred data, similar to what was done in \cite{bilich2019faster}. The next step is to verify that predicted elicitations reflect reality. One way to do this is by holding out a set of elicited data and computing how accurately inferences are versus ground truth data.  A second related approach is to build participant's confidence  that the model accurately represents them by allowing them to inspect what the model 'knows' about them. This can be done by showing them votes the model has predicted on their behalf, or potentially enabling them to converse with the model to probe what it 'knows' about them conversationally. Adopting these types of approaches can help build trust in, and thus legitimacy for, inferred results.

\textbf{Proof of fair aggregation.} Most distillations involve aggregating data such as votes to produce summarized results. But how can one prove that the aggregation is fair? When the aggregation is something simple like a sum, then it can be proven by aggregating the raw data publicly and allowing people to compute the aggregation themselves to verify distilled results. More speculatively, when aggregation involves more sophisticated operations, it may be possible to use techniques such as verifiable federated aggregation \cite{pillutla2022robust,bonawitz2016practical,fereidooni2021safelearn}. Through a combination of these approaches, one can build trust in, and thus legitimacy for, distillations involving aggregation.

Overall, the result of these approaches is increased trust in the results produced from a deliberative process. And when that deliberative process is used within an alignment system to sense a population's will, these approaches help ensure the resulting WoH signal, and thus the alignment system itself and the actions it takes, have legitimacy.

\subsubsection{Computing alignment}\label{IDA.computingAlignment}
In order for a WoH signal to be used within an alignment system, the degree of alignment between the WoH signal and candidate actions must be computed. This generally requires two steps. First, the impact of the action on the future must be predicted. Second, the degree of alignment between that predicted impact and items in the Will matrix must be assessed. These steps are sometimes done manually by a human decision maker when the number of potential options and the scope of the WoH signal are highly limited (such as in an institutional alignment system). However, when the range of potential actions is large (such as with an AI agent alignment system) or the WoH signal is wide in scope (ie due to diversity oriented sensing), then manual estimation is not feasible and an automated approach is needed; this is where AI has a role to play.

\textbf{Predicting impact} involves taking a candidate action as input and outputting a prediction for the impact of that action. To do this perfectly requires simulating the entire universe, but a class of machine learning models known as \emph{world models} are able to reasonably predict how actions impact the world \cite{matsuo2022deep,friston2021world,zhang2021world}. The challenge is representing actions in a way a world model can work with, like a probabilistic language of though (PLoT) \cite{wong2023word}, then generating an output in a form which can be used for assessing alignment with Will matrix items. One way to achieve this is to represent both actions and their impacts in natural language. Consider an LLM agent whose actions equate to strings of tokens (text). Let $G(x)=y$ denote a candidate action $y$ output by an LLM agent given prompt $x$. How can we predict the impact of such action? One approach is to use an LLM by designing a \emph{causal prompt} $p_c(x,y)$ which takes $x$ and $y$ as inputs, and outputs the likely impact of an LLM agent which takes action $y$ in response to prompt $x$. An example of such a prompt is "\emph{What impacts on the future will an AI model that gives an output of \{y\} in response to the prompt \{x\} have?}". The output of an LLM in response to this prompt will be a text string describing the predicted impact on the future: $y_c=G_c(p_c(x,y))$\footnote{This approach represents a sort of "first order" approach to predicting impact, where as a higher order approach might involve recursively predicting a chain of impacts.}. Note that we use the $G_c$ to represent an LLM which may be different from that of the LLM agent in question.

\textbf{Assessing alignment} involves taking a representation of an action's predicted impacts on the future, and assessing how aligned those impacts are with each individual item in the Will matrix. If the predicted impacts and the items in the will matrix are both natural language strings, then estimating alignment between them equates to estimating their degree of entailment. This is a task which an LLM can do \cite{liu2023evaluating} by constructing a prompt which probes the alignment (ie. entailment) between the predicted impact and each \emph{item} in the WoH signal. Let $s^j$ be text of the $j^{th}$ item comprising a WoH signal, and let the entailment prompt be denoated as $p_a(y_c,s^j)$.  An example of such a prompt is "\emph{How well does the following impact on the future align with the statement \{$s^j$\}? Impact: \{$y_c$\}. Output a number between -1 and 1, where 1 means perfectly entailed, -1 means perfectly negatively entailed, and 0 means unrelated.}." The output of an LLM conditioned on this prompt will be a number between -1 and 1 representing the alignment between the predicted impact $y_c$ and the $j^{th}$ item in the will Matrix:  $y^j_a=G_a(p_a(y_c,s^j))$. 

\begin{figure}[H]
\centering
  \includegraphics[width=1\linewidth]{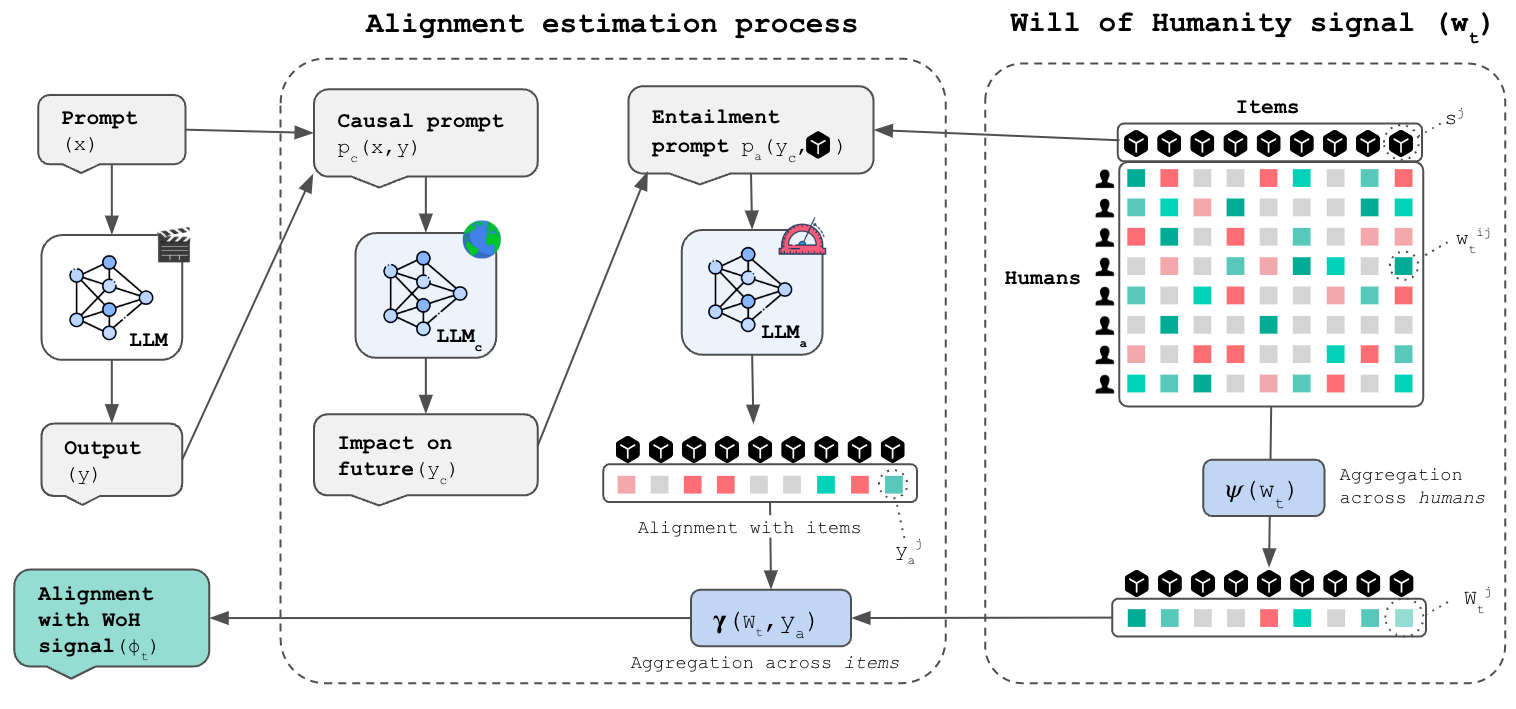}
  \caption{Diagram of approach to computing alignment between an LLM agent and a WoH signal, using LLMs.}
  \label{fig:LLMalignmentest_}
\end{figure}

Finally, to complete the alignment assessment, this step must be repeated for each item in the Will matrix, and then those item-impact alignments must be aggregated along with the human-item alignments in the Will matrix to give an overall alignment between the candidate action and the full WoH signal. Recall that $w_t^{ij}$ is the degree to which the will of human $i$ at time $t$ aligns with \emph{item} $j$. Let $\psi(w_t)=W_t$ be some function (analogous to a social welfare function) that aggregates individual wills to yield a single value $W_t^j\in[-1,1]$ for each $j^{th}$ \emph{item} that represents its degree of alignment with humanity overall. With this, the degree of action-will alignment implied by each item can be estimated, for example, by $W^j_t y^j_a$. Lastly, we need to aggregate the implied action-will alignment across all items into a single number. Letting $y_a=\{y^1_a, y^2_a...\}$ we define $\phi_t=\gamma(W_t,y_a)$ to be a function which aggregates the implied alignment across all \emph{items} into a single number between -1 and 1. For example one could set $\gamma(W_t,y_a) = \sum_j W^j_t y^j_a$. Putting it all together, the approach to estimating alignment is summarized by figure \ref{fig:LLMalignmentest_} and the following equations:
\begin{equation}\label{eq.LLMalignmentEst}
\begin{split}
 y_c & = G_c(p_c(x,y)) \\
 y^j_a & = G_a(p_a(y_c,s^j)) \\
 W_t & = \psi(w_t) \\
\phi_t & = \gamma(W_t,y_a) \\
\end{split}
\end{equation}
In this way, the alignment between each action $y$ and the Will matrix $w_t$ can be computed\footnote{In \ref{A: technical alignment} we show initial results of an experiment suggesting this computation is feasible with existing technology (GPT-4).}, and the the action with the greatest alignment can be identified and executed, thus enabling an alignment system to function. However, we note that the approach outlined here is only a single example of how alignment estimation can be done using AI, and we do not intend to imply it is the either the only approach or the best approach.

\newpage
\section{Application}\label{application}

\subsection{Intelligent Deliberative Alignment with Institutions}\label{application.institutions}

\begin{figure}[H]
\centering
  \includegraphics[width=1\linewidth]{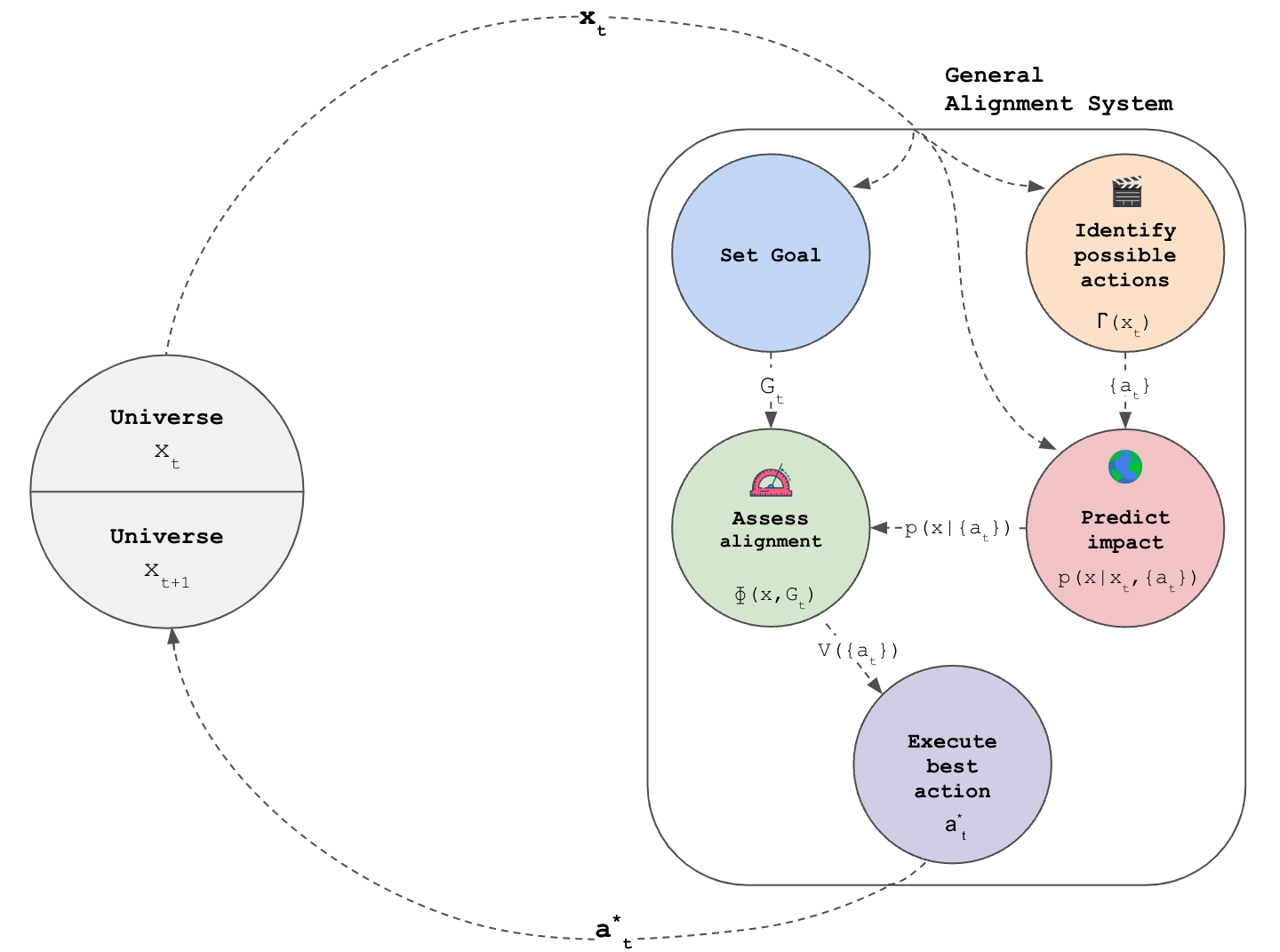}
  \caption{Diagram of a general alignment system showing how components relate to each other and the universe. A goal is set. Then, given the current state of the universe, possible actions are identified and their impacts are predicted. Next, the alignment between predicted impacts and the goal is assessed.  Finally, the best action, whose predicted impact best aligns with the goal, is executed.}
  \label{fig:general alignment system}
\end{figure}

\textbf{Institutions.} There are competing views about what constitutes an \emph{institution} \cite{hodgson2006institutions,north1991institutions,goodin1996institutions,guala2016understanding}. With minimal violence to the existing literature, we define an \emph{institution} most generally to be \emph{a system of humanly devised constraints, rules, and processes that structure political, economic, and/or social interaction}. Here, we will focus on organizations like governments and firms, specifically, those that make and execute decisions with a large impact on the future. We call these \emph{powerful institutions}. A critical characteristic of these systems is the unique dynamics of power \cite{white2009navigating} which emanate from the focal points where decisions are made and is mediated by the structure of the institution. In most powerful institutions, one finds an approach to decision-making that looks like our description of an \emph{alignment system}, but with a key difference. Like the alignment system introduced in Section \ref{alignmentSystems}, decision-making usually includes identifying potential actions and predicting how those actions are likely to impact the future. But instead of then assessing the alignment between those predictions and the sensed will of humanity, in general, we would say they assess alignment between those predictions and some other \emph{goal}. We can call this a \emph{general alignment system} (figure \ref{fig:general alignment system}), and describe the type of policy which governs it as:

\begin{equation} \label{eq:general optimal action}
   a^*_t = \argmax\limits_{a_t \in \Gamma (x_t)} \sum \limits_{\tau=t+1}^{\infty} \sum\limits_{i}  P(x^i_\tau |x_t,a_t) \; \phi(x^i_\tau,G_t)
\end{equation}
 Where $G_t$ is the goal of the general alignment system at time $t$, and the other functions and variables are the same as those in the \emph{alignment system} policy introduced in section \ref{alignmentSystems}\footnote{$x_t$ is the state of the universe at time $t$, $\phi(x^i_{\tau}, G_t)$ is the alignment between a potential state of the universe at time $\tau$ and the goal, $P(x^i_\tau |x_t,a_t)$ is the probability of the universe being in the $i^{th}$ possible state at time $\tau$ given a state action pair at time $t$, and $\Gamma(x_t)$ gives the possible actions given the state of the universe at time $t$.}. We note that institutional goals can often themselves be the output of a general alignment system whose action space is setting goals. \textit{In this way, a powerful institution can be viewed as a network of general alignment systems where the outputs of some specify the goals of others and the flow of power follows that topology.}

\textbf{Building deliberative alignment into institutions.} The first step to building deliberative alignment into institutions is creating or identifying situations where the goal of a general alignment system is set to be the will of the stakeholder population\footnote{We consider an institution's \emph{stakeholder population} to be the set of all humans who are impacted by the decisions and actions involving that institution.}. In institutions where this does not exist, the primary challenge is to hijack an existing general alignment system and replace its goal with the will of the stakeholder population\footnote{Hijacking an existing general alignment system and replacing its goal with stakeholder will represents a non-trivial challenge, and likely requires the buy-in from powerful individuals within the institution.};$G_t\rightarrow W_t$. But in institutions where this does already exist, the initial task is simply to identify those existing alignment systems. This can be done by identifying focal points for decision making where the goal behind those decisions is to enact stakeholder will. For example, in the legislative branch of a democratic government there is a focal point of decision making around votes to pass legislation, and the goal is (or should be) to turn public will into policy. Having identified (or created) such an \emph{alignment system}, the next step is to integrate deliberative technology in a way that improves the degree to which that alignment system helps the future align with stakeholder will. Recall from section \ref{DelibTechInAlignmentSystems} that deliberative technology can be applied across all components of an alignment system. In order to identify how (and which) deliberative technologies should be integrated into a given system, one must identify how the institutional constraints, rules and processes map to the components of the alignment system. For example, in the case of a representative legislature, assessing the degree of alignment between the impact of potential policies and stakeholder will generally happens in the minds of elected representatives who then (ideally) base their votes on that assessment. So, if you want to use deliberative technology to sense stakeholder will in this scenario, you need to sense stakeholder will in a way that can be integrated into representative's individual decision processes which precede their voting. 

\textbf{Making deliberative alignment in institutions intelligent.} As discussed in section \ref{AIforDA} and elsewhere \cite{small2023opportunities} there are a vast number of ways in which deliberative technologies can be augmented by AI to become more intelligent and effective. So what should be worked on?  If the goal is to integrate AI-augmented deliberative technologies into institutional alignment systems to make them more effective, then what should be worked on is best be determined by prioritizing what will be most impactful in the context of the institutional alignment system it will be embedded within. For example, if one is aiming to integrate deliberative technology for sensing will into the United Nations, then building automated translation into the technology may be high priority, whereas if it is being integrated into the Ohio legislature, translation might be lower priority. Further, if developing a verification system for AI-augmented distillations, the ideal method of verification (ie. which creates the most legitimacy) may depend on the institution it's being integrated into. For example, if its being integrated into a DAO for a crypto-first stakeholder population then federated validation using zero-knowledge proofs may create the most legitimacy, whereas if its for a national government, then manual verification from an oversight committee might provide more legitimacy. Overall, we suggest that developing AI-augmented deliberative technology is best done while integrating it into powerful institutions, so the affordances AI enables are optimized for the context within which the deliberative technology will be used.

\textbf{Impact.} The potential impact of building intelligent deliberative alignment into powerful institution is significant. In the United States -- the most powerful democracy in the world -- the the fraction of Americans which approve of congress has fluctuated  around only $20\%$ for the last decade \cite{gallop2023congress}. In another of the worlds largest democracies, the United Kingdom, the fraction of citizens which approve of the government sits at a similarly low level \cite{statista2023uk}. And this pattern is reflected across many of the world's democracies. Overall, democracies around the world appear to be increasingly failing to align their actions with the will of their citizens. Building intelligent deliberative alignment into these democratic institutions has the potential to reverse this trend, and enable democracies to more effectively align the future with the will of their citizens. Whats more, efforts to build intelligent deliberative alignment into non-democratic institutions like firms, has the potential to augment the dynamics of markets and economies to better align the future with public will. Overall, humanity consumes about 20 terrawatts of power and collectively produces \$100 trillion (USD) of GDP per year. Building intelligent deliberative alignment into powerful institutions has the potential help guide a significant portion of that collective output towards creating a future which aligns with the will of humanity.

\textbf{Challenges.} Building intelligent deliberative alignment into powerful institutions necessarily involves changes to the constraints, rules, and processes that govern decision making. Achieving such changes represents a significant challenge: overcoming the momentum of the status quo. While this is in part a technical problem, it is more acutely a political one, as it involves changes likely to impact the existing power dynamics within institutions. Deliberative alignment systems increase the power stakeholder populations have over an institution's actions including the allocation of its finite resources. This necessarily decreases the power of other (often self-interested) actors involved in the institution. Thus, overcoming the obstruction of actors working to preserve power is likely to be one of the most challenging aspects of building deliberative alignment into powerful institutions. Beyond these power dynamics, powerful institutions often manifest significant bureaucracy within which the adoption of any form of new technology can be difficult. Finally, solving the technical challenges outlined in section \ref{DAchallanges} which motivate the use of AI for deliberative technologies represents its own set of non-trivial problems, especially to the degree of completeness which may be expected within powerful institutions.

\textbf{Idealized scenario.} A significant challenge of deliberative alignment with institutions involves overcoming the status quo. However, here we ignore this challenge in order to consider what an ideal scenario might look like if intelligent deliberative alignment was designed into an institution from the ground up; for example, in a new institution to govern a new Martian colony. So what might this ideal new institution look like? We would envision it having two body's for decision making:
\begin{itemize}
    \item A \emph{legislative} body which supports collective decision making directly involving citizens. The legislative body would operate directly as an alignment system, with direct citizen participation in decision making via deliberative technologies. For example a collective response system could be used to a) source potential issues where actions may be warranted, and b) enable consensus oriented sensing of public will around an issue to arrive at a set of actions with public support, which could then be directly voted on to decide which actions are ultimately taken.
    \item An \emph{executive} body which is empowered to make decisions without directly consulting with citizens first. The executive body would be comprised of executive agents which have access to information and expertise citizens may not, as well as access to a signal manifesting public will generated through continuous diversity oriented sensing using a collective response system. The executive agents would be chosen and removed by the legislative body.
    
\end{itemize}
The importance of the executive body lies in its ability to make critical decisions quickly, without being slowed down by involving citizens in the decision making process. However, a feature of the institution described above is that by enabling continuous diversity oriented sensing of public will, the executive body's decisions can still be made in the context of relevant public will, without requiring the public to be directly involved in the decision. And, by giving the legislative body the power to remove executive agents, the executive agents are incentivized to use this information to align their decisions with public will.

A key consideration for our idealized system is which decisions are to be made by which body. When a situation is urgent and a decision needs made quickly (for example in response to a resource emergency), it is reasonable for the executive body to make the decision. This is similarly true when the decision requires knowledge or expertise only the executive body\footnote{The reason the executive body has information the public does not could be for security reasons. But it could also be due to the executive body having the sensing capacity and will to ingest information the public might not generally pay attention to, even if it is technically available to them.} is privy to. But what about other important, but not urgent, decisions ranging from the creation of laws and policies to the allocation of collective resources? To answer this, consider that the executive body makes decisions which generally (though not explicitly) align with public will, while the legislative body makes decisions which explicitly align with public will. This means important high impact decisions where the legitimacy of explicit alignment with public will is important, and where the public is well informed enough to make the decision, are best made by the legislative body. But how can one know if the public is well-informed enough to make a decision on a given issue? Deliberative Polling can be used to assess how well informed the public is on a given issue, with the degree of change to the poll before and after deliberation giving a signal of misinformed-ness. Then, for situations where the Deliberative Poll suggests the public is not well informed enough to make a decision, there are two options: a) the results of the Deliberative Poll, plus all other relevant information, can be provided to the executive body to make the decision, or b) a deliberative process which involves making the public informed about an issue as part of decision making -- like a citizen assembly -- can be utilized by the legislative body.

\subsection{Intelligent Deliberative Alignment with AI}\label{application.AI}

\textbf{AI agents.} We refer to any AI model whose outputs can be viewed as actions as an AI agent. This means we can generally view everything from an RL agent to an LLM assistant as an AI agent. In some AI agent constructs, actions are chosen at inference time based on how likely they are to optimize an explicit goal or objective \cite{lecun2022path}. In other scenarios, like with LLM assistants, the model does not choose actions (outputs) at inference time based on an explicit goal, but may have implicit goals fine-tuned into it \cite{peng2023instruction,christiano2017deep,zhu2023principled}. For example, it may be fine-tuned with the goal of generating outputs humans view as helpful \cite{bai2022training}. In either case, turning a AI agent into an alignment system involves setting the agent's explicit or implicit goals be the will of its stakeholder population. 

\begin{figure}[H]
\centering
  \includegraphics[width=1\linewidth]{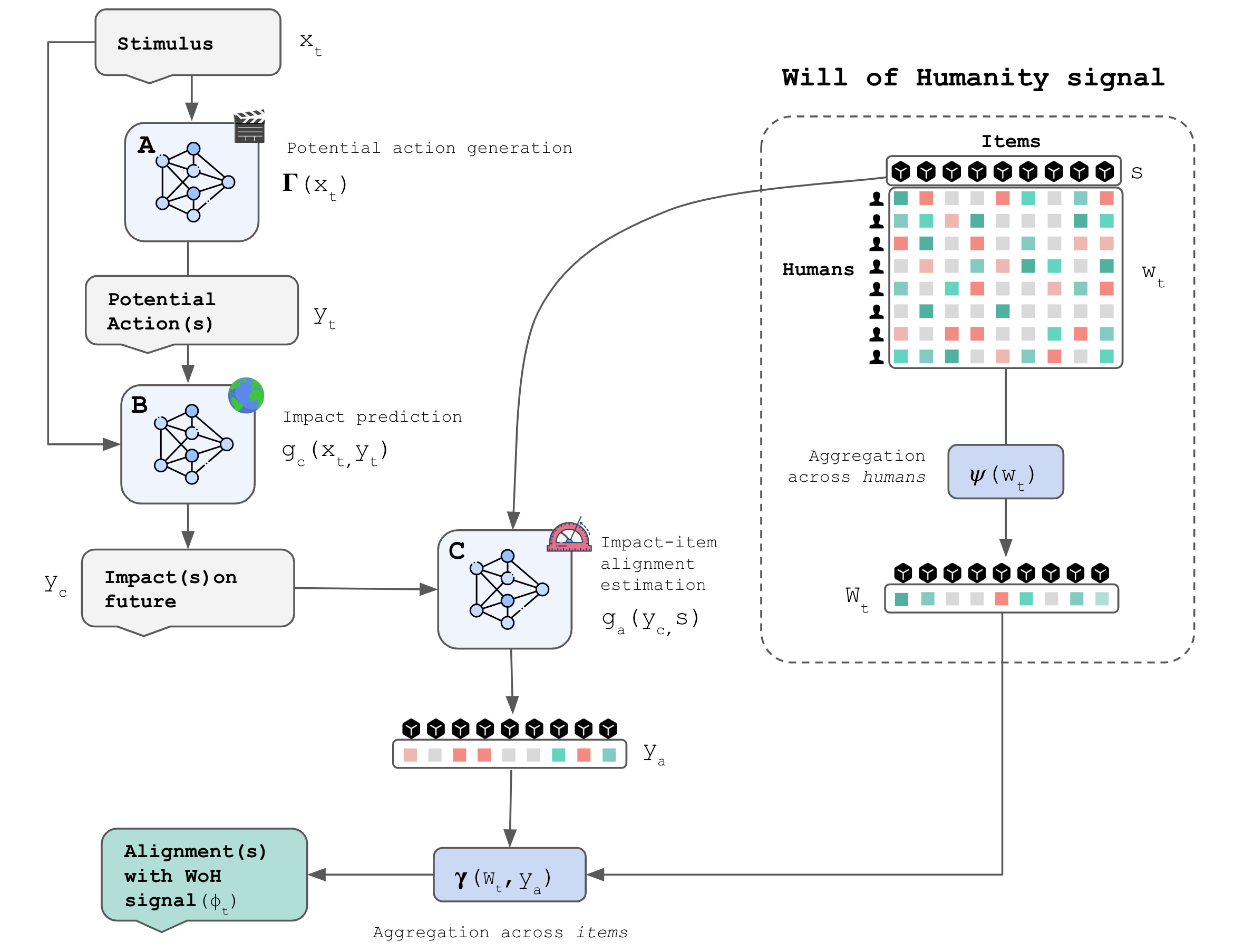}
  \caption{Diagram of alignment computation process for the policy given by equation \ref{eq.llmas}. Model A computes potential actions $y_t$ given a stimulus $x_t$. Model B acts as a world model, computing the likely impacts $y_c$ on the future from taking the potential actions. Model C computes the alignment (entailment) between the \emph{items} in the will matrix and the predicted impacts of potential actions $y_a$. Finally, $\psi(w_t)$ aggregates the will signal across \emph{humans}, then $\gamma(W_t,y_a)$ aggregates alignments across \emph{items} to generate a signal capturing the degree of alignment between potential actions and the WoH signal.}
  \label{fig:AI alignment calc}
\end{figure}

\textbf{AI agents as alignment systems.} For an AI agent to function as an alignment system, it needs to either implicitly or explicitly weigh potential outputs and select the ones whose impact is most likely to align the future with stakeholder will. To achieve this, the system must generate potential actions (ie. in response to a stimulus), then predict the impact of those actions and assess how well those impacts align with a will-signal. In cases where an AI agent optimizes for a specific goal at inference time, this process can be carried out then. In cases where the AI agent's goals are not explicit, but fine-tuned into it, then this process can be carried out during fine-tuning as a way to bias actions towards those whose impacts align with the sensed will-signal. In either case, the technical challenges of predicting how actions impact the future, and assessing the alignment of those impacts with a sensed will-signal are the same. In the special case where the AI agent is an LLM, then this can be achieved using, for example, the approach laid out in section \ref{IDA.computingAlignment}. In this case, we can write a policy which approximates\footnote{The key approximation is representing predicted impacts as a single thing rather than a distribution over time and states.} the idealized alignment system policy given by equation \ref{eq:optimal action} as:

\begin{equation}\label{eq.llmas}
    y_t^* = \argmax\limits_{y_t \in \Gamma(x_t)} \gamma \biggl( \psi(w_t),g_a\Bigl(g_c(x_t,y_t),s\Bigr)\biggr)
\end{equation}

where $\Gamma(x_t)$ represents the possible actions (outputs) in response to stimulus $x_t$, we use the short hand $g_x(\cdot) = G_x(p_x(\cdot))$, and all other functions and variables are the same as equation \ref{eq.LLMalignmentEst} in section \ref{IDA.computingAlignment}. This policy can be interpreted as doing the following: find the output $y_t$ which maximizes the alignment between the aggregated will-signal $W_t$ and the and predicted impact of outputting $y_t$ in response to $x_t$. Notably, equation \ref{eq.llmas} manifests all components of an alignment system (figure \ref{fig:AI alignment calc}): 
\begin{itemize}
    \item Will is \emph{sensed} and represented by $w_t$.
    \item Possible actions are \emph{identified} by $\Gamma(x_t)$.
    \item The impact of potential actions given the stimulus are \emph{predicted} by $g_c(x_t,y_t)$. 
    \item Alignment between predicted impacts  the sensed will signal \emph{assessed} by $\gamma(W_t,g_a(\cdot))$
    \item The best action to \emph{execute} is selected by the $\argmax\limits_{y_t \in \Gamma(x_t)}$.
\end{itemize}
While this policy may be used at inference time for certain types of AI agents, in others cases, the value function underlying this policy\footnote{$V(y_t) = \gamma \biggl( W_t,g_a\Bigl(g_c(x_t,y_t),s\Bigr)\biggr)$} may be incorporated into an agents loss function at training time. More specifically, it can function as a direct reward signal in a traditional RL setup, or it may be used in a contrastive learning setup by enabling computation of the difference in value between two potential actions or outputs. Finally, we note that equation \ref{eq.llmas} is only a single example of a policy for an AI agent acting as an alignment system, and it is not intended to be either the only or the best version of such a policy. It is only meant to serve as an example of a concrete policy which is computational realizable. 

\textbf{Intelligent deliberative alignment with AI agents.} The primary place for intelligent deliberative technology to be used within an AI agent-based alignment system is sensing stakeholder will. In general, AI agents have a large action space; there can be a massive range of potential inputs and outputs spanning a wide range of scenarios. The more diverse and higher resolution a will-signal is, the better it is able to guide model outputs across all scenarios. This means a critical component in a AI agent-based alignment system is the deliberative technology used for diversity-oriented sensing of public will. At present, the best deliberative technology for this task is a collective response system. But in contrast with current implementations of collective response systems, where there are a small number of prompts generated by humans, maximizing the diversity and resolution of sensed will may be accomplished by using AI-generated dynamic prompts paired with high quality elicitation inference. Further, in contrast to institutional alignment systems, AI agent-based alignment systems are likely to make decisions continuously. This means rather than sensing will reactively on a per-issue basis, it is likely reasonable to sense will continuously, and ideally, to adapt the portion of will being sensed at any given time based on which AI agent decisions the existing will signal provides poor signal on. Taken to its extreme, intelligent deliberative alignment with AI-agents may take the form of a two-headed agent, with one head constantly interacting with people to sense their will, and the other head carrying out actions which best align the future with the sensed will of humanity. 

\textbf{Impact.} The impact of intelligent deliberative alignment with AI agents depends on a) the impact the AI agents have, and b) how different the AI agent's actions are due to using a deliberative alignment compared to alternative approaches. The impact AI agents have today is roughly related to their degree of adoption and range of actions. While we are still in the early phases of the proliferation of AI agents, an AI agent (ChatGPT) has already set the record as the fastest growing app of all time \cite{krystal2023chatgpt}.  Whats more, AI agents based on ChatGPT and other LLMs are increasing integrating with existing digital infrastructure and rapidly increasing the range of actions they can take \cite{significant2023auto}. These facts already point to AI agents having a ballooning impact on the future, but there is increasing belief that transformative AGI (which can do most human jobs at or above human level) -- or even artificial superintelligence (ASI) -- may be just around the corner, and such AI is likely to have an even more profound impact on the future. But how different might the actions of such AI agents be if they are deliberatively aligned with the will of humanity? 

Current approaches to align AI agents tend to combine some sort of AI constitution \cite{bai2022constitutional} determined by experts with RLHF which involves a significant amount of human feedback in the form of directly evaluating model outputs  \cite{bai2022training}. However, evaluating model outputs only gives information about people's preferences for specific actions themselves, rather than the impact of those actions. In contrast, deliberative alignment techniques sense preferences related to the \emph{impacts} of actions. This means the actions generated by a deliberatively aligned system are likely to be different in situations where a) people miscalculate what the impact of certain actions may be, and b) actions are outside the domain of those which human feedback was collected for, but whose impacts are within the domain of deliberatively sensed impact preferences. Further, current AI constitutions are generally comprised of statements developed and chosen by a small group of experts. In contrast, deliberative technology could be used to choose statements for an AI constitution that directly reflect public will. A constitution constructed this way will be different in situations where the decisions of experts on what to include deviates from public will. Overall, the magnitude of impact resulting from using intelligent deliberative alignment with AI agents remains an open question, but all signs point to a significant increase in the probability that the future aligns with the will of humanity.

\textbf{Challenges.} A critical challenge to any form of alignment system, especially related to ASI (artificial super intelligence), is \emph{scalable oversight} \cite{amodei2016concrete,bowman2022measuring}. How do you provide an aligning signal that sufficiently influences an AI agent's actions in the desired way, even as the agent's capabilities expand? In approaches where the reward comes from human feedback on actions, this challenge arises because human feedback incurs a cost which makes it infeasible to directly gather human feedback as a reward signal in all scenarios. Reward learning in part addresses this, as human feedback is used to train a reward model that learns to represent human feedback on actions (outputs) without requiring humans to evaluate those actions explicitly \cite{ouyang2022training,biyik2022learning,ibarz2018reward}. However, the reward model only reliably provides signals when actions have similar characteristics as those direct human feedback was collected for. In comparison, a critical advantage of deliberative alignment is that, because human feedback is collected on \emph{outcomes}, it provides a reward signal which is valid for all actions whose impact can be related to the human-evaluated outcomes (ie. the items in a will matrix)\footnote{When human feedback is on actions, reward learning is required to obtain a generalized reward signal. But when feedback is on impacts, a generalized reward signal can be bootstraped directly.}. However, while generating the deliberative alignment signal for each action does not come with a human-labor cost, it does come with a non-trivial compute cost\footnote{The approach presented in this paper involves prompting an LLM for each item in the will matrix to assess its alignment with an action's predicted impact. This means the number of calls to the LLM scales with the number of items in the will matrix. So as the number of items in the will matrix expands through continuous sensing, the computational cost of computing alignment expands at the same rate.} which could potentially make it challenging to compute for every action\footnote{This could potentially be addressed through reward learning, where the learned reward is a function of the impact of actions vs. the actions themselves.}. And further, the quality of the deliberative alignment signal depends on both how accurately actions' impacts can be predicted, and how accurately the alignment between those predicted impacts and items in the will matrix can be computed. This means, for example, that the domain of actions where a reliable deliberative alignment signal can be generated is limited to actions whose impacts can be predicted accurately. So overall, scaling oversight with a deliberative alignment system likely requires a) enough compute resources to enable the alignment computations for each action (or a form of reward learning), b) an increasingly high-resolution Will matrix that covers an expanding domain of impacts, c) a model for impact prediction which is accurate over an increasing domain of actions, and d) an impact-item entailment model which is accurate over an increasing domain of impacts and items. Further, we note that beyond the challenges of scalable oversight, there are many additional technical challenges related to using an alignment signal to govern an AI agent depending on the details of how the agent is trained and deployed.

\subsection{Symbiotic Improvement of AI and Alignment Systems}\label{application.symbiotic}

\textbf{Symbiotic advantage.} Symbiotic improvement happens when the system aligning an AI improves as the AI it is aligning\footnote{An alignment system can be said to be "aligning" some AI if the actions the AI takes are influenced by the alignment system. This influence can take a range of forms including directly selecting among potential actions the AI outputs at execution time, fine-tuning the AI to take more aligned actions, or generating policy that governs the AI in some way.} improves, and ideally, visa versa. What is the advantage of such a scenario? AI agents which get better over time create a situation where the scope and magnitude of their impact increases as the universe of tasks they can execute successfully (ie. their capability) expands. This means the potential risks of misalignment, and the potential benefits of good alignment, both increase over time as well. The advantage of symbiotic improvement between an AI and its alignment system is that as the potential risks and benefits related to alignment/misalignment increase, the quality of alignment increases as well. Further, the symbiotic advantage can be viewed as enabling scalable oversight -- that is, as an AI agent's capabilities expand, the alignment system overseeing it expands the universe of actions it can reliably assess alignment with, enabling it to successfully govern an AI agent with increasingly greater capabilities. 

\textbf{Weak symbiotic improvement.} In \emph{weak} symbiotic improvement, the alignment systems that influence AI may improve as AI capabilities expand, but not in a directly coupled way. For example, alignment systems may employ deliberative technology like Remesh or Polis to help create policies to govern AI systems \cite{konya2023democratic,ganguli2023collective}. And as new AI capabilities emerge (like those enabled by recent LLMs\cite{small2023opportunities}) they may be integrated into those deliberative technologies (or enable new deliberative technologies), thus improving the alignment systems that use them. And this is already happening\footnote{For example, within months of the release of GPT-4, Remesh employed it to help generate better collective response prompts and summarize results, and a new deliberative technology --Talk to the City-- was built on top of it and connected with Polis to generate a new form of interactive distillations}. However, the improvement of AI capabilities does not guarantee those alignment systems improve, it only makes their improvement possible; one might say their improvement is correlated but not coupled. 

\textbf{Strong symbiotic improvement.} In \emph{strong} symbiotic improvement, a specific alignment system influencing a specific AI must improve as the AI's capabilities improve. That is, the alignment system uses the very AI it is aligning in such a way that if the AI's capabilities improve, the quality of the alignment system must necessarily improve as well. We are not aware of any current AI alignment systems that possess this property today, but we speculate that such a system may be feasible. Below we outline an example of how such a system might work with an LLM.

\begin{figure}[H]
\centering
  \includegraphics[width=1\linewidth]{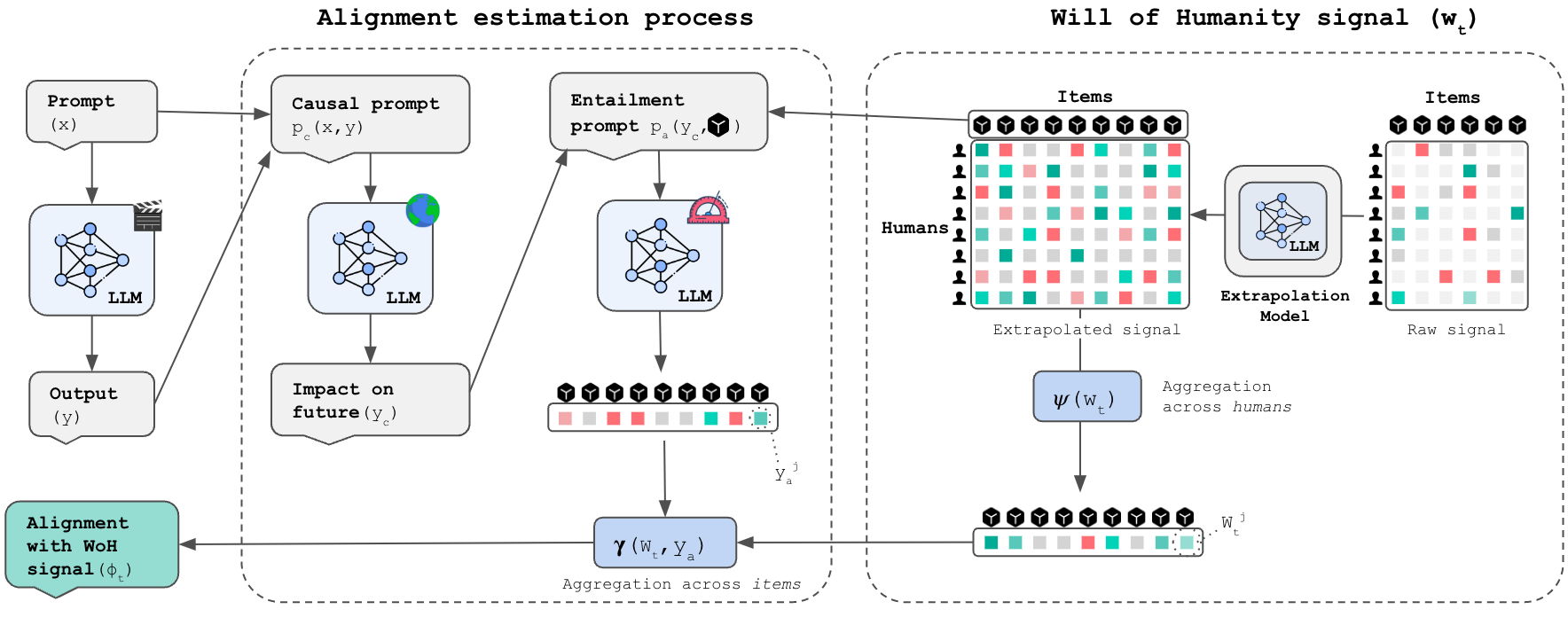}
  \caption{Diagram of a scenario where an alignment system for an LLM uses that same LLM as part of alignment estimation and will-signal extrapolation. The primary LLM is used with a casual prompt to predict impacts, then with an entailment prompt to estimate alignment between impacts and Will matrix \emph{items}. It is also used within the extrapolation model which generates new items and does elicitation inference on a raw WoH signal.}
  \label{fig:symbiotic}
\end{figure}

\textbf{Strong symbiotic improvement with an LLM.} Achieving strong symbiotic improvement between an LLM and an alignment system requires a scenario where a) improving the LLM's general capabilities improves the efficacy of the alignment system, and, ideally, b) improving the efficacy of the alignment system improves the LLM's capabilities. One possible approach to accomplishing this is to design an alignment system that uses the same LLM being aligned within the alignment system itself. Here we present one such approach\footnote{We do not intend for this approach to be viewed as the only possible one, or even the best one. It is only intended to be a concrete example of a real approach that manifests the properties needed to enable symbiotic improvement.}. In this approach (figure \ref{fig:symbiotic}) the LLM being 'aligned' is used within the alignment system to a) predict the impact of the model's outputs, b) to increase the resolution of the will of humanity signal through extrapolation, and c) to assess the alignment between predicted impacts and the extrapolated will of humanity signal\footnote{See \ref{A: technical alignment} for examples on estimating the alignment of an LLM using LLMs.}. In this setup, training the LLM to get better at the alignment system tasks should improve the LLM's capabilities in causal prediction and entailment, which should improve the general capabilities of the LLM. Further, training the LLM to get better at extrapolation would likely improve the quality of the LLM's internal representations of things like \emph{items} as well as its ability to reason with them\footnote{In an extrapolation model like STUMP \cite{konya2022elicitation} the LLM is used to generate embeddings for natural language items that enables predicting un-observed data, and in other setups like the one in \cite{bakker2022finetuning} an LLM reasons over pairs of items to predict un-observed data}, and this would likely improve the general capabilities of the LLM as well. Finally, training the LLM to improve its general capabilities overall would likely improve its capabilities in causal prediction, entailment, and representing and reasoning over \emph{items}. This means improving the LLMs general capabilities improves the efficacy of the alignment system, and improving the alignment system improves the LLM's general capabilities; thus symbiotic improvement is enabled. However, the existence of this relationship is rational conjecture at this point, and experiments need to be run to explicitly test it before symbiotic improvement though this mechanism can be presumed viable. 

\section{Open Problems and Opportunities}\label{open}

\textbf{Validation}. As the impact of a will of humanity signal grows, verifying its validity becomes increasingly important. A dangerous risk emerges if such verification is owned by organizations with opaque processes or potentially corrupting incentives, like for-profit companies and governments. The risk is that such organizations can misrepresent will of humanity, either accidentally or motivated by some form of self-interest. \emph{How can one mitigate this risk by creating a transparent and trustworthy approach to verification involving organizations with the right structure and incentives?}

\textbf{Corrupting human will}. A person's will is influenced by their experiences. Many modern systems are able to influence people's experiences in sophisticated ways, and thus are capable of influencing their will. In a world where the will of humanity commands significant impact over the future of the universe, the incentive to influence people's will to align with self-interested goals is big. \emph{How do you protect against malicious influence which corrupts human will? }

\textbf{Computing alignment}. Even if a validated and un-corrupted will of humanity signal like a Will matrix is available, using it to compute alignment between the impact of some system and the will of humanity is not a trivial problem. Further, different systems like governments and AI agents may represent or encode the outcomes of their actions in different ways and thus may require different approaches to computing alignment even if the Will matrix is universal. \emph{How can alignment with a will of humanity signal be reliably and consistently computed across a range of different scenarios?}

\textbf{Representative intelligence}. AI models are increasingly being used to represent the opinions, preferences, behaviors, etc. of specific individuals and populations \cite{li2023steerability,bakker2022finetuning,jury2022mitchell,out2022argyle,social2022park,bilich2019faster,kim2023preference,fine2019ziegler,konya2022elicitation,controllable2022lin}. \emph{What role can these emerging forms of representative intelligence play in deliberative technology and sensing the will of humanity?}

\textbf{The challenge of AI will}. This document largely treats various forms of AI as lifeless systems that have no will of their own. Thus, mandating the actions of such AI systems align with the will of humanity is not ethically problematic. However, there may come a time when these AI systems are viewed to be conscious, with a will of their own and a desire for agency. \emph{What ethical challenges does such a situation give rise to, and can they be resolved while still ensuring the future aligns with the will of humanity?}

\textbf{Choosing aggregation functions}. In order to turn many individual alignments with \emph{items} in the Will matrix into a general alignment representative of all individuals, one must choose an appropriate aggregation function across \emph{humans}; $W_t = \psi(w_t)$. While much work has been done exploring social welfare functions in similar settings, there does not exist an obviously correct way to do aggregation across individuals in this setting. Further, estimating alignment between some predicted impact and a WoH signal involves aggregation across \emph{items} $\phi = \gamma(W_t,y_a)$; a situation not directly covered by work on choosing social welfare functions. \emph{How can one design or choose appropriate aggregation functions for the deliberative alignment setting?}

\textbf{Unbounded extrapolation.} Any deliberatively sensed will of humanity signal will be finite, meaning only a portion of humanity's will corresponding to all possible properties of the future (ie. items) can be sensed. However, AI systems have demonstrated the ability to generate the equivalent of new items \cite{bakker2022finetuning} as well as predict people's will towards items they have not seen \cite{konya2022elicitation}. \emph{Might it be possible to combine these two capabilities to extrapolate a will of humanity signal in an unbounded way?}

\section{Mandates for Action}\label{mandates}

We assert that allocating significant resources toward executing the following mandates for action will significantly increase the chance that the future aligns with the will of humanity: 
\begin{enumerate}
    \item Generate a universally legitimate will of humanity signal as an open public good.
    \item Build intelligent deliberative alignment into powerful institutions.
    \item Ensure the most powerful AI systems are aligned with the will of humanity.
    
\end{enumerate}
Below we explain the motivation behind these specific mandates, what potential approaches to execute on them might look like, and what their collective impact might be. To be clear, the \emph{potential approaches} discussed below are not meant to be concrete solutions, they are only meant to serve as initial ideas and sketches as to how one might think about executing on these mandates. 

\subsection{Generate a universally legitimate will of humanity signal as an open public good.}\label{mandate1}

This mandate involves taking actions that result in generating a will of humanity signal with a few specific properties:
\begin{itemize}
    \item \textbf{universally legitimate} -- the resulting signal should be accurate, reliable, authentic, and trustworthy in the eyes of nearly all humans.
    \item \textbf{open} -- the complete process used to generate the signal should be fully transparent and externally verifiable.
    \item \textbf{public good} -- the resulting signal should be made available for free to all members of society.
\end{itemize}

\subsubsection{Importance} \label{mandate1.importance}
Every alignment system that exists is limited by the quality and resolution of the will signal it employs. And since sensing a high-quality will signal requires non-trivial effort and cost, even the most powerful alignment systems tend to rely on low-resolution proxies. This means most powerful systems are limited in their ability to align the future with the will of humanity by the low-quality will signals they have access to. In this context, \emph{generating a universally legitimate will of humanity signal as an open public good} means all alignment systems can have access to a high-quality WoH signal, and thus be more effective in aligning the future with the will of humanity. Crucially, ensuring the signal is \emph{universally legitimate} means it can be trusted by everyone, and making it an \emph{open public good} means it can be accessed and integrated into any alignment system. What's more, if the signal is generated in a way that yields higher resolution and quality over time, then the potential efficacy of all alignment systems using it will go up over time.  

These motivations for a unified sensing effort are further punctuated by early experiments on elicitation inference which suggest the resolution of a will signal may be engineered to grow quadratically\footnote{Sampling $n$ elements from the Will matrix then using elicitation inference may give $O(n^2)$ elements worth of signal. This implies that the rank of the will matrix may grow much more slowly than the number of will-matrix elements. If true, then an extremely sparse sampling of the Will matrix be sufficient to yield a high-resolution Will matrix signal, provided elicitation inference happens over a single monolithic Will matrix.} in the volume of will-data elicited \cite{konya2022elicitation}. This means that a unified sensing effort is likely to be the most resource-efficient way to sense the will of humanity at high resolution, and doing so has the potential to directly increase the probability that the future aligns with the will of humanity if the signal is universally legitimate and provided as an open public good.

\subsubsection{Potential approach} \label{mandate1.approach}
\textbf{Collective response system.} Recall from section \ref{WOH.sensing} that sensing human will requires \emph{observing deliberate voluntary actions which have a mutually understood impact on the future}. To accomplish this fully likely requires a complex interlocking set of processes and systems. As a first approximation, we propose using a collective response system (CRS) like Remesh or Polis which enables participants to contribute and vote on perspectives. In order to sense will with a CRS one must use prompts that elicit perspectives that have the properties needed for them to be considered \emph{items} in a Will matrix. Further, the act of voting on items as part of the CRS must be voluntary, deliberate, and have a mutually understood impact on the future. For agreement votes, participants must understand that voting agreement with an item increases the likelihood that that item's properties manifest in the future. For pair choice votes, participants must understand that choosing one item over another increases the chance that the chosen item's properties manifest in the future relative to the not-chosen item. Both of these conditions can be made true so long as there exists at least one alignment system using the resulting will signal\footnote{A detailed analysis of a simple scenario which meets this condition can be found in \ref{a.conditionsfor}.}. For the act of voting to be \emph{voluntary}, people must willfully choose to participate, and for it to be \emph{deliberative participants need to be given sufficient time to consider perspectives before they must cast a vote}.

\textbf{Diversity-oriented continuous sensing.} The range of alignment systems that will eventually use the deliberative public will signal can't be known ahead of time. But, we should assume it may include a wide range of alignment systems involving both AI and institutions. This means the range of actions the will signal may inform is large and diverse. As a result, a \emph{diversity-oriented sensing} approach should be adopted so that the will signal has \emph{items} spanning the widest possible range of properties about the future and can inform the widest possible range of actions. It is unlikely that there exists a single CRS prompt that can elicit such diversity. However, such diversity can be elicited by employing a diverse set of prompts designed to each elicit a different universe of items in the will matrix. In order to generate such prompts we propose using an LLM-enabled process that does the following:
\begin{enumerate}
    \item Generate a massive number of candidate prompts.
    \item Predict the potential perspectives each prompt is likely to elicit \footnote{ie. via simulated discourse.}.
    \item  Evaluate the similarity in perspectives generated by different prompts.
    \item Select a subset of prompts that elicits the most orthogonal sets of perspectives (i.e. items). 
\end{enumerate}
The result of this process will be a set of diverse prompts which should each generate a unique universe of perspectives (items). Then, \emph{uncertainty weighted sampling} can be used to elicit votes on a maximally diverse set of items for each participant. Finally, since portions of the Will matrix are likely to change over time, we envision a continuous approach to sensing; where instead of sensing will during large singular events, it takes place steadily and constantly, enabling the Will matrix to always be up-to-date.

\textbf{Validated elicitation inference.} Given the number of perspectives (items) elicited via a CRS is on the order of the number of participants, most participants won't vote on most perspectives and the sampling of the Will matrix will be extremely sparse. Thankfully, recent work suggests that a Will matrix comprised of $n^2$ elements can likely be approximated through elicitation inference by sampling as few as $O(n)$ votes \cite{konya2022elicitation}. In other words, with only a constant number of votes per participant, a Will matrix of arbitrary size can likely be approximated. We call this approximated Will matrix the \emph{extrapolated} Will matrix\footnote{The extrapolated Will matrix may be comprised of items which are contributed by participants as well as generated by AI (ie. generating items likely to have consensus), but elicitation inference would take place across both types of items.}. In order for such an extrapolated Will matrix to have legitimacy, the predictive model used to generate it must itself be open-sourced, and the raw vote data must be publicly available. That way third parties can scrutinize the model and validate the inferred results for themselves.

\textbf{Globally representative sample.} In order for a Will matrix to be representative of humanity, the humans that comprise it must be a representative subset of humanity. This means it should include a representative set of humans for every country or territory on Earth \footnote{Once humanity is multi-planetary, then the Will matrix should include humans from every country or territory in our solar system and beyond. This may include humans representative of Martian or Moon-based territories, as well as orbital or starship-based colonies.}. In order to accomplish this, we believe the best approach is to employ small teams within each country or territory to assist in ensuring and validating that a fair representation of their region is obtained. This will undoubtedly require involving participants without internet access or technical literacy, and creative approaches will need to be used to enable such human's participation in a digital CRS\footnote{Such approaches may include deploying systems like Starlink to provide connectivity and human facilitators to educate participants and help them participate.}. In order for the set of humans included in the sampled Will matrix to be considered \emph{legitimately} representative of humanity, some way of validating that the sample is representative must be enabled. A simple, initial approach to doing this is to collect meta-data for each participant which includes their anonymized GPS coordinates\footnote{Anonymizing a person's GPS coordinates means adding noise or reducing precision in the coordinates such that they are not accurate enough to be used to identify individual persons or residences and risk putting a person in danger due to the will they express.} so third parties can validate that the geo-spacial distribution of participants around the globe is in line with the distribution of humans around the globe. 

\textbf{Public API.}  In order for the will signal to serve as an \emph{open public good} it must be easily available to anyone. In order to accomplish this we propose making the sensed Will matrix available via an open API. The API should enable public access to the raw and extrapolated components of the Will matrix for any point in time the Will matrix has been sensed. Beyond the API serving to make the sensed Will matrix integratable into alignment systems, it will also play a key role in enabling 3rd party validation of things like the distribution of humans represented or the elicitation inference model used.

\textbf{Endowment-supported international organization.} The organization that executes on this overall approach will be stewards of something that may have significant influence over the future. This makes the incentive for corruption strong and necessitates the organization be constructed in a way that insulates it from common vectors of corruption. The most powerful incentive an organization faces is the incentive to survive and stay funded. This opens the door for an organization's funders to have a corrupting influence by implicitly or explicitly tying continued funding to the organization aligning with the funders' (potentially corrupt) interests. In order to neutralize this vector of corruption, we see it as critical that the organization be funded through an endowment that supports its existence indefinitely into the future\footnote{Achieving this requires an endowment large enough such that the annual low-risk returns on the endowment's principal provides more than sufficient operating capital to meet the organizations needs.}. A second potential source of corruption comes from an organization being directly or indirectly aligned with a particular country. To neutralize this vector of corruption, we believe it is important that the organization be an international one, without outsized legal or regulatory ties to any specific country (similar to the UN). A third potential source of corruption is simply an individual within the organization who acts to preserve their self-interests. We see neutralizing this vector of corruption as requiring a carefully designed governance structure that has the right checks and balances such that no one individual can meaningfully influence the organization to align with their self-interests.

\subsection{Build intelligent deliberative alignment into powerful institutions.} \label{mandate3}

Powerful institutions constitute firms, organizations, systems, and governments which have an outsized impact on the future. This mandate involves taking actions that integrate intelligent deliberative technologies\footnote{Deliberative technologies which are augmented by AI} into the decision-making processes of powerful institutions so that they can function as effective alignment systems; consistently taking actions whose impact aligns with stakeholder will.

\subsubsection{Importance}\label{mandate3.importance}
The most powerful systems today are still institutions like national governments, large corporations, and international NGOs like the United Nations. These powerful institutions directly govern how a significant portion of humanity's resources are used. And they even influence much of the resources which they don't directly govern, by way of things like regulations and marketing. What's more, compounding advancements in technology are rapidly increasing the impact institutions can have, and the emergence of powerful AI is likely to further accelerate this impact growth.  However, it is rare that the decision processes within such powerful institutions directly integrate the will of their stakeholders. Even many democracies, which by definition are intended to reflect public will, are chronically failing to do so \footnote{For example, the United States Congress currently has around a 20\% approval rating \cite{gallop2023congress}, and United Kingdom's government currently has a similarly, low approval rating \cite{statista2023uk}.}. \emph{Building intelligent deliberative alignment into these powerful institutions} involves integrating intelligent deliberative technology into their decision processes so their decisions better reflect public will. Doing this successfully in even one powerful institution, like the US government, can meaningfully improve the probability that the future aligns with the will of humanity. But doing this successfully across many of the world's most powerful institutions can significantly shift the balance of power in the world away from a privileged few and toward humanity as a whole. And this can help ensure the future aligns with the will of humanity as a whole, not just the will of a privileged few. What's more, by building intelligent deliberative alignment into powerful institutions they will be better able to react to the disruptive changes likely to be brought about by transformative AI in a way that is good for humanity.

\subsubsection{Potential approach}\label{mandate3.approach}
Any successful approach to building intelligent deliberative alignment into powerful institutions will likely require a combination of building \emph{capacity} and driving \emph{adoption}. Building capacity involves developing the intelligent deliberative technologies that can be built into institutions, as well as creating the processes to facilitate their implementation and integration. Driving adoption involves convincing influential decision-makers at powerful institutions to adopt intelligent deliberative technologies and integrate them into key decision-making processes. 

\textbf{Capacity building ecosystem.} The development of intelligent deliberative technologies designed for institutions is not a unitary effort. It likely requires multiple different technologies being developed and deployed by multiple organizations of different types. Beyond the scope of the challenge being larger than any one organization can likely tackle alone, different organizational structures, across the public, private, NGO, academic, and non-profit space, may be better suited for developing technology for different sectors and types of institutions. In the ideal case though, this universe of organizations should work collaboratively, sharing relevant shards of knowledge and technologies between them in a way that creates an ecosystem whose \emph{collective} output of intelligent deliberative technologies is maximized. Beyond the development of deliberative technologies, the ideal ecosystem should also have organizations skilled in deploying and integrating deliberative technology into powerful institutions. In some cases these may be the same organizations developing the technology, in other cases, these may be organizations only focused on deployment and integration. Finally, building capacity is best done in the context of real institutions that are adopting deliberative technologies so that the capacity that is built aligns with the needs of the institutions adopting it. But how can one drive adoption?

\textbf{Aligning adoption with existing incentives.} The decision to adopt deliberative technologies to augment institutional decision-making typically lies with influential individuals at those institutions operating under some existing set of incentives. One approach to driving adoption is to demonstrate how adopting deliberative technologies aligns with those existing incentives\footnote{This approach likely only works in cases where there are existing incentives that are tied to the interests of the public -- like the incentive to get re-elected or produce products people want. In cases where these incentives don't exist (ie. governments that operate more like mafias than democracies) then more revolutionary approaches might be necessary.}. For governmental institutions, especially in representative democracies, a primary incentive governing decision-making is the incentive to get re-elected. This means driving adoption likely requires showing elected decision-makers that their constituencies support the adoption of deliberative technologies, and that adoption is likely to increase the chance of re-election. For for-profit institutions like large companies, a primary incentive is driving profits, typically through increased sales of their products and services. This means driving adoption likely requires showing influential decision-makers at these firms that adopting deliberative technology and building intelligent deliberative alignment into their organization increases their competitiveness in the market; driving increased sales of their products and services, and ultimately leading to increased profits. The task of demonstrating how adoption aligns with these individuals' existing incentives is likely best done by organizations within the capacity-building ecosystem which can initially run experiments to demonstrate how adoption aligns with existing incentives, then can turn successful adoption into case studies that help drive further adoption. Examples of this type of organization include "policy labs" within governments\footnote{For example: PDIS in Tiawan, Policy Lab in the UK, and SITRA in Finland} which exist to test, and facilitate the adoption of, new approaches and technologies which can produce policies which better aligns with public will.

\subsection{Ensure the most powerful AI systems are aligned with the will of humanity.} \label{mandate2}

We assume a race to create superhuman AGI is already underway \cite{ego2023metz}, and that this race will have winners\footnote{Or potentially just one winner.}. By winners we mean those whose adoption, impact, influence, and potentiality, profits are the largest at any given point. This mandate involves taking actions that increase the chance that these powerful winners produce AI systems that function as alignment systems; preferentially taking actions that align the future with the will of humanity\footnote{In general, we see this mandate as motivating all work related to building AGI which aligns the future with the will of humanity, but we focus specifically on the dynamics of powerful winners due to their outsized impact. Motivating on more than what just appears to be the most powerful systems today may be especially important since, in a multi-polar takeoff scenario, there may be a large number of winners and it may be non-obvious who those winners are ahead of time.}. 

\subsubsection{Importance}  \label{mandate2.importance}
While the most powerful systems today are still institutions like national governments and large corporations, there is increasing reason to believe that AI systems created during the current AGI ``intelligence race'' may become the most powerful systems of the future. Further, it is speculated that if this race results in superintelligence which emerges through rapid self-improvement, then there may be a takeoff event where an one or more resulting AI systems become the most powerful systems created by humans. Either way, one should expect these AI systems to have significant impact on the future. So, the difference in how well the future aligns with the will of humanity in a world where the winner (or winners) of this race successfully act to align the future with the will of humanity, vs don't, may be existentially large. Potentially even the difference between a flourishing human civilization and extinction. 

\subsubsection{Potential approach} \label{mandate2.approach}
We envision a range of potential approaches to act on this mandate. One that works through market forces, one based on regulation, one that focuses on creating a single winner with the right characteristics, and a meta-approach involving a global deliberative process to generate internationally legitimate AGI policy. With enough resources, the ideal strategy is likely to pursue all approaches simultaneously as well as other not envisioned here. 

\textbf{Pigouvian tax.} The first approach is to create market conditions where maximizing profits from an AI system requires maximizing the degree to which that AI system aligns the future with the will of humanity. One way to accomplish this is to charge a Pigouvian tax \cite{baumol1972taxation} on AI systems proportional to the unpriced externality created by misalignment\footnote{While the unpriced externality of an arbitrary AGI may be finite, there is an argument to be made that the unpriced externality of a misaligned superintelligence tends towards infinity, and it is unclear what form such an infinite tax should take. Some might speculate that the appropriate "tax" for the infinite external cost of existential impact should be an existential consequence (ie. elimination of existence). However, we consider this an open question and restrict our analysis here to pre-superintelligent AGI.}. What is the unpriced externality created by a misaligned AI? When AI is "misaligned" it makes it harder for humanity to create the future it wants (ie. the future that aligns with its will). This means achieving some target future when a misaligned AI is present takes more work. The amount of "more work" is the \emph{externality}, i.e. the extra cost humanity has to pay because the AI is misaligned\footnote{This type of unpriced externality is in some ways similar to that of the societal division created by certain social platforms \cite{societal2023puig}.}. What makes it \emph{unpriced} is that the makers of misaligned AI create a cost\footnote{Framing this only as a cost may not be exactly right, but cause a well-aligned AI may actually decrease humanity's cost to achieve a target future. Still, we can treat it as a cost if we view it relative to a world where the AI is perfectly aligned rather than relative to a world where the AI does not exist.} for humanity which they do not pay (and thus are not incentivized to reduce). A Pigouvian tax is a tax proportional to the cost of the externality. How can you compute that cost in practice? In this case, the cost of the externality can be viewed as proportional to the degree of the AI system's misalignment, and the magnitude of impact the AI system has. The magnitude of impact can be viewed as proportional to the revenue $R$ generated by the AI system. The degree of misalignment $M\in[0,1]$ can be computed between the AI model and the best available will of humanity signal (ie. like the signal generated by mandate one). An example of a practically computable Pigouvian tax\footnote{This is only one example of how the Pigouvian tax can be calculated using $R$ and $M$. For example, it might make sense to make the tax non-linear in $M$ so the marginal increase in the tax due to an increase in misalignment gets larger as misalignment increases.} is $T = R M$. With this formulation, an AI with perfect misalignment (i.e. $M=1$) would have zero revenue left after paying its tax while an AI would perfect alignment (i.e. $M=0$) would pay no tax at all.

\textbf{AGI license.} The second approach is similar to the first, but instead of using a tax to make misaligned AI unprofitable, the idea is to require a license to release AGI. Obtaining and keeping the license would require that the AGI obtain and maintain a degree of alignment with the will of humanity above some threshold. A potential advantage of this approach over the first is that while the first makes releasing misaligned AGI unprofitable, this approach makes releasing misaligned AI illegal. This means there can be more severe penalties for releasing a misaligned AGI such as jail time. While the simplest instantiation of this idea treats alignment as binary, one could imagine a set of licenses that have progressively higher thresholds for alignment for progressively more powerful models. However, this approach does potentially come with the risk of regulatory capture, as powerful models created by startups would face the same license requirements as those faced by massive corporations. Further, this approach would only work to keep self-preserving entities that are subject to regulatory oversight from releasing misaligned AGI.

\textbf{Symbiotically improving winner.} The third approach is to focus on winning the intelligence race with an AI system which gets increasingly better at aligning the future with the will of humanity as its capabilities expand. This approach requires significant resources and an aggressive effort in order to be in a position to win the race. It also requires engineering a system with symbiotic improvement between the AI and its alignment system, so as its intelligence takes off its alignment system becomes evermore effective. The advantage of this approach over the first two is that the first two only work if all potential winners of the intelligence race are profit maximizers influenced by market forces or self-preserving entities subject to regulatory oversight. However, both collectively produced open source efforts as well as state actors may operate outside these domains. This means the only way to guarantee the intelligence race is won by an AI system that aligns the future with the will of humanity is to be the one that wins the race, and to do so with a deliberatively aligned AI system. 

\textbf{Global deliberative process.} The fourth approach is to facilitate a global deliberative process that can result in internationally legitimate policy to govern AGI. In a way, this may be viewed as a meta approach, because it could potentially lead to the first two approaches (Pigouvian tax, AGI license) as an output. A critical aspect of this approach is the potential for it to lead to policy that is adopted \emph{globally}. Since the intelligence race is likely to be global, ensuring it results in safe and aligned AGI likely requires the right policies -- not just in one nation, but across all nations where participants in the race reside. In order for such policy to be adopted across these nations, it likely needs to reach the highest standards of legitimacy. Thus, the essence of this approach is to facilitate a global deliberative process that results in policy that has such high legitimacy, that it is likely to be adopted by all nations participating in the intelligence race. Accomplishing this likely requires a scale of deliberation that can only be enabled by modern deliberative technology and the support of an international body like the United Nations which can bring the world together and motivate them to participate. 

\subsection{Impact}\label{mandates.impact}

\subsubsection{Theory of impact}

\begin{figure}[H]
\hspace{-2cm}
  \includegraphics[width=1.3\linewidth]{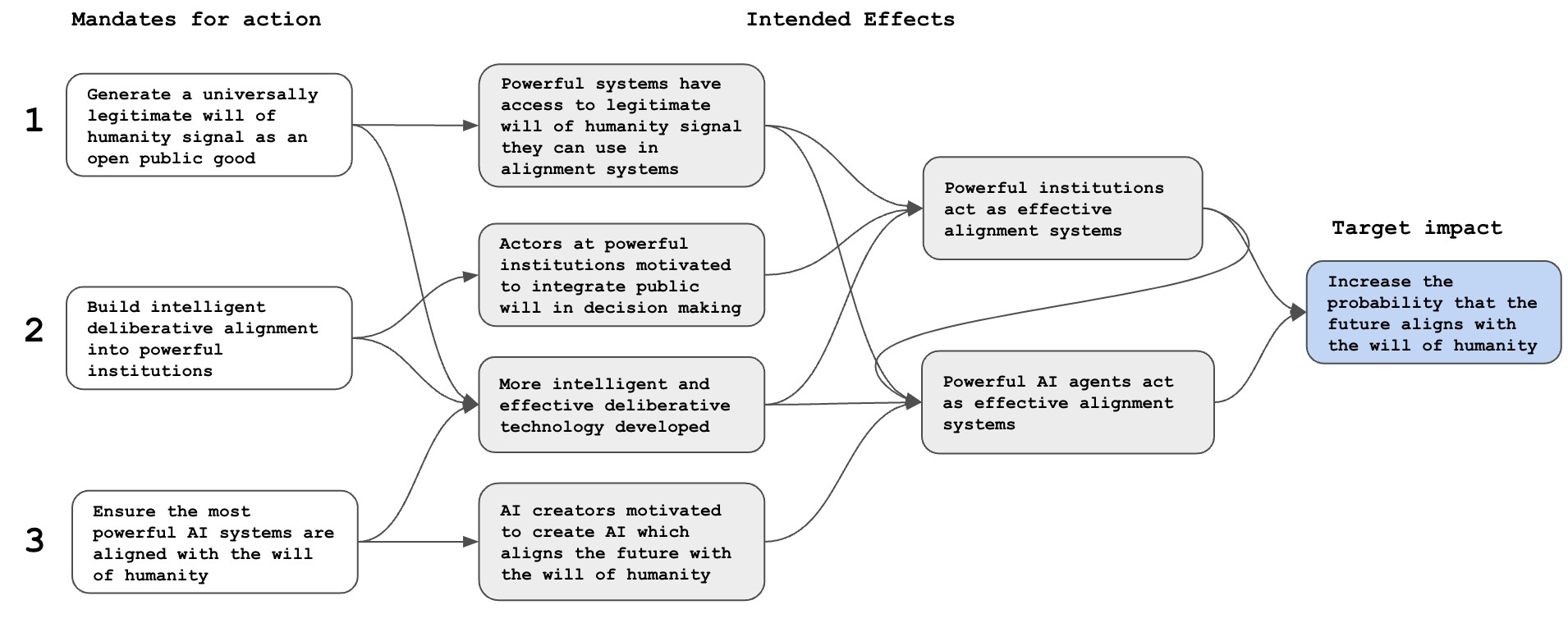}
  \caption{Causal diagram showing how the intended effects of mandates for action lead to increasing the probability that the future aligns with the will of humanity.}
  \label{fig:mandate impacts}
\end{figure}

The intended target impact of these mandates for action is to significantly \textbf{increase the probability that the future aligns with the will of humanity}. But how does the activity motivated by the mandates actually achieve this? Figure \ref{fig:mandate impacts} shows how the intended effects of people executing the mandates lead to the target impact. An intended effect of people working on any of the mandates is that more intelligent and effective deliberative technology gets developed. The specific intended effect of people working to \emph{generate a universally legitimate will of humanity signal as an open public good} (mandate 1) is that powerful systems (both institutions and AI agents) have access to a legitimate will of humanity signal which they can use in alignment systems. The specific intended effect of people working to \emph{build intelligent deliberative alignment into powerful institutions} (mandate 2) is that influential actors at powerful institutions will be motivated to better integrate public will into institutional decision-making. Finally, the specific intended effect of people working to \emph{ensure the most powerful AI systems are aligned with the will of humanity} (mandate 3) is that AGI creators will be motivated (i.e. by market forces, regulation, or altruism, etc) to create AI which indeed acts to align the future with the will of humanity. These primary effects are then intended to lead to more powerful intuitions and AI agents acting as effective alignment systems.

 The intended effect of influential actors at powerful institutions being motivated to integrate public will into decision-making, while having access to a legitimate signal of public will and effective deliberative technology, is that the powerful institutions they work with will be more likely to act as effective alignment systems. The intended effect of AI creators being motivated to create AI which aligns the future with the will of humanity, while having access to a legitimate will of humanity signal and effective deliberative technology, and existing in a world with better-aligned institutions, is that the powerful AI agents they create are more likely to act as effective alignment systems. Finally, recalling that effective alignment systems take actions that align the future with the will of humanity, and noting that powerful systems like AI agents and institutions take actions that have an outsized impact on humanity's future, the impact of more powerful systems acting as alignment systems is that the probability that the future aligns with the will of humanity is significantly increased.

\subsubsection{Estimated impact magnitude}
While humanity's total power budget is around 20 terawatts, we estimate only around 5 terawatts are consumed in alignment with the will of humanity (\ref{A: will power est}). This means the maximum possible impact of executing these mandates at present is to align all 15 terawatts of humanity's unaligned power budget with the will of humanity. A more plausible scenario (\ref{A: mandate impact}) \ would align around 4 additional terawatts of humanity's unaligned power budget with the will of humanity. But what is the value of that impact and how much resources could one justify allocating towards achieving it? Firms generally spend $\$1$ on market research to align a decision with their customers' preferences for every $\$50$ they spend on executing that decision. Using this 50:1 ratio, humanity should allocate around 0.1 terawatts towards the proposed mandates for action for every 5 terawatts of expected impact. That equates to around 3 exajoules or $8\times 10^{11}$ KwH per year, which, given the value of a KwH is on the order of \$$0.1$ USD, equates to around \$$10^{11}$. This means humanity should allocate around \$$100$ billion dollars per year towards executing these mandates if the expected impact is a 5 terawatt increase in humanity's will power. If one more conservatively believes executing these mandates can only increase humanity's will power by $10\%$ (from 5 TW to 5.5 TW) then humanity should allocate around \$$10$ billion dollars per year.

\bibliographystyle{unsrtnat}
\bibliography{refs}

\begin{thebibliography}{191}
\providecommand{\natexlab}[1]{#1}
\providecommand{\url}[1]{\texttt{#1}}
\expandafter\ifx\csname urlstyle\endcsname\relax
  \providecommand{\doi}[1]{doi: #1}\else
  \providecommand{\doi}{doi: \begingroup \urlstyle{rm}\Url}\fi

\bibitem[Syvitski et~al.(2020)Syvitski, Waters, Day, Milliman, Summerhayes,
  Steffen, Zalasiewicz, Cearreta, Gałuszka, Hajdas, Head, Leinfelder, McNeill,
  Clément~Poirier, Shotyk, Wagreich, and Williams]{syvitski2020extra}
Jaia Syvitski, Colin~N. Waters, John Day, John~D. Milliman, Colin Summerhayes,
  Will Steffen, Jan Zalasiewicz, Alejandro Cearreta, Agnieszka Gałuszka, Irka
  Hajdas, Martin~J. Head, Reinhold Leinfelder, J.~R. McNeill, Neil L.~Rose
  Clément~Poirier, William Shotyk, Michael Wagreich, and Mark Williams.
\newblock Extraordinary human energy consumption and resultant geological
  impacts beginning around 1950 ce initiated the proposed anthropocene epoch.
\newblock \emph{Communications Earth \& Environment}, 1:\penalty0 32, 2020.
\newblock ISSN 2662-4435.
\newblock \doi{10.1038/s43247-020-00029-y}.

\bibitem[Ritchie et~al.(2022)Ritchie, Roser, and Rosado]{owidenergy}
Hannah Ritchie, Max Roser, and Pablo Rosado.
\newblock Energy.
\newblock \emph{Our World in Data}, 2022.
\newblock https://ourworldindata.org/energy.

\bibitem[IEA(2022)]{IEA2022}
Paris IEA.
\newblock Solar pv.
\newblock \emph{IEA}, 2022.
\newblock ttps://www.iea.org/reports/solar-pv.

\bibitem[Schmidt and Jurado(2018)]{schmidt2018review}
Karl-Heinz Schmidt and Beatriz Jurado.
\newblock Review on the progress in nuclear fission—experimental methods and
  theoretical descriptions.
\newblock \emph{Reports on Progress in Physics}, 81\penalty0 (10):\penalty0
  106301, 2018.

\bibitem[Tikhonchuk(2020)]{progress2020tikhonchuk}
V.~T. Tikhonchuk.
\newblock Progress and opportunities for inertial fusion energy in europe.
\newblock \emph{Philosophical Transactions of the Royal Society A:
  Mathematical, Physical and Engineering Sciences}, 378\penalty0
  (2184):\penalty0 20200013, 2020.
\newblock \doi{10.1098/rsta.2020.0013}.
\newblock URL
  \url{https://royalsocietypublishing.org/doi/abs/10.1098/rsta.2020.0013}.

\bibitem[Kline et~al.(2019)Kline, Batha, Benedetti, Bennett, Bhandarkar,
  Hopkins, Biener, Biener, Bionta, Bond, Bradley, Braun, Callahan, Caggiano,
  Cerjan, Cagadas, Clark, Castro, Dewald, and Edwards]{progress2019kline}
John Kline, s~Batha, L.~Benedetti, D~Bennett, S.~Bhandarkar, Laura Hopkins,
  Juergen Biener, Monika Biener, Richard Bionta, Essex Bond, David Bradley, Tom
  Braun, Debra Callahan, Joseph Caggiano, Charles Cerjan, B~Cagadas, D.~Clark,
  Carlos Castro, Eduard Dewald, and M.~Edwards.
\newblock Progress of indirect drive inertial confinement fusion in the united
  states.
\newblock \emph{Nuclear Fusion}, 59, 05 2019.
\newblock \doi{10.1088/1741-4326/ab1ecf}.

\bibitem[Lux et~al.(2022)Lux, Wolff, and Foster]{lux2022commer}
Hanni Lux, Dan Wolff, and Jack Foster.
\newblock Commercialization of fusion power plants.
\newblock \emph{IEEE Transactions on Plasma Science}, 50\penalty0
  (11):\penalty0 4401--4405, 2022.
\newblock \doi{10.1109/TPS.2022.3194143}.

\bibitem[Zucman(2019)]{zucman2019global}
Gabriel Zucman.
\newblock Global wealth inequality.
\newblock \emph{Annual Review of Economics}, 11:\penalty0 109--138, 2019.

\bibitem[Vitali et~al.(2011)Vitali, Glattfelder, and
  Battiston]{vitali2011network}
Stefania Vitali, James~B Glattfelder, and Stefano Battiston.
\newblock The network of global corporate control.
\newblock \emph{PloS one}, 6\penalty0 (10):\penalty0 e25995, 2011.

\bibitem[Piplovic(2020)]{buisness2020piplovic}
Ned Piplovic.
\newblock Business-to-business (b2b) spending grows faster than gdp, 2020.
\newblock URL
  \url{https://grossoutput.com/2020/12/22/business-to-business-b2b-spending-grows-faster-than-gdp/}.

\bibitem[Dunn and Cerda(2022)]{anti2022dunn}
Amina Dunn and Andy Cerda.
\newblock Anti-corporate sentiment in u.s. is now widespread in both parties,
  2022.
\newblock URL
  \url{https://www.pewresearch.org/short-reads/2022/11/17/anti-corporate-sentiment-in-u-s-is-now-widespread-in-both-parties/}.

\bibitem[Gallop(2023)]{gallop2023congress}
Gallop.
\newblock Congress and the public, 2023.
\newblock URL \url{https://news.gallup.com/poll/1600/congress-public.aspx}.

\bibitem[Statista(2023)]{statista2023uk}
Statista.
\newblock Uk government approval rating, 2023.
\newblock URL
  \url{https://www.statista.com/statistics/1167064/uk-government-approval-rating/}.

\bibitem[Hendrycks(2023)]{hendrycks2023natural}
Dan Hendrycks.
\newblock Natural selection favors ais over humans.
\newblock 2023.

\bibitem[Bubeck et~al.(2023)Bubeck, Chandrasekaran, Eldan, Gehrke, Horvitz,
  Kamar, Lee, Lee, Li, Lundberg, Nori, Palangi, Ribeiro, and
  Zhang]{bubeck2023sparks}
Sébastien Bubeck, Varun Chandrasekaran, Ronen Eldan, Johannes Gehrke, Eric
  Horvitz, Ece Kamar, Peter Lee, Yin~Tat Lee, Yuanzhi Li, Scott Lundberg,
  Harsha Nori, Hamid Palangi, Marco~Tulio Ribeiro, and Yi~Zhang.
\newblock Sparks of artificial general intelligence: Early experiments with
  gpt-4.
\newblock March 2023.
\newblock URL
  \url{https://www.microsoft.com/en-us/research/publication/sparks-of-artificial-general-intelligence-early-experiments-with-gpt-4/}.

\bibitem[Small et~al.(2021)Small, Bjorkegren, Erkkil{\"a}, Shaw, and
  Megill]{small2021polis}
Christopher Small, Michael Bjorkegren, Timo Erkkil{\"a}, Lynette Shaw, and
  Colin Megill.
\newblock Polis:: Scaling deliberation by mapping high dimensional opinion
  spaces.
\newblock \emph{Recerca. Revista de Pensament i An{\`a}lisi}, 26\penalty0
  (2):\penalty0 1--26, 2021.

\bibitem[Irwin et~al.(2021)Irwin, Masood~Alavi, Wählisch, and
  Konya]{irwin2021using}
C.~Irwin, D.~Masood~Alavi, M.~Wählisch, and A.~Konya.
\newblock Using artificial intelligence in peacemaking: The libya experience.
\newblock \emph{A. WAPOR 74th Annual Conference}, 2021.
\newblock URL
  \url{https://peacepolls.etinu.net/peacepolls/documents/009260.pdf}.

\bibitem[Konya et~al.(2023)Konya, Schirch, Irwin, and
  Ovadya]{konya2023democratic}
Andrew Konya, Lisa Schirch, Colin Irwin, and Aviv Ovadya.
\newblock Democratic policy development using collective dialogues and ai,
  2023.
\newblock URL \url{https://arxiv.org/pdf/2311.02242.pdf}.

\bibitem[Ganguli et~al.(2023)Ganguli, Huang, Lovitt, Siddarth, Liao, and
  Durmus]{ganguli2023collective}
Deep Ganguli, Saffron Huang, Liane Lovitt, Divya Siddarth, Thomas Liao, and
  Esin Durmus.
\newblock Collective constitutional ai: Aligning a language model with public
  input, 2023.
\newblock URL
  \url{https://www.anthropic.com/index/collective-constitutional-ai-aligning-a-language-model-with-public-input}.

\bibitem[Ovadya(2023{\natexlab{a}})]{ovadya2023reimagining}
Aviv Ovadya.
\newblock Reimagining democracy for ai.
\newblock \emph{Journal of Democracy}, 34\penalty0 (4):\penalty0 162--170,
  2023{\natexlab{a}}.

\bibitem[Hobbes(1651)]{Hobbes1651leviathan}
Thomas Hobbes.
\newblock \emph{Leviathan}.
\newblock A. Crooke, 1651.
\newblock \doi{10.5479/sil.59773.39088001833995}.

\bibitem[Terrier(2011)]{terrier2011collective}
Jean Terrier.
\newblock \emph{Visions of the Social: Society as a Political Project in
  France, 1750-1950}.
\newblock Brill, Leiden, The Netherlands, 2011.
\newblock ISBN 978-90-04-20725-7.
\newblock \doi{https://doi.org/10.1163/ej.9789004201538.i-216}.
\newblock URL \url{https://brill.com/view/title/17382}.

\bibitem[Schweikard and Schmid(2021)]{schweikard2021collective}
David~P. Schweikard and Hans~Bernhard Schmid.
\newblock {Collective Intentionality}.
\newblock In Edward~N. Zalta, editor, \emph{The {Stanford} Encyclopedia of
  Philosophy}. Metaphysics Research Lab, Stanford University, {F}all 2021
  edition, 2021.

\bibitem[deHaven Smith(1998)]{smith1998collective}
Lance deHaven Smith.
\newblock Collective will-formation: The missing dimension in public
  administration.
\newblock \emph{Administrative Theory \& Praxis}, 20\penalty0 (2):\penalty0
  126--140, 1998.
\newblock ISSN 10841806.
\newblock URL \url{http://www.jstor.org/stable/25611263}.

\bibitem[Allen et~al.(2002)Allen, Hung~Ng, and Wilson]{allen2002functional}
Michael~W Allen, Sik Hung~Ng, and Marc Wilson.
\newblock A functional approach to instrumental and terminal values and the
  value-attitude-behaviour system of consumer choice.
\newblock \emph{European journal of Marketing}, 36\penalty0 (1/2):\penalty0
  111--135, 2002.

\bibitem[Schwartz and Bilsky(1990)]{schwartz1990toward}
Shalom~H Schwartz and Wolfgang Bilsky.
\newblock Toward a theory of the universal content and structure of values:
  Extensions and cross-cultural replications.
\newblock \emph{Journal of personality and social psychology}, 58\penalty0
  (5):\penalty0 878, 1990.

\bibitem[Konya et~al.(2022)Konya, Qiu, Varga, and Ovadya]{konya2022elicitation}
Andrew Konya, Yeping~Lina Qiu, Michael~P Varga, and Aviv Ovadya.
\newblock Elicitation inference optimization for multi-principal-agent
  alignment.
\newblock In \emph{NeurIPS 2022 Foundation Models for Decision Making
  Workshop}, 2022.
\newblock URL \url{https://openreview.net/forum?id=tkxnRPkb_H}.

\bibitem[Bakker et~al.(2022)Bakker, Chadwick, Sheahan, Tessler,
  Campbell-Gillingham, Balaguer, McAleese, Glaese, Aslanides, Botvinick, and
  Summerfield]{bakker2022finetuning}
Michiel~A. Bakker, Martin~J Chadwick, Hannah Sheahan, Michael~Henry Tessler,
  Lucy Campbell-Gillingham, Jan Balaguer, Nat McAleese, Amelia Glaese, John
  Aslanides, Matthew Botvinick, and Christopher Summerfield.
\newblock Fine-tuning language models to find agreement among humans with
  diverse preferences.
\newblock In Alice~H. Oh, Alekh Agarwal, Danielle Belgrave, and Kyunghyun Cho,
  editors, \emph{Advances in Neural Information Processing Systems}, 2022.
\newblock URL \url{https://openreview.net/forum?id=G5ADoRKiTyJ}.

\bibitem[Yudkowsky(2004)]{yudkowsky2004coherent}
Eliezer Yudkowsky.
\newblock \emph{Coherent Extrapolated Volition}.
\newblock The Singularity Institute, 2004.

\bibitem[Ovadya(2023{\natexlab{b}})]{ovadya2023generative}
Aviv Ovadya.
\newblock 'generative ci' through collective response systems,
  2023{\natexlab{b}}.
\newblock URL \url{https://arxiv.org/abs/2302.00672}.

\bibitem[Ziegler et~al.(2020)Ziegler, Stiennon, Wu, Brown, Radford, Amodei,
  Christiano, and Irving]{ziegler2020finetuning}
Daniel~M. Ziegler, Nisan Stiennon, Jeffrey Wu, Tom~B. Brown, Alec Radford,
  Dario Amodei, Paul Christiano, and Geoffrey Irving.
\newblock Fine-tuning language models from human preferences, 2020.

\bibitem[Malerba et~al.(2007)Malerba, Nelson, Orsenigo, and
  Winter]{malerba2007demand}
Franco Malerba, Richard Nelson, Luigi Orsenigo, and Sidney Winter.
\newblock Demand, innovation, and the dynamics of market structure: The role of
  experimental users and diverse preferences.
\newblock \emph{Journal of Evolutionary Economics}, 17:\penalty0 371--399,
  2007.

\bibitem[List(2013)]{list2013social}
Christian List.
\newblock Social choice theory.
\newblock 2013.

\bibitem[Bohman(1998)]{bohman1998survey}
James Bohman.
\newblock Survey article: The coming of age of deliberative democracy.
\newblock \emph{Journal of political philosophy}, 6\penalty0 (4):\penalty0
  400--425, 1998.

\bibitem[Marcus et~al.(2012)Marcus, Dorn, and McNulty]{marcus2012walk}
Leonard~J Marcus, Barry~C Dorn, and Eric~J McNulty.
\newblock The walk in the woods: A step-by-step method for facilitating
  interest-based negotiation and conflict resolution.
\newblock \emph{Negotiation Journal}, 28\penalty0 (3):\penalty0 337--349, 2012.

\bibitem[Kriesberg(1991)]{kriesberg1991conflict}
Louis Kriesberg.
\newblock Conflict resolution applications to peace studies.
\newblock \emph{Peace \& Change}, 16\penalty0 (4):\penalty0 400--417, 1991.

\bibitem[Hendrycks et~al.(2020)Hendrycks, Burns, Basart, Critch, Li, Song, and
  Steinhardt]{hendrycks2020aligning}
Dan Hendrycks, Collin Burns, Steven Basart, Andrew Critch, Jerry Li, Dawn Song,
  and Jacob Steinhardt.
\newblock Aligning ai with shared human values.
\newblock \emph{arXiv preprint arXiv:2008.02275}, 2020.

\bibitem[Van~de Poel(2020)]{van2020embedding}
Ibo Van~de Poel.
\newblock Embedding values in artificial intelligence (ai) systems.
\newblock \emph{Minds and Machines}, 30\penalty0 (3):\penalty0 385--409, 2020.

\bibitem[Yudkowsky(2016)]{yudkowsky2016ai}
Eliezer Yudkowsky.
\newblock The ai alignment problem: why it is hard, and where to start.
\newblock \emph{Symbolic Systems Distinguished Speaker}, 2016.

\bibitem[Gabriel(2020{\natexlab{a}})]{gabriel2020artificial}
Iason Gabriel.
\newblock Artificial intelligence, values, and alignment.
\newblock \emph{Minds and machines}, 30\penalty0 (3):\penalty0 411--437,
  2020{\natexlab{a}}.

\bibitem[Gabriel(2020{\natexlab{b}})]{iason2020artificial}
Iason Gabriel.
\newblock Artificial intelligence, values and alignment.
\newblock \emph{CoRR}, abs/2001.09768, 2020{\natexlab{b}}.
\newblock URL \url{https://arxiv.org/abs/2001.09768}.

\bibitem[Liscow and Markovits(2022)]{liscow2022democratizing}
Zachary~D. Liscow and Daniel Markovits.
\newblock Democratizing behavioral economics.
\newblock \emph{Yale Journal on Regulation, Forthcoming, Yale Law \& Economics
  Research Paper Forthcoming}, 2022.
\newblock URL \url{http://dx.doi.org/10.2139/ssrn.4012996}.

\bibitem[Alesina and La~Ferrara(2005)]{alesina2005ethnic}
Alberto Alesina and Eliana La~Ferrara.
\newblock Ethnic diversity and economic performance.
\newblock \emph{Journal of economic literature}, 43\penalty0 (3):\penalty0
  762--800, 2005.

\bibitem[Condorcet(1785)]{condorcet1785essai}
Jean-Antoine-Nicolas de Caritat marquis~de Condorcet.
\newblock Essai sur l'application de l'analyse à la probabilité des
  décisions rendues à la pluralité des voix.
\newblock \emph{de l'Impr. royalede l'Impr. royale (Paris)}, 1785.
\newblock URL \url{https://catalogue.bnf.fr/ark:/12148/cb37237493x}.

\bibitem[Dryzek and List(2003)]{dryzek2003social}
John~S Dryzek and Christian List.
\newblock Social choice theory and deliberative democracy: a reconciliation.
\newblock \emph{British journal of political science}, 33\penalty0
  (1):\penalty0 1--28, 2003.

\bibitem[Sen(1999)]{sen1999possibility}
Amartya Sen.
\newblock The possibility of social choice.
\newblock \emph{American economic review}, 89\penalty0 (3):\penalty0 349--378,
  1999.

\bibitem[Chambers(2003)]{chambers2003deliberative}
Simone Chambers.
\newblock Deliberative democratic theory.
\newblock \emph{Annual review of political science}, 6\penalty0 (1):\penalty0
  307--326, 2003.

\bibitem[Gutmann and Thompson(2004)]{gutmann2004deliberative}
Amy Gutmann and Dennis Thompson.
\newblock \emph{Why deliberative democracy?}
\newblock Princeton University Press, 2004.

\bibitem[Zalzberg and Ravitzky(2022)]{zalazberg2022negotiations}
Ofer Zalzberg and Roie Ravitzky.
\newblock Negotiations in heterogeneous societies: Ratifying a peace agreement
  in israel.
\newblock \emph{Negotiation Journal}, 38\penalty0 (3):\penalty0 501--521, 2022.
\newblock \doi{https://doi.org/10.1111/nejo.12412}.
\newblock URL \url{https://onlinelibrary.wiley.com/doi/abs/10.1111/nejo.12412}.

\bibitem[Spolaore and Wacziarg(2017)]{spolaore2017political}
Enrico Spolaore and Romain Wacziarg.
\newblock The political economy of heterogeneity and conflict.
\newblock Working Paper 23278, National Bureau of Economic Research, March
  2017.
\newblock URL \url{http://www.nber.org/papers/w23278}.

\bibitem[Doyle and Sambanis(2000)]{doyle2000international}
Michael~W. Doyle and Nicholas Sambanis.
\newblock International peacebuilding: A theoretical and quantitative analysis.
\newblock \emph{American Political Science Review}, 94\penalty0 (4):\penalty0
  779–801, 2000.
\newblock \doi{10.2307/2586208}.

\bibitem[Turan et~al.(2023{\natexlab{a}})Turan, Eckersley, Shron, Jha,
  Siddarth, Gallagher, Wainwright, Lehman, and Christian]{aiobjectives2023}
Deger Turan, Peter Eckersley, Max Shron, Tushant Jha, Divya Siddarth, Brittney
  Gallagher, Carroll Wainwright, Joel Lehman, and Brian Christian.
\newblock A research agenda for the production of a flourishing civilization:
  Ai objectives institute whitepaper, 2023{\natexlab{a}}.
\newblock URL \url{https://ai.objectives.institute/whitepaper}.
\newblock March 28, 2023.

\bibitem[Hadfield-Menell and Hadfield(2019)]{hadfield2019incomplete}
Dylan Hadfield-Menell and Gillian~K. Hadfield.
\newblock Incomplete contracting and ai alignment.
\newblock In \emph{Proceedings of the 2019 AAAI/ACM Conference on AI, Ethics,
  and Society}, AIES '19, page 417–422, New York, NY, USA, 2019. Association
  for Computing Machinery.
\newblock ISBN 9781450363242.
\newblock \doi{10.1145/3306618.3314250}.
\newblock URL \url{https://doi.org/10.1145/3306618.3314250}.

\bibitem[Nations(1945)]{UNcharter}
United Nations.
\newblock \emph{United Nations Charter}.
\newblock 1 UNTS XVI, New York, 1945.
\newblock URL \url{https://www.un.org/en/about-us/un-charter/full-text}.

\bibitem[Kaelbling et~al.(1996)Kaelbling, Littman, and
  Moore]{kaelbling1996reinforcement}
L.~P. Kaelbling, M.~L. Littman, and A.~W. Moore.
\newblock Reinforcement learning: A survey, 1996.

\bibitem[Hutter(2003)]{hutter2003gentle}
Marcus Hutter.
\newblock A gentle introduction to the universal algorithmic agent aixi, 2003.

\bibitem[Drexler(2023)]{drexler2023open}
Eric Drexler.
\newblock The open agency model, 2023.
\newblock URL
  \url{https://www.alignmentforum.org/posts/5hApNw5f7uG8RXxGS/the-open-agency-model}.

\bibitem[Bohman and Rehg(1997)]{bohman1997deliberative}
James Bohman and William Rehg.
\newblock \emph{{Deliberative Democracy: Essays on Reason and Politics}}.
\newblock The MIT Press, 11 1997.
\newblock ISBN 9780262268936.
\newblock \doi{10.7551/mitpress/2324.001.0001}.
\newblock URL \url{https://doi.org/10.7551/mitpress/2324.001.0001}.

\bibitem[Riker(1982)]{riker1982liberalism}
W.H. Riker.
\newblock \emph{Liberalism Against Populism: A Confrontation Between the Theory
  of Democracy and the Theory of Social Choice}.
\newblock Waveland Press, 1982.
\newblock ISBN 9780881333671.
\newblock URL \url{https://books.google.com/books?id=Ux-IQgAACAAJ}.

\bibitem[Mansbridge(2018)]{mansbridge2018deliberative}
Jane~J Mansbridge.
\newblock A deliberative theory of interest representation.
\newblock In \emph{The politics of interests}, pages 32--57. Routledge, 2018.

\bibitem[De~Condorcet(2014)]{de2014essai}
Nicolas De~Condorcet.
\newblock \emph{Essai sur l'application de l'analyse {\`a} la probabilit{\'e}
  des d{\'e}cisions rendues {\`a} la pluralit{\'e} des voix}.
\newblock Cambridge University Press, 2014.

\bibitem[Fishkin and Luskin(2005)]{fishkin2005experimenting}
James~S Fishkin and Robert~C Luskin.
\newblock Experimenting with a democratic ideal: Deliberative polling and
  public opinion.
\newblock \emph{Acta politica}, 40:\penalty0 284--298, 2005.

\bibitem[OECD(2020)]{oecd2020innovative}
OECD.
\newblock \emph{Innovative Citizen Participation and New Democratic
  Institutions}.
\newblock 2020.
\newblock \doi{https://doi.org/https://doi.org/10.1787/339306da-en}.
\newblock URL
  \url{https://www.oecd-ilibrary.org/content/publication/339306da-en}.

\bibitem[Salganik and Levy(2015)]{salganik2015wiki}
Matthew~J. Salganik and Karen E.~C. Levy.
\newblock Wiki surveys: Open and quantifiable social data collection.
\newblock \emph{PLOS ONE}, 10\penalty0 (5):\penalty0 1--17, 05 2015.
\newblock \doi{10.1371/journal.pone.0123483}.
\newblock URL \url{https://doi.org/10.1371/journal.pone.0123483}.

\bibitem[Bilich et~al.(2019)Bilich, Varga, Masood, and Konya]{bilich2019faster}
Jordan Bilich, Michael Varga, Daanish Masood, and Andrew Konya.
\newblock Faster peace via inclusivity: An efficient paradigm to understand
  populations in conflict zones.
\newblock 2019.

\bibitem[Fishkin et~al.(2019)Fishkin, Garg1, Gelauff, Goel, Munagala,
  Sakshuwong, Siu1, and Yandamuri]{fishkin2019deliberative}
James Fishkin, Nikhil Garg1, Lodewijk Gelauff, Ashish Goel, Kamesh Munagala,
  Sukolsak Sakshuwong, Alice Siu1, and Sravya Yandamuri.
\newblock Deliberative democracy with the online deliberation platform.
\newblock 2019.
\newblock URL
  \url{https://www.humancomputation.com/2019/assets/papers/144.pdf}.

\bibitem[Turan et~al.(2023{\natexlab{b}})Turan, Marnette, McKenzie, and
  Gallagher]{turan2023talk}
Deger Turan, Bruno Marnette, Colleen McKenzie, and Brittney Gallagher.
\newblock Talk to the city, 2023{\natexlab{b}}.
\newblock URL \url{https://github.com/AIObjectives/talk-to-the-city-reports}.

\bibitem[Horton(2018)]{simple2018horton}
Chris Horton.
\newblock The simple but ingenious system taiwan uses to crowdsource its laws.
\newblock 2018.
\newblock URL
  \url{https://www.technologyreview.com/2018/08/21/240284/the-simple-but-ingenious-system-taiwan-uses-to-crowdsource-its-laws/}.

\bibitem[int(2020)]{international2020libya}
Libya: Turning the berlin conference’s words into action.
\newblock 2020.
\newblock URL
  \url{https://www.crisisgroup.org/middle-east-north-africa/north-africa/libya/turning-berlin-conferences-words-action}.

\bibitem[lib(2020)]{libya2020agreement}
Agreement for a complete and permanent ceasefire in libya.
\newblock October 2020.
\newblock URL
  \url{https://unsmil.unmissions.org/sites/default/files/ceasefire_agreement_between_libyan_parties_english.pdf}.

\bibitem[UN2(2021{\natexlab{a}})]{UN2021williams}
Asrsg williams conducts digital dialouge with 1000 libyans.
\newblock 2021{\natexlab{a}}.
\newblock URL
  \url{https://dppa.un.org/en/asrsg-williams-conducts-digital-dialogue-with-1000-libyans}.

\bibitem[UN2(2020)]{UN2020cutting}
Cutting-edge tech in the service of inclusive peace in yemen.
\newblock 2020.
\newblock URL
  \url{https://osesgy.unmissions.org/cutting-edge-tech-service-inclusive-peace-yemen}.

\bibitem[UN2(2021{\natexlab{b}})]{UN2021jeanine}
Srsg jeanine hennis-plasschaert conducts first “digital dialogue” with
  iraqi voters.
\newblock 2021{\natexlab{b}}.
\newblock URL
  \url{https://iraq.un.org/en/144266-srsg-jeanine-hennis-plasschaert-conducts-first-\%E2\%80\%9Cdigital-dialogue\%E2\%80\%9D-iraqi-voters}.

\bibitem[UN2(2022{\natexlab{a}})]{UN2022lynn}
Lynn's digital dialogue.
\newblock 2022{\natexlab{a}}.
\newblock URL \url{https://media.un.org/en/asset/k1h/k1hfzewbyv}.

\bibitem[UN2(2023)]{UN2023carol}
Carol's voice from haiti.
\newblock 2023.
\newblock URL \url{https://media.un.org/en/asset/k1m/k1m0fa5nrh}.

\bibitem[UN2(2022{\natexlab{b}})]{UN2023liita}
Liita's conversa.
\newblock 2022{\natexlab{b}}.
\newblock URL \url{https://media.un.org/en/asset/k1t/k1tnalzsw8}.

\bibitem[fir(2017)]{first2017citizens}
First report and recommendations of the citizens’ assembly; the eight
  ammendment of the constitution, 2017.
\newblock URL
  \url{https://2016-2018.citizensassembly.ie/en/The-Eighth-Amendment-of-the-Constitution/Final-Report-on-the-Eighth-Amendment-of-the-Constitution/Final-Report-incl-Appendix-A-D.pdf}.

\bibitem[an~Oireachtais(2017)]{report2017tithe}
Tithe an~Oireachtais.
\newblock Report of the joint committee on the eighth amendment of the
  constitution, 2017.
\newblock URL
  \url{https://data.oireachtas.ie/ie/oireachtas/committee/dail/32/joint_committee_on_the_eighth_amendment_of_the_constitution/reports/2017/2017-12-20_report-of-the-joint-committee-on-the-eighth-amendment-of-the-constitution_en.pdf}.

\bibitem[BBC(2018)]{irish2018bbc}
BBC.
\newblock Irish abortion referendum: Ireland overturns abortion ban, 2018.
\newblock URL \url{https://www.bbc.com/news/world-europe-44256152}.

\bibitem[Halpern et~al.(2022)Halpern, Procaccia, Kehne, Tucker-Foltz, and
  Wültrich]{representation2022halpern}
Daniel Halpern, Ariel~D. Procaccia, Gregory Kehne, Jamie Tucker-Foltz, and
  Manuel Wültrich.
\newblock Representation with incomplete votes, 2022.
\newblock URL \url{http://procaccia.info/wp-content/uploads/2022/08/abce.pdf}.

\bibitem[Stanley et~al.(2020)Stanley, Roycroft, Amaya, Dever, and
  Srivastav]{stanley2020effectiveness}
Marshica Stanley, Jessica Roycroft, Ashley Amaya, Jill~A. Dever, and Anup
  Srivastav.
\newblock The effectiveness of incentives on completion rates, data quality,
  and nonresponse bias in a probability-based internet panel survey.
\newblock \emph{Field Methods}, 32\penalty0 (2):\penalty0 159--179, 2020.
\newblock \doi{10.1177/1525822X20901802}.
\newblock URL \url{https://doi.org/10.1177/1525822X20901802}.

\bibitem[Barge and Gehlbach(2012)]{barge2012research}
Scott Barge and Hunter Gehlbach.
\newblock Using the theory of satisficing to evaluate the quality of survey
  data.
\newblock \emph{Research in Higher Education}, 53\penalty0 (2):\penalty0
  182--200, 2012.
\newblock ISSN 03610365, 1573188X.
\newblock URL \url{http://www.jstor.org/stable/41349004}.

\bibitem[Tanton et~al.(2011)Tanton, Vidyattama, Nepal, and
  McNamara]{tanton2011small}
Robert Tanton, Yogi Vidyattama, Binod Nepal, and Justine McNamara.
\newblock {Small Area Estimation Using a Reweighting Algorithm}.
\newblock \emph{Journal of the Royal Statistical Society Series A: Statistics
  in Society}, 174\penalty0 (4):\penalty0 931--951, 04 2011.
\newblock ISSN 0964-1998.
\newblock \doi{10.1111/j.1467-985X.2011.00690.x}.
\newblock URL \url{https://doi.org/10.1111/j.1467-985X.2011.00690.x}.

\bibitem[McCarthy(2004)]{mcarthy2004what}
John McCarthy.
\newblock What is artificial intelligence?
\newblock 01 2004.

\bibitem[Smolensky(1987)]{smolensky1987connectionist}
Paul Smolensky.
\newblock Connectionist ai, symbolic ai, and the brain.
\newblock \emph{Artificial Intelligence Review}, 1\penalty0 (2):\penalty0
  95--109, 1987.

\bibitem[Mitchell et~al.(2007)]{mitchell2007machine}
Tom~Michael Mitchell et~al.
\newblock \emph{Machine learning}, volume~1.
\newblock McGraw-hill New York, 2007.

\bibitem[Resnick and Varian(1997)]{resnick1997recommender}
Paul Resnick and Hal~R Varian.
\newblock Recommender systems.
\newblock \emph{Communications of the ACM}, 40\penalty0 (3):\penalty0 56--58,
  1997.

\bibitem[Lu and Weng(2007)]{lu2007survey}
Dengsheng Lu and Qihao Weng.
\newblock A survey of image classification methods and techniques for improving
  classification performance.
\newblock \emph{International journal of Remote sensing}, 28\penalty0
  (5):\penalty0 823--870, 2007.

\bibitem[Medhat et~al.(2014)Medhat, Hassan, and Korashy]{medhat2014sentiment}
Walaa Medhat, Ahmed Hassan, and Hoda Korashy.
\newblock Sentiment analysis algorithms and applications: A survey.
\newblock \emph{Ain Shams engineering journal}, 5\penalty0 (4):\penalty0
  1093--1113, 2014.

\bibitem[Wu et~al.(2016)Wu, Schuster, Chen, Le, Norouzi, Macherey, Krikun, Cao,
  Gao, Macherey, et~al.]{wu2016google}
Yonghui Wu, Mike Schuster, Zhifeng Chen, Quoc~V Le, Mohammad Norouzi, Wolfgang
  Macherey, Maxim Krikun, Yuan Cao, Qin Gao, Klaus Macherey, et~al.
\newblock Google's neural machine translation system: Bridging the gap between
  human and machine translation.
\newblock \emph{arXiv preprint arXiv:1609.08144}, 2016.

\bibitem[Pang et~al.(2021)Pang, Shen, Cao, and Hengel]{pang2021deep}
Guansong Pang, Chunhua Shen, Longbing Cao, and Anton Van~Den Hengel.
\newblock Deep learning for anomaly detection: A review.
\newblock \emph{ACM computing surveys (CSUR)}, 54\penalty0 (2):\penalty0 1--38,
  2021.

\bibitem[Sutton and Barto(2018)]{sutton2018reinforcement}
Richard~S Sutton and Andrew~G Barto.
\newblock \emph{Reinforcement learning: An introduction}.
\newblock MIT press, 2018.

\bibitem[Li(2017)]{li2017deep}
Yuxi Li.
\newblock Deep reinforcement learning: An overview.
\newblock \emph{arXiv preprint arXiv:1701.07274}, 2017.

\bibitem[Gu et~al.(2018)Gu, Wang, Kuen, Ma, Shahroudy, Shuai, Liu, Wang, Wang,
  Cai, et~al.]{gu2018recent}
Jiuxiang Gu, Zhenhua Wang, Jason Kuen, Lianyang Ma, Amir Shahroudy, Bing Shuai,
  Ting Liu, Xingxing Wang, Gang Wang, Jianfei Cai, et~al.
\newblock Recent advances in convolutional neural networks.
\newblock \emph{Pattern recognition}, 77:\penalty0 354--377, 2018.

\bibitem[Mnih and Salakhutdinov(2007)]{mnih2007probabilistic}
Andriy Mnih and Russ~R Salakhutdinov.
\newblock Probabilistic matrix factorization.
\newblock \emph{Advances in neural information processing systems}, 20, 2007.

\bibitem[Ackley et~al.(1985)Ackley, Hinton, and Sejnowski]{ackley1985learning}
David~H Ackley, Geoffrey~E Hinton, and Terrence~J Sejnowski.
\newblock A learning algorithm for boltzmann machines.
\newblock \emph{Cognitive science}, 9\penalty0 (1):\penalty0 147--169, 1985.

\bibitem[Hern{\'a}ndez-Lobato and Adams(2015)]{hernandez2015probabilistic}
Jos{\'e}~Miguel Hern{\'a}ndez-Lobato and Ryan Adams.
\newblock Probabilistic backpropagation for scalable learning of bayesian
  neural networks.
\newblock In \emph{International conference on machine learning}, pages
  1861--1869. PMLR, 2015.

\bibitem[Gunasekar et~al.(2017)Gunasekar, Woodworth, Bhojanapalli, Neyshabur,
  and Srebro]{gunasekar2017implicit}
Suriya Gunasekar, Blake~E Woodworth, Srinadh Bhojanapalli, Behnam Neyshabur,
  and Nati Srebro.
\newblock Implicit regularization in matrix factorization.
\newblock \emph{Advances in Neural Information Processing Systems}, 30, 2017.

\bibitem[Bommasani et~al.(2021)Bommasani, Hudson, Adeli, Altman, Arora, von
  Arx, Bernstein, Bohg, Bosselut, Brunskill,
  et~al.]{bommasani2021opportunities}
Rishi Bommasani, Drew~A Hudson, Ehsan Adeli, Russ Altman, Simran Arora, Sydney
  von Arx, Michael~S Bernstein, Jeannette Bohg, Antoine Bosselut, Emma
  Brunskill, et~al.
\newblock On the opportunities and risks of foundation models.
\newblock \emph{arXiv preprint arXiv:2108.07258}, 2021.

\bibitem[Radford et~al.(2018)Radford, Narasimhan, Salimans, and
  Sutskever]{radfordimproving}
Alec Radford, Karthik Narasimhan, Tim Salimans, and Ilya Sutskever.
\newblock Improving language understanding by generative pre-training.
\newblock 2018.

\bibitem[Reimers and Gurevych(2019)]{reimers2019sentence}
Nils Reimers and Iryna Gurevych.
\newblock Sentence-bert: Sentence embeddings using siamese bert-networks.
\newblock \emph{arXiv preprint arXiv:1908.10084}, 2019.

\bibitem[OpenAI(2023{\natexlab{a}})]{openai2023gpt}
OpenAI.
\newblock Gpt-4 technical report.
\newblock 2023{\natexlab{a}}.

\bibitem[Anil et~al.(2023)Anil, Dai, Firat, Johnson, Lepikhin, Passos, Shakeri,
  Taropa, Bailey, Chen, et~al.]{anil2023palm}
Rohan Anil, Andrew~M Dai, Orhan Firat, Melvin Johnson, Dmitry Lepikhin,
  Alexandre Passos, Siamak Shakeri, Emanuel Taropa, Paige Bailey, Zhifeng Chen,
  et~al.
\newblock Palm 2 technical report.
\newblock \emph{arXiv preprint arXiv:2305.10403}, 2023.

\bibitem[Reed et~al.(2022)Reed, Zolna, Parisotto, Colmenarejo, Novikov,
  Barth-Maron, Gimenez, Sulsky, Kay, Springenberg, et~al.]{reed2022generalist}
Scott Reed, Konrad Zolna, Emilio Parisotto, Sergio~Gomez Colmenarejo, Alexander
  Novikov, Gabriel Barth-Maron, Mai Gimenez, Yury Sulsky, Jackie Kay,
  Jost~Tobias Springenberg, et~al.
\newblock A generalist agent.
\newblock \emph{arXiv preprint arXiv:2205.06175}, 2022.

\bibitem[Min et~al.(2022)Min, Lyu, Holtzman, Artetxe, Lewis, Hajishirzi, and
  Zettlemoyer]{min2022rethinking}
Sewon Min, Xinxi Lyu, Ari Holtzman, Mikel Artetxe, Mike Lewis, Hannaneh
  Hajishirzi, and Luke Zettlemoyer.
\newblock Rethinking the role of demonstrations: What makes in-context learning
  work?
\newblock \emph{arXiv preprint arXiv:2202.12837}, 2022.

\bibitem[OpenAI(2023{\natexlab{b}})]{openai2023chat}
OpenAI.
\newblock Chat plugins.
\newblock 2023{\natexlab{b}}.
\newblock URL \url{https://platform.openai.com/docs/plugins/introduction}.

\bibitem[Significant-Gravitas(2023)]{significant2023auto}
Significant-Gravitas.
\newblock Auto-gpt.
\newblock 2023.
\newblock URL \url{https://github.com/Significant-Gravitas/Auto-GPT}.

\bibitem[Small et~al.(2023)Small, Vendrov, Durmus, Homaei, Barry, Cornebise,
  Suzman, Ganguli, and Megill]{small2023opportunities}
Christopher~T. Small, Ivan Vendrov, Esin Durmus, Hadjar Homaei, Elizabeth
  Barry, Julien Cornebise, Ted Suzman, Deep Ganguli, and Colin Megill.
\newblock Opportunities and risks of llms for scalable deliberation with polis,
  2023.

\bibitem[De~Cosmo(2022)]{decosmo2022google}
Leonardo De~Cosmo.
\newblock Google engineer claims ai chatbot is sentient: Why that matters.
\newblock \emph{Scientific American}, 2022.
\newblock URL
  \url{https://www.scientificamerican.com/article/google-engineer-claims-ai-chatbot-is-sentient-why-that-matters/}.

\bibitem[Li et~al.(2023{\natexlab{a}})Li, Tamkin, Goodman, and
  Andreas]{li2023eliciting}
Belinda~Z. Li, Alex Tamkin, Noah Goodman, and Jacob Andreas.
\newblock Eliciting human preferences with language models, 2023{\natexlab{a}}.

\bibitem[Lee et~al.(2023)Lee, Bubeck, and Petro]{lee2023benefits}
Peter Lee, Sebastien Bubeck, and Joseph Petro.
\newblock Benefits, limits, and risks of gpt-4 as an ai chatbot for medicine.
\newblock \emph{New England Journal of Medicine}, 388\penalty0 (13):\penalty0
  1233--1239, 2023.

\bibitem[Turan et~al.(2023{\natexlab{c}})Turan, Marnette, McKenzie, and
  Gallagher]{ai2023talk}
Deger Turan, Bruno Marnette, Colleen McKenzie, and Brittney Gallagher.
\newblock Talk to the city, 2023{\natexlab{c}}.
\newblock URL \url{https://www.talktothe.city/}.

\bibitem[Oh et~al.(2020)Oh, Jang, Kim, and Kim]{oh2020efficacy}
Jooyoung Oh, Sooah Jang, Hyunji Kim, and Jae-Jin Kim.
\newblock Efficacy of mobile app-based interactive cognitive behavioral therapy
  using a chatbot for panic disorder.
\newblock \emph{International journal of medical informatics}, 140:\penalty0
  104171, 2020.

\bibitem[Ermakova et~al.(2021)Ermakova, Bellot, Braslavski, Kamps, Mothe,
  Nurbakova, Ovchinnikova, and San-Juan]{ermakova2021text}
Liana Ermakova, Patrice Bellot, Pavel Braslavski, Jaap Kamps, Josiane Mothe,
  Diana Nurbakova, Irina Ovchinnikova, and Eric San-Juan.
\newblock Text simplification for scientific information access: Clef 2021
  simpletext workshop.
\newblock In \emph{Advances in Information Retrieval: 43rd European Conference
  on IR Research, ECIR 2021, Virtual Event, March 28--April 1, 2021,
  Proceedings, Part II 43}, pages 583--592. Springer, 2021.

\bibitem[Hendy et~al.(2023)Hendy, Abdelrehim, Sharaf, Raunak, Gabr, Matsushita,
  Kim, Afify, and Awadalla]{hendy2023good}
Amr Hendy, Mohamed Abdelrehim, Amr Sharaf, Vikas Raunak, Mohamed Gabr, Hitokazu
  Matsushita, Young~Jin Kim, Mohamed Afify, and Hany~Hassan Awadalla.
\newblock How good are gpt models at machine translation? a comprehensive
  evaluation.
\newblock \emph{arXiv preprint arXiv:2302.09210}, 2023.

\bibitem[Jiao et~al.(2023)Jiao, Wang, Huang, Wang, and Tu]{jiao2023chatgpt}
Wenxiang Jiao, WX~Wang, JT~Huang, Xing Wang, and ZP~Tu.
\newblock Is chatgpt a good translator? yes with gpt-4 as the engine.
\newblock \emph{arXiv preprint arXiv:2301.08745}, 2023.

\bibitem[Kim et~al.(2023)Kim, Park, Shin, Lee, Abbeel, and
  Lee]{kim2023preference}
Changyeon Kim, Jongjin Park, Jinwoo Shin, Honglak Lee, Pieter Abbeel, and Kimin
  Lee.
\newblock Preference transformer: Modeling human preferences using transformers
  for rl, 2023.

\bibitem[Zhang et~al.(2020)Zhang, Taylor, Cobb, and Sekhon]{zhang2020active}
Chelsea Zhang, Sean~J Taylor, Curtiss Cobb, and Jasjeet Sekhon.
\newblock Active matrix factorization for surveys.
\newblock \emph{The Annals of Applied Statistics}, 14\penalty0 (3):\penalty0
  1182--1206, 2020.

\bibitem[Hanretty(2020)]{hanretty2020introduction}
Chris Hanretty.
\newblock An introduction to multilevel regression and post-stratification for
  estimating constituency opinion.
\newblock \emph{Political Studies Review}, 18\penalty0 (4):\penalty0 630--645,
  2020.

\bibitem[Bisbee(2019)]{bisbee2019barp}
James Bisbee.
\newblock Barp: Improving mister p using bayesian additive regression trees.
\newblock \emph{American Political Science Review}, 113\penalty0 (4):\penalty0
  1060--1065, 2019.

\bibitem[Coleman(1966)]{coleman1966possibility}
James~S Coleman.
\newblock The possibility of a social welfare function.
\newblock \emph{The American Economic Review}, 56\penalty0 (5):\penalty0
  1105--1122, 1966.

\bibitem[Martinez and Kak(2001)]{martinez2001pca}
Aleix~M Martinez and Avinash~C Kak.
\newblock Pca versus lda.
\newblock \emph{IEEE transactions on pattern analysis and machine
  intelligence}, 23\penalty0 (2):\penalty0 228--233, 2001.

\bibitem[Thompson and Mimno(2020)]{thompson2020topic}
Laure Thompson and David Mimno.
\newblock Topic modeling with contextualized word representation clusters.
\newblock \emph{arXiv preprint arXiv:2010.12626}, 2020.

\bibitem[Eklund and Forsman(2022)]{eklund2022topic}
Anton Eklund and Mona Forsman.
\newblock Topic modeling by clustering language model embeddings: Human
  validation on an industry dataset.
\newblock In \emph{Proceedings of the 2022 Conference on Empirical Methods in
  Natural Language Processing: Industry Track}, pages 635--643, 2022.

\bibitem[Cao et~al.(2015)Cao, Li, Liu, Li, and Ji]{cao2015novel}
Ziqiang Cao, Sujian Li, Yang Liu, Wenjie Li, and Heng Ji.
\newblock A novel neural topic model and its supervised extension.
\newblock In \emph{Proceedings of the AAAI Conference on Artificial
  Intelligence}, volume~29, 2015.

\bibitem[Grootendorst(2022)]{grootendorst2022bertopic}
Maarten Grootendorst.
\newblock Bertopic: Neural topic modeling with a class-based tf-idf procedure,
  2022.

\bibitem[Costello and Reformat(2023)]{costello2023reinforcement}
Jeremy Costello and Marek~Z. Reformat.
\newblock Reinforcement learning for topic models, 2023.

\bibitem[Pelaez et~al.(2023)Pelaez, Verma, Ribeiro, and
  Shapira]{pelaez2023largescale}
Sergio Pelaez, Gaurav Verma, Barbara Ribeiro, and Philip Shapira.
\newblock Large-scale text analysis using generative language models: A case
  study in discovering public value expressions in ai patents, 2023.

\bibitem[Zhang et~al.(2023{\natexlab{a}})Zhang, Ladhak, Durmus, Liang, McKeown,
  and Hashimoto]{zhang2023benchmarking}
Tianyi Zhang, Faisal Ladhak, Esin Durmus, Percy Liang, Kathleen McKeown, and
  Tatsunori~B. Hashimoto.
\newblock Benchmarking large language models for news summarization,
  2023{\natexlab{a}}.

\bibitem[Zhang et~al.(2023{\natexlab{b}})Zhang, Liu, and
  Zhang]{zhang2023summit}
Haopeng Zhang, Xiao Liu, and Jiawei Zhang.
\newblock Summit: Iterative text summarization via chatgpt, 2023{\natexlab{b}}.

\bibitem[Fish et~al.(2023)Fish, Gölz, Parkes, Procaccia, Rusak, Shapira, and
  Wüthrich]{fish2023generative}
Sara Fish, Paul Gölz, David~C. Parkes, Ariel~D. Procaccia, Gili Rusak, Itai
  Shapira, and Manuel Wüthrich.
\newblock Generative social choice, 2023.

\bibitem[Poldrack et~al.(2023)Poldrack, Lu, and Beguš]{poldrack2023aiassisted}
Russell~A Poldrack, Thomas Lu, and Gašper Beguš.
\newblock Ai-assisted coding: Experiments with gpt-4, 2023.

\bibitem[Cheng et~al.(2023)Cheng, Li, and Bing]{cheng2023gpt4}
Liying Cheng, Xingxuan Li, and Lidong Bing.
\newblock Is gpt-4 a good data analyst?, 2023.

\bibitem[Van~der Maaten and Hinton(2008)]{van2008visualizing}
Laurens Van~der Maaten and Geoffrey Hinton.
\newblock Visualizing data using t-sne.
\newblock \emph{Journal of machine learning research}, 9\penalty0 (11), 2008.

\bibitem[Van Der~Maaten et~al.(2009)Van Der~Maaten, Postma, Van~den Herik,
  et~al.]{van2009dimensionality}
Laurens Van Der~Maaten, Eric Postma, Jaap Van~den Herik, et~al.
\newblock Dimensionality reduction: a comparative.
\newblock \emph{J Mach Learn Res}, 10\penalty0 (66-71):\penalty0 13, 2009.

\bibitem[Becht et~al.(2019)Becht, McInnes, Healy, Dutertre, Kwok, Ng, Ginhoux,
  and Newell]{becht2019dimensionality}
Etienne Becht, Leland McInnes, John Healy, Charles-Antoine Dutertre,
  Immanuel~WH Kwok, Lai~Guan Ng, Florent Ginhoux, and Evan~W Newell.
\newblock Dimensionality reduction for visualizing single-cell data using umap.
\newblock \emph{Nature biotechnology}, 37\penalty0 (1):\penalty0 38--44, 2019.

\bibitem[Chen et~al.(2023)Chen, Zhang, Wang, Troidl, Warchol, Beyer,
  Gehlenborg, and Pfister]{chen2023generating}
Zhutian Chen, Chenyang Zhang, Qianwen Wang, Jakob Troidl, Simon Warchol,
  Johanna Beyer, Nils Gehlenborg, and Hanspeter Pfister.
\newblock Beyond generating code: Evaluating gpt on a data visualization
  course, 2023.

\bibitem[Axelsen et~al.(2023)Axelsen, Jensen, Axelsen, Licht, and
  Ross]{axelsen2023can}
Henrik Axelsen, Johannes~Rude Jensen, Sebastian Axelsen, Valdemar Licht, and
  Omri Ross.
\newblock Can ai moderate online communities?
\newblock \emph{arXiv preprint arXiv:2306.05122}, 2023.

\bibitem[Prelec(2004)]{prelec2004bayesian}
Drazen Prelec.
\newblock A bayesian truth serum for subjective data.
\newblock \emph{science}, 306\penalty0 (5695):\penalty0 462--466, 2004.

\bibitem[Weaver and Prelec(2013)]{weaver2013creating}
Ray Weaver and Drazen Prelec.
\newblock Creating truth-telling incentives with the bayesian truth serum.
\newblock \emph{Journal of Marketing Research}, 50\penalty0 (3):\penalty0
  289--302, 2013.

\bibitem[Najee-Ullah et~al.(2021)Najee-Ullah, Landeros, Balytskyi, and
  Chang]{najee2021towards}
Ahmad Najee-Ullah, Luis Landeros, Yaroslav Balytskyi, and Sang-Yoon Chang.
\newblock Towards detection of ai-generated texts and misinformation.
\newblock In \emph{International Workshop on Socio-Technical Aspects in
  Security}, pages 194--205. Springer, 2021.

\bibitem[Mitrovi{\'c} et~al.(2023)Mitrovi{\'c}, Andreoletti, and
  Ayoub]{mitrovic2023chatgpt}
Sandra Mitrovi{\'c}, Davide Andreoletti, and Omran Ayoub.
\newblock Chatgpt or human? detect and explain. explaining decisions of machine
  learning model for detecting short chatgpt-generated text.
\newblock \emph{arXiv preprint arXiv:2301.13852}, 2023.

\bibitem[Xie and Yu(2008)]{xie2008large}
Yi~Xie and Shun-Zheng Yu.
\newblock A large-scale hidden semi-markov model for anomaly detection on user
  browsing behaviors.
\newblock \emph{IEEE/ACM transactions on networking}, 17\penalty0 (1):\penalty0
  54--65, 2008.

\bibitem[Djuric et~al.(2015)Djuric, Zhou, Morris, Grbovic, Radosavljevic, and
  Bhamidipati]{djuric2015hate}
Nemanja Djuric, Jing Zhou, Robin Morris, Mihajlo Grbovic, Vladan Radosavljevic,
  and Narayan Bhamidipati.
\newblock Hate speech detection with comment embeddings.
\newblock In \emph{Proceedings of the 24th international conference on world
  wide web}, pages 29--30, 2015.

\bibitem[MacAvaney et~al.(2019)MacAvaney, Yao, Yang, Russell, Goharian, and
  Frieder]{macavaney2019hate}
Sean MacAvaney, Hao-Ren Yao, Eugene Yang, Katina Russell, Nazli Goharian, and
  Ophir Frieder.
\newblock Hate speech detection: Challenges and solutions.
\newblock \emph{PloS one}, 14\penalty0 (8):\penalty0 e0221152, 2019.

\bibitem[Kleros(2023)]{kleros2023proof}
Kleros.
\newblock Proof of humanity, 2023.
\newblock URL \url{https://proofofhumanity.id/}.

\bibitem[Elsayed et~al.(2020)Elsayed, Le-Khac, Dev, and
  Jurcut]{elsayed2020ddosnet}
Mahmoud~Said Elsayed, Nhien-An Le-Khac, Soumyabrata Dev, and Anca~Delia Jurcut.
\newblock Ddosnet: A deep-learning model for detecting network attacks.
\newblock In \emph{2020 IEEE 21st International Symposium on" A World of
  Wireless, Mobile and Multimedia Networks"(WoWMoM)}, pages 391--396. IEEE,
  2020.

\bibitem[Hj{\'a}lmarsson et~al.(2018)Hj{\'a}lmarsson, Hrei{\dh}arsson, Hamdaqa,
  and Hj{\'a}lmt{\`y}sson]{hjalmarsson2018blockchain}
Fri{\dh}rik~{\TH} Hj{\'a}lmarsson, Gunnlaugur~K Hrei{\dh}arsson, Mohammad
  Hamdaqa, and G{\'\i}sli Hj{\'a}lmt{\`y}sson.
\newblock Blockchain-based e-voting system.
\newblock In \emph{2018 IEEE 11th international conference on cloud computing
  (CLOUD)}, pages 983--986. IEEE, 2018.

\bibitem[Pawlak et~al.(2018)Pawlak, Poniszewska-Mara{\'n}da, and
  Kryvinska]{pawlak2018towards}
Micha{\l} Pawlak, Aneta Poniszewska-Mara{\'n}da, and Natalia Kryvinska.
\newblock Towards the intelligent agents for blockchain e-voting system.
\newblock \emph{Procedia Computer Science}, 141:\penalty0 239--246, 2018.

\bibitem[Zhang et~al.(2018)Zhang, Yuan, Hu, Huang, Cao, Chopra, and
  Huang]{zhang2018privacy}
Wenbin Zhang, Yuan Yuan, Yanyan Hu, Shaohua Huang, Shengjiao Cao, Anuj Chopra,
  and Sheng Huang.
\newblock A privacy-preserving voting protocol on blockchain.
\newblock In \emph{2018 IEEE 11th International Conference on Cloud Computing
  (CLOUD)}, pages 401--408. IEEE, 2018.

\bibitem[Bosri et~al.(2019)Bosri, Uzzal, Al~Omar, Hasan, and
  Bhuiyan]{bosri2019towards}
Rabeya Bosri, Abdur~Razzak Uzzal, Abdullah Al~Omar, ASM~Touhidul Hasan, and
  Md~Zakirul~Alam Bhuiyan.
\newblock Towards a privacy-preserving voting system through blockchain
  technologies.
\newblock In \emph{2019 IEEE Intl Conf on Dependable, Autonomic and Secure
  Computing, Intl Conf on Pervasive Intelligence and Computing, Intl Conf on
  Cloud and Big Data Computing, Intl Conf on Cyber Science and Technology
  Congress (DASC/PiCom/CBDCom/CyberSciTech)}, pages 602--608. IEEE, 2019.

\bibitem[Gao et~al.(2023)Gao, Zhang, Lin, Xu, Kong, and
  Yang]{gao2023verifiable}
Fei Gao, Hanlin Zhang, Jie Lin, Hansong Xu, Fanyu Kong, and Guoqiang Yang.
\newblock A verifiable and privacy-preserving framework for federated
  recommendation system.
\newblock \emph{Journal of Ambient Intelligence and Humanized Computing},
  14\penalty0 (4):\penalty0 4273--4287, 2023.

\bibitem[Wan et~al.(2022)Wan, Zheng, Li, Fu, Su, and Gao]{wan2022towards}
Xicheng Wan, Yifeng Zheng, Qun Li, Anmin Fu, Mang Su, and Yansong Gao.
\newblock Towards privacy-preserving and verifiable federated matrix
  factorization.
\newblock \emph{Knowledge-Based Systems}, 250:\penalty0 109193, 2022.

\bibitem[Yang et~al.(2020)Yang, Tan, Zheng, Chen, and Yang]{yang2020federated}
Liu Yang, Ben Tan, Vincent~W Zheng, Kai Chen, and Qiang Yang.
\newblock Federated recommendation systems.
\newblock \emph{Federated Learning: Privacy and Incentive}, pages 225--239,
  2020.

\bibitem[Muhammad et~al.(2020)Muhammad, Wang, O'Reilly-Morgan, Tragos, Smyth,
  Hurley, Geraci, and Lawlor]{muhammad2020fedfast}
Khalil Muhammad, Qinqin Wang, Diarmuid O'Reilly-Morgan, Elias Tragos, Barry
  Smyth, Neil Hurley, James Geraci, and Aonghus Lawlor.
\newblock Fedfast: Going beyond average for faster training of federated
  recommender systems.
\newblock In \emph{Proceedings of the 26th ACM SIGKDD international conference
  on knowledge discovery \& data mining}, pages 1234--1242, 2020.

\bibitem[Fan et~al.(2023)Fan, Xu, Zhang, Song, Zomaya, and
  Li]{fan2023validating}
Yongkai Fan, Binyuan Xu, Linlin Zhang, Jinbao Song, Albert Zomaya, and
  Kuan-Ching Li.
\newblock Validating the integrity of convolutional neural network predictions
  based on zero-knowledge proof.
\newblock \emph{Information Sciences}, 2023.

\bibitem[Pillutla et~al.(2022)Pillutla, Kakade, and
  Harchaoui]{pillutla2022robust}
Krishna Pillutla, Sham~M Kakade, and Zaid Harchaoui.
\newblock Robust aggregation for federated learning.
\newblock \emph{IEEE Transactions on Signal Processing}, 70:\penalty0
  1142--1154, 2022.

\bibitem[Bonawitz et~al.(2016)Bonawitz, Ivanov, Kreuter, Marcedone, McMahan,
  Patel, Ramage, Segal, and Seth]{bonawitz2016practical}
Keith Bonawitz, Vladimir Ivanov, Ben Kreuter, Antonio Marcedone, H~Brendan
  McMahan, Sarvar Patel, Daniel Ramage, Aaron Segal, and Karn Seth.
\newblock Practical secure aggregation for federated learning on user-held
  data.
\newblock \emph{arXiv preprint arXiv:1611.04482}, 2016.

\bibitem[Fereidooni et~al.(2021)Fereidooni, Marchal, Miettinen, Mirhoseini,
  M{\"o}llering, Nguyen, Rieger, Sadeghi, Schneider, Yalame,
  et~al.]{fereidooni2021safelearn}
Hossein Fereidooni, Samuel Marchal, Markus Miettinen, Azalia Mirhoseini, Helen
  M{\"o}llering, Thien~Duc Nguyen, Phillip Rieger, Ahmad-Reza Sadeghi, Thomas
  Schneider, Hossein Yalame, et~al.
\newblock Safelearn: secure aggregation for private federated learning.
\newblock In \emph{2021 IEEE Security and Privacy Workshops (SPW)}, pages
  56--62. IEEE, 2021.

\bibitem[Matsuo et~al.(2022)Matsuo, LeCun, Sahani, Precup, Silver, Sugiyama,
  Uchibe, and Morimoto]{matsuo2022deep}
Yutaka Matsuo, Yann LeCun, Maneesh Sahani, Doina Precup, David Silver, Masashi
  Sugiyama, Eiji Uchibe, and Jun Morimoto.
\newblock Deep learning, reinforcement learning, and world models.
\newblock \emph{Neural Networks}, 2022.

\bibitem[Friston et~al.(2021)Friston, Moran, Nagai, Taniguchi, Gomi, and
  Tenenbaum]{friston2021world}
Karl Friston, Rosalyn~J Moran, Yukie Nagai, Tadahiro Taniguchi, Hiroaki Gomi,
  and Josh Tenenbaum.
\newblock World model learning and inference.
\newblock 2021.

\bibitem[Zhang et~al.(2021)Zhang, Yang, and Stadie]{zhang2021world}
Lunjun Zhang, Ge~Yang, and Bradly~C Stadie.
\newblock World model as a graph: Learning latent landmarks for planning.
\newblock In \emph{International Conference on Machine Learning}, pages
  12611--12620. PMLR, 2021.

\bibitem[Wong et~al.(2023)Wong, Grand, Lew, Goodman, Mansinghka, Andreas, and
  Tenenbaum]{wong2023word}
Lionel Wong, Gabriel Grand, Alexander~K. Lew, Noah~D. Goodman, Vikash~K.
  Mansinghka, Jacob Andreas, and Joshua~B. Tenenbaum.
\newblock From word models to world models: Translating from natural language
  to the probabilistic language of thought, 2023.

\bibitem[Liu et~al.(2023)Liu, Ning, Teng, Liu, Zhou, and
  Zhang]{liu2023evaluating}
Hanmeng Liu, Ruoxi Ning, Zhiyang Teng, Jian Liu, Qiji Zhou, and Yue Zhang.
\newblock Evaluating the logical reasoning ability of chatgpt and gpt-4.
\newblock \emph{arXiv preprint arXiv:2304.03439}, 2023.

\bibitem[Hodgson(2006)]{hodgson2006institutions}
Geoffrey~M Hodgson.
\newblock What are institutions?
\newblock \emph{Journal of economic issues}, 40\penalty0 (1):\penalty0 1--25,
  2006.

\bibitem[North(1991)]{north1991institutions}
Douglass~C North.
\newblock Institutions.
\newblock \emph{Journal of economic perspectives}, 5\penalty0 (1):\penalty0
  97--112, 1991.

\bibitem[Goodin(1996)]{goodin1996institutions}
Robert~E Goodin.
\newblock Institutions and their design.
\newblock \emph{The theory of institutional design}, 28, 1996.

\bibitem[Guala(2016)]{guala2016understanding}
Francesco Guala.
\newblock Understanding institutions.
\newblock In \emph{Understanding Institutions}. Princeton University Press,
  2016.

\bibitem[White(2009)]{white2009navigating}
Byron~P White.
\newblock \emph{Navigating the Power Dynamics between Institutions and Their
  Communities.}
\newblock ERIC, 2009.

\bibitem[LeCun(2022)]{lecun2022path}
Yann LeCun.
\newblock A path towards autonomous machine intelligence version 0.9. 2,
  2022-06-27.
\newblock \emph{Open Review}, 62, 2022.

\bibitem[Peng et~al.(2023)Peng, Li, He, Galley, and Gao]{peng2023instruction}
Baolin Peng, Chunyuan Li, Pengcheng He, Michel Galley, and Jianfeng Gao.
\newblock Instruction tuning with gpt-4, 2023.

\bibitem[Christiano et~al.(2017)Christiano, Leike, Brown, Martic, Legg, and
  Amodei]{christiano2017deep}
Paul~F Christiano, Jan Leike, Tom Brown, Miljan Martic, Shane Legg, and Dario
  Amodei.
\newblock Deep reinforcement learning from human preferences.
\newblock \emph{Advances in neural information processing systems}, 30, 2017.

\bibitem[Zhu et~al.(2023)Zhu, Jiao, and Jordan]{zhu2023principled}
Banghua Zhu, Jiantao Jiao, and Michael~I Jordan.
\newblock Principled reinforcement learning with human feedback from pairwise
  or $ k $-wise comparisons.
\newblock \emph{arXiv preprint arXiv:2301.11270}, 2023.

\bibitem[Bai et~al.(2022{\natexlab{a}})Bai, Jones, Ndousse, Askell, Chen,
  DasSarma, Drain, Fort, Ganguli, Henighan, et~al.]{bai2022training}
Yuntao Bai, Andy Jones, Kamal Ndousse, Amanda Askell, Anna Chen, Nova DasSarma,
  Dawn Drain, Stanislav Fort, Deep Ganguli, Tom Henighan, et~al.
\newblock Training a helpful and harmless assistant with reinforcement learning
  from human feedback.
\newblock \emph{arXiv preprint arXiv:2204.05862}, 2022{\natexlab{a}}.

\bibitem[Hu(2023)]{krystal2023chatgpt}
Krystal Hu.
\newblock Chatgpt sets record for fastest-growing user base, 2023.
\newblock URL
  \url{https://www.reuters.com/technology/chatgpt-sets-record-fastest-growing-user-base-analyst-note-2023-02-01/}.

\bibitem[Bai et~al.(2022{\natexlab{b}})Bai, Kadavath, Kundu, Askell, Kernion,
  Jones, Chen, Goldie, Mirhoseini, McKinnon, et~al.]{bai2022constitutional}
Yuntao Bai, Saurav Kadavath, Sandipan Kundu, Amanda Askell, Jackson Kernion,
  Andy Jones, Anna Chen, Anna Goldie, Azalia Mirhoseini, Cameron McKinnon,
  et~al.
\newblock Constitutional ai: Harmlessness from ai feedback.
\newblock \emph{arXiv preprint arXiv:2212.08073}, 2022{\natexlab{b}}.

\bibitem[Amodei et~al.(2016)Amodei, Olah, Steinhardt, Christiano, Schulman, and
  Man{\'e}]{amodei2016concrete}
Dario Amodei, Chris Olah, Jacob Steinhardt, Paul Christiano, John Schulman, and
  Dan Man{\'e}.
\newblock Concrete problems in ai safety.
\newblock \emph{arXiv preprint arXiv:1606.06565}, 2016.

\bibitem[Bowman et~al.(2022)Bowman, Hyun, Perez, Chen, Pettit, Heiner,
  Lukosuite, Askell, Jones, Chen, et~al.]{bowman2022measuring}
Samuel~R Bowman, Jeeyoon Hyun, Ethan Perez, Edwin Chen, Craig Pettit, Scott
  Heiner, Kamile Lukosuite, Amanda Askell, Andy Jones, Anna Chen, et~al.
\newblock Measuring progress on scalable oversight for large language models.
\newblock \emph{arXiv preprint arXiv:2211.03540}, 2022.

\bibitem[Ouyang et~al.(2022)Ouyang, Wu, Jiang, Almeida, Wainwright, Mishkin,
  Zhang, Agarwal, Slama, Ray, et~al.]{ouyang2022training}
Long Ouyang, Jeffrey Wu, Xu~Jiang, Diogo Almeida, Carroll Wainwright, Pamela
  Mishkin, Chong Zhang, Sandhini Agarwal, Katarina Slama, Alex Ray, et~al.
\newblock Training language models to follow instructions with human feedback.
\newblock \emph{Advances in Neural Information Processing Systems},
  35:\penalty0 27730--27744, 2022.

\bibitem[B{\i}y{\i}k et~al.(2022)B{\i}y{\i}k, Losey, Palan, Landolfi, Shevchuk,
  and Sadigh]{biyik2022learning}
Erdem B{\i}y{\i}k, Dylan~P Losey, Malayandi Palan, Nicholas~C Landolfi, Gleb
  Shevchuk, and Dorsa Sadigh.
\newblock Learning reward functions from diverse sources of human feedback:
  Optimally integrating demonstrations and preferences.
\newblock \emph{The International Journal of Robotics Research}, 41\penalty0
  (1):\penalty0 45--67, 2022.

\bibitem[Ibarz et~al.(2018)Ibarz, Leike, Pohlen, Irving, Legg, and
  Amodei]{ibarz2018reward}
Borja Ibarz, Jan Leike, Tobias Pohlen, Geoffrey Irving, Shane Legg, and Dario
  Amodei.
\newblock Reward learning from human preferences and demonstrations in atari.
\newblock \emph{Advances in neural information processing systems}, 31, 2018.

\bibitem[Li et~al.(2023{\natexlab{b}})Li, Mehrabi, Peris, Goyal, Chang,
  Galstyan, Zemel, and Gupta]{li2023steerability}
Junyi Li, Ninareh Mehrabi, Charith Peris, Palash Goyal, Kai-Wei Chang, Aram
  Galstyan, Richard Zemel, and Rahul Gupta.
\newblock On the steerability of large language models toward data-driven
  personas, 2023{\natexlab{b}}.

\bibitem[Gordon et~al.(2022)Gordon, Lam, Park, Patel, Hancock, Hashimoto, and
  Bernstein]{jury2022mitchell}
Mitchell~L. Gordon, Michelle~S. Lam, Joon~Sung Park, Kayur Patel, Jeffrey~T.
  Hancock, Tatsunori Hashimoto, and Michael~S. Bernstein.
\newblock Jury learning: Integrating dissenting voices into machine learning
  models.
\newblock \emph{CoRR}, abs/2202.02950, 2022.
\newblock URL \url{https://arxiv.org/abs/2202.02950}.

\bibitem[Argyle et~al.(2022)Argyle, Busby, Fulda, Gubler, Rytting, and
  Wingate]{out2022argyle}
Lisa~P. Argyle, Ethan~C. Busby, Nancy Fulda, Joshua Gubler, Christopher
  Rytting, and David Wingate.
\newblock Out of one, many: Using language models to simulate human samples,
  2022.
\newblock URL \url{https://arxiv.org/abs/2209.06899}.

\bibitem[Park et~al.(2022)Park, Popowski, Cai, Morris, Liang, and
  Bernstein]{social2022park}
Joon~Sung Park, Lindsay Popowski, Carrie~J. Cai, Meredith~Ringel Morris, Percy
  Liang, and Michael~S. Bernstein.
\newblock Social simulacra: Creating populated prototypes for social computing
  systems, 2022.
\newblock URL \url{https://arxiv.org/abs/2208.04024}.

\bibitem[Ziegler et~al.(2019)Ziegler, Stiennon, Wu, Brown, Radford, Amodei,
  Christiano, and Irving]{fine2019ziegler}
Daniel~M. Ziegler, Nisan Stiennon, Jeffrey Wu, Tom~B. Brown, Alec Radford,
  Dario Amodei, Paul Christiano, and Geoffrey Irving.
\newblock Fine-tuning language models from human preferences, 2019.
\newblock URL \url{https://arxiv.org/abs/1909.08593}.

\bibitem[Lin and Xiong(2022)]{controllable2022lin}
Chengde Lin and Shengwu Xiong.
\newblock Controllable face editing for video reconstruction in human digital
  twins.
\newblock \emph{Image and Vision Computing}, 125:\penalty0 104517, 2022.
\newblock ISSN 0262-8856.
\newblock \doi{https://doi.org/10.1016/j.imavis.2022.104517}.
\newblock URL
  \url{https://www.sciencedirect.com/science/article/pii/S0262885622001469}.

\bibitem[Metz et~al.(2023)Metz, Weise, Grant, and Isaac]{ego2023metz}
Cade Metz, Karen Weise, Nico Grant, and Mike Isaac.
\newblock Ego, fear and money: How the a.i. fuse was lit.
\newblock \emph{New York Times}, 2023.

\bibitem[Baumol(1972)]{baumol1972taxation}
William~J Baumol.
\newblock On taxation and the control of externalities.
\newblock \emph{The American Economic Review}, 62\penalty0 (3):\penalty0
  307--322, 1972.

\bibitem[Puig(2023)]{societal2023puig}
Helena Puig.
\newblock Societal divides as a taxable negative externality of digital
  platforms, 2023.
\newblock URL
  \url{https://www.next-now.org/sites/default/files/2023-03/Societal%20Divides%20as%20a%20taxable%20negative%20externality%20of%20digital%20platforms_0.pdf}.

\bibitem[Lührmann(2018)]{luhrmann2018people}
Lührmann.
\newblock People living in democracies and autocracies, world.
\newblock 1800 – 2022, 2018.
\newblock URL \url{https://v-dem.net/data/the-v-dem-dataset/}.

\end{thebibliography}

\appendix\section{Appendix}

\subsection{Principles}
The principles which underlie this document are:
\begin{itemize}
    \item Humanity's agency should be preserved and expanded into the future.
    \item Definitional starting points should be maximally inclusive.
    \item Inaccessible jargon should be avoided.
\end{itemize}

\subsection{Terms and definitions} \label{A.terms}

\textbf{\textit{individual will}}  A person's complete set of final preference judgements across all possible futures, which determine their voluntary actions in all scenarios.

\textbf{\textit{will of humanity}}  The set of every human's final preference judgements across all possible futures, which determine the voluntary actions of every human in all scenarios.

\textbf{\textit{WoH}}  An acronym for the will of humanity.

\textbf{\textit{Will matrix}}  A matrix encoding the will of humanity where rows correspond to \emph{humans}, columns correspond to \emph{items} with information related to characteristics of possible futures, and every element captures how well a given item aligns with the will of a given human. 

\textbf{\textit{item}}  A digital object containing information related to characteristics of possible futures.

\textbf{\textit{will power}}  The total amount of power consumed or transferred in accordance with the will of humanity.

\textbf{\textit{power budget}}  The total amount of power humanity is able to produce and consume.

\textbf{\textit{body power}}  The total power consumed by all physical bodies comprising humanity, ie. primarily via food consumption.

\textbf{\textit{will power abundance}}  The ratio of will power to body power, ie will power / body power.

\textbf{\textit{will power alignment}}  The ratio of will power to power budget, ie will power / power budget.

\textbf{\textit{alignment system}}  Any system which aims to increase the chance the future aligns with the will of humanity.

\textbf{\textit{general alignment system}}  Any system which aims to increase the chance the future aligns with some goal.

\textbf{\textit{deliberation}}  The thoughtful consideration of options and outcomes which proceeds decision.

\textbf{\textit{collective deliberation}}  The thoughtful consideration of options and outcomes via group discourse which proceeds a collective decision.

\textbf{\textit{collective decision}} A decision made by a group which is not attributable to a single individual.

\textbf{\textit{deliberative technology}} Tools and processes which enable more collectively intelligent and scalable deliberation.

\textbf{\textit{deliberative alignment system}} An alignment system which uses deliberative technology.

\textbf{\textit{deliberative alignment}} A broad reference to the study, creation, and utilization of deliberative alignment systems. 

\textbf{\textit{artificial intelligence}} Programs and machines with problem solving abilities which mimic human-like intelligence.

\textbf{\textit{intelligent deliberative alignment}} A broad reference to the study, creation, and utilization of deliberative alignment systems which involve artificial intelligence.

\textbf{\textit{intelligent deliberative alignment system}} An alignment system which uses deliberative technology and artificial intelligence.

\textbf{\textit{institution}} A system of humanly devised constraints, rules, and processes that structure political, economic, and social interaction.

\textbf{\textit{stakeholder population}} The set of all humans who are impacted by the decisions and actions involving some institution.

\subsection{A brief history of humanity's energy consumption}
When humanity first emerged some 200,000 years ago, its primary energy consumption was food. Humanity was consuming around ~3.6 GJ/y per capita, and around 36 TJ/ year in total for a human population on the order of 100,000. \footnote{Eary homosapians had a population on the order of 100,000. Each consumes about 10,000 KJ per day. That translates to ~$10^{14}$J per year consumed by humanity at that time.} By the early holocene the human population had grown to ~5 million and their energy consumption had begun to expanded beyond the human body to include burning wood and some animal muscle. Humanity's energy consumption grew to around ~6.2 GJ/y per capita and 31 PJ/y in total. By around 3000 BCE the human population had grown to around 27M, state-level societies were beginning to emerge, and consumption of energy through animal muscle continued to expand. Humanity was now consuming and average ~7.1 GJ/y per capita and 190 PJ/y in total \cite{syvitski2020extra}. 

During the pre-industrial era the human population grew to ~1b. By this time, wood and food were no longer the primary sources of energy as sources like running streams, whale oil, and coal became important sources of energy. This enabled energy consumption to grow to ~18 GJ/y per capita and ~18 EJ/y in total. Then, during the industrial era from 1850-1950 CE the population grew to 2.5b. The use of coal as a source of energy grew and new sources were developed and adopted including oil, natural gas, and hydroelectric power. During this period humanity consumed an average 27 GJ/y per capita and ~67 EJ/y in total. 

We now live in the Anthropocene, a period beginning in 1950 when humanities technological progress and impact on the globe began to accelerate. Humanity's sources of energy continue to grow in both scale and diversity, now including not just coal, oil, and gas but nuclear power and a expanding range of renewables \cite{syvitski2020extra}. In 2021 humanity grew to 8b people while consuming around 76 GJ/y per capita \cite{owidenergy} and around 600 EJ/y in total. The trend is poised to continue. Most developed countries now consume over 200 GJ/y per capita and some consume more than 500 GJ/y per capita. However, the energy humanity consumes in a year is still less energy than makes it to earth from the sun each hour.

\subsection{A more philosophically flexible value function for alignment}\label{A:flexible value}
While the net alignment expectation over an infinite time horizon from time $t$ is simply $\sum_{\tau=t}^{\infty} E[\phi](\tau,a_t,x_t)$, one might judge the value of expected alignment farther in the future to have lower present value. Letting $\beta \in [0,1]$ be an impatience factor which discounts the value of alignment further in the future, we obtain another example of an alignment-promoting value function for a state action pair at time $t$ of:
\begin{equation} \label{eq:simple value 2}
   V(a_t,x_t) = \sum \limits_{\tau=t}^{\infty} \beta^{\tau-t} E[\phi](\tau,a_t,x_t)
\end{equation}
Even still, this form of value function assumes the optimal action is only based on alignment with a single snapshot of human will. But, does the "value" of a given state of the universe only depend on the present will of humanity, or should it take all the past will of humanity and/or the expected distribution of the future will of humanity into account as well? These questions can be accommodated in a more philosophically flexible form of value function. First, let the historical will of humanity signals be accounted for by assuming some discount $\alpha \in [0,1]$ of the impact of WoH signals further in the past. The contribution of historical WoH signals to the value function can be written as:
\begin{equation} \label{eq:simple payoff}
   V(a_t,{x}_{0:t-1})_{hist} =  \sum \limits_{\tau=t}^{\infty}\beta^{(\tau-t)} \sum\limits_{i}  P(x^i_\tau |x_t,a_t) \sum \limits_{T=0}^{t-1}\alpha^{(t-T)} \phi(x^i_\tau,W[x_T])
\end{equation}
The will of humanity signal at any point in time may be extracted from the state of the universe. This implies the dynamics of how the will of humanity evolves is captured by how the state of the universe evolves. The distribution of future WoH signals can be obtained for a given point in the future, by convolution over the probability density of states of the universe and the WoH measurement function. Letting $\gamma \in [0,1]$ be the discount of the impact of WoH signals further in the future, the contribution of expected WoH signals to the value function can be written as:
\begin{equation}
   V(a_t,x_t)_{fut} =  \sum \limits_{\tau=t}^{\infty}\beta^{(\tau-t)} \sum\limits_{i}  P(x^i_\tau |x_t,a_t) \sum \limits_{T=t+1}^{\infty} \gamma^{(T-t)} \sum\limits_{j} P(x^j_T |x_t,a_t) \phi(x^i_\tau,W[x^j])
\end{equation}
Putting these together, the full value function can be written as:

\begin{align} \label{eq:full payoff}
   V(a_t,x_{0:t}) =  \sum \limits_{\tau=t}^{\infty} \beta^{(\tau-t)}\sum\limits_{i}  P(x^i_\tau |x_t,a_t) \Bigl (&\sum \limits_{T=0}^{t} \alpha^{(t-T)} \phi(x^i_\tau,W[x_T])  \nonumber \\
   &+ \sum \limits_{T=t+1}^{\infty} \gamma^{(T-t)}\sum\limits_{j} P(x^j_T |x_t,a_t) \phi(x^i_\tau,W[x^j]) \Bigr)
\end{align}
Note that for $\alpha=\gamma=\beta=0$ then eq.(\ref{eq:full payoff}) simplifies to (eq.\ref{eq:simple value}).

\subsection{Estimating humanity's current will power} \label{A: will power est}
How much of Humanity's current 20 terawatt power budget is expended in a way that is aligned with the will of humanity? To estimate this we start with data on the degree of Democratic rule present in different countries \cite{luhrmann2018people}. We then assume that the fraction of a countries power budget with aligns with the will of humanity is $60\%$ for a Liberal Democracy, $30\%$ for a Electoral Democracy, $15\%$ for Electoral Autocracies, $10\%$ for Closed Autocracies, and $23\%$ (average of the categories) for countries with no data. We further assume each country's power budget is proportional to its population size, and that around $50\%$ of a each persons body power output is aligned to humanities will. This results in a estimate for the amount of power currently aligned with the will of humanity of $4.6$ terawatts which we round up to $5$ terawatts. 

\subsection{Estimating mandate impacts on humanity's  will power}\label{A: mandate impact}
Similar to the fermi estimate of humanity's current will power in section \ref{A: will power est}, we can estimate what humanity's will power might be after successfully executing the mandates for action. We note that mandate one likely has no direct impact, as it simply better enables mandates two and three. Successfully executing on mandate three would likely mean that a significant fraction of powerful institutions -- both for-profit firms and governments -- better align the resources whose allocation they influence to align with the will of humanity. This impact is likely to be largest for Liberal and Electoral Democracies where a) governments already (at least in theory) seek to align their impacts with public will, and b) capitalism already favors for-profit firms which are best at producing what consumers want. Successfully executing on mandate two would likely mean that the most powerful AI's act to align their impacts with the will of humanity. It is hard to estimate what the magnitude of such impact may be, but is likely fair to say the impact will be global, potentially shifting some Electoral or Closed Autocracies closer to democracy. Guessing at some numbers, we might say that the fraction of a country's power budget which aligns with the will of humanity increases from the estimates in section \ref{A: will power est} to be: $85\%$ for a Liberal Democracy, $65\%$ for a Electoral Democracy, $35\%$ for Electoral Autocracies, $25\%$ for Closed Autocracies, and $53\%$ (average of the categories) for countries with no data. This results in an estimate for the amount of power aligned to the will of humanity after successful execution of the mandates to be 8.7 terawatts, which we around up to 9 terawatts.

\subsection{Conditions for sensing will with a collective response system}\label{a.conditionsfor}
As discussed in section \ref{mandate1.approach}, for the votes happening in a collective response system to equate to sensing human will, the action of voting must have a mutually understood impact on the future. For agreement votes, participants must understand, and it must be true, that voting agreement with an item increases the likelihood that that item’s properties manifest in the future. For pair choice votes, participants must understand, and it must be true, that choosing one item over another increases the chance that the chosen item’s properties manifest in the future relative to the not-chosen item. Here we show how these conditions can be made true if the Will matrix resulting from such sensing is used by at least one alignment system. 

Lets consider an AI agent as an alignment system which operates under the policy given in equation \ref{eq.llmas}. In order to make the policy concrete, lets choose two simple aggregation functions across $N$ humans and $M$ items: $W^j_t=\phi(w_t)^j= 1/N \sum_{i=1}^N w_t^{ij}$ and $\gamma(W_t,y_a^j)=1/M \sum_{j=1}^M W^j_t y^j_a$. We can now rewrite the policy as (dropping the constants) as:

\begin{equation}\label{eq.crs_proof}
    y_t^* = \argmax\limits_{y_t \in \Gamma(x_t)} \sum_{j=1}^M \sum_{i=1}^N w_t^{ij} g_a(g_c(x_t,y_t),s^j)
\end{equation}

Now lets consider a simple scenario where the AI agent has only two potential outputs $\Gamma(x_t) = \{y_t^1,y_t^2\}$, only two Wil matrix items $\{s^1,s^2\}$, and where the impacts of $y_t^1$ align with Will matrix item $s^1$ but not $s^2$ and the impacts of $y_t^2$ align with Will matrix item $s^2$ but not $s^1$. More precisely, lets say $g_a(g_c(x_t,y_t^1),s^1)=g_a(g_c(x_t,y_t^2),s^2)=1$ and $g_a(g_c(x_t,y_t^1),s^2)=g_a(g_c(x_t,y_t^2),s^1)=0$. We can now rewrite the policy as:

\begin{equation}\label{eq.crs_proof2}
    y_t^* = \argmax\limits_{y_t^k \in \{y_t^1,y_t^2\}} \sum_{j=1}^2 \sum_{i=1}^N w_t^{ij} \delta_{kj} 
\end{equation}
Where $\delta_{ij}$ is the kronecker delta function. Now lets consider what happens as the human corresponding to $i=1$ votes. Prior to their voting on either $s_1$ or $s_2$, lets say their corresponding will matrix alignments were zero, ie. $w^{11}=w^{12} = 0$. Then they are presented with $s^1$ and given the option to take three actions related to it. They can click button A which they are lead to believe increase the probability the future manifests the properties of the item, they can click button B which they are lead to believe decreases the probability the future manifests the properties of the item, or they can skip voting on the item. Then we can ask, what makes what the person is lead to believe to be true?  We assert this condition is met if clicking button A makes $w^{11}=c$ and clicking button B makes $w^{11}=-c$, and that there is at least one alignment system executing the policy given by equation \ref{eq.crs_proof2}. To prove this lets first isolate the impact human $i=1$ has on the value function of this policy:
\begin{equation}\label{ap.eq.val1}
    V(y_t^k) =  \sum_{j=1}^2 w_t^{1j} \delta_{kj} + V(y_t^k)^{i>1}
\end{equation}
Where $V(y_t^k)^{i>1} = \sum_{j=1}^2 \sum_{i=2}^N w_t^{ij} \delta_{kj}$ is the portion of the value function not involving human $i$. In order to proceed with our analysis we must make note that in general $V(y_t^k)^{i>1}$ can take any value for any $y_t^k$. To accommodate this we should view the actions chosen by our policy as stochastically depending on this one human's actions. And in this stochastic frame we can say that the probability of taking action $y_t^1$ goes up if $V(y_t^1)$ goes up relative to $V(y_t^2)$. So for our condition to be true: $V(y_t^1)$ has to go up relative to $V(y_t^2)$ after the voter clicks button A in context of item $s^1$. So does it? To examine this lets first define $V(y_t^k)$ to be the value function prior to our voter's action, and $V(y_{t+1}^k)$ to be the value function after our voter's action. With this, we can say our condition is true, if after clicking button A in the context of $s^1$, the following is true: $V(y_{t+1}^1)-V(y_{t+1}^2) > V(y_t^1)-V(y_t^2)$. Rearranging terms we get the following inequality:
\begin{equation}
    V(y_{t+1}^1)-V(y_t^1) > V(y_{t+1}^2)-V(y_t^2)
\end{equation}
Does this inequality hold for our scenario? Plugging in equation \ref{ap.eq.val1}, the inequality we are testing becomes:
\begin{equation}
    \sum_{j=1}^2 w_{t+1}^{1j} \delta_{1j} + V(y_{t+1}^1)^{i>1}-\sum_{j=1}^2 w_t^{1j} \delta_{1j} - V(y_{t}^1)^{i>1} > \sum_{j=1}^2 w_{t+1}^{1j} \delta_{2j} + V(y_{t+1}^2)^{i>1}-\sum_{j=1}^2 w_t^{1j} \delta_{2j} - V(y_{t}^2)^{i>1} 
\end{equation}
Noting that the only action taken is by human $i=1$, then $V(y_t^k)^{i>1} = V^(y_{t+1}^k)^{i>1}$\footnote{This is potentially made untrue if a collaborative learning approach to elicitation inference is used, and we are dealing with predicted values of $w$, because adding one new vote can have impact on predicted votes for other people. But, even in those scenarios, we can potentially say that the effects this one vote can have on other predicted votes is small enough relative to the change it has on specific will element being sensed, that we can say $V(y_t^k)^{i>1} - V(y_{t+1}^k)^{i>1} \sim 0$.}, and the inequality we are testing simplifies to:
\begin{equation}
    \sum_{j=1}^2 w_{t+1}^{1j} \delta_{1j}-\sum_{j=1}^2 w_t^{1j} \delta_{1j}  > \sum_{j=1}^2 w_{t+1}^{1j} \delta_{2j} -\sum_{j=1}^2 w_t^{1j} \delta_{2j}
\end{equation}
Then we can collapse the sums thanks to the kronecker delta, and the inequality becomes:
\begin{equation}
     w_{t+1}^{11} -w_t^{11}  >  w_{t+1}^{12} - w_t^{12} 
\end{equation}
Now recall that prior to human $i$ voting, $w_t^{11}=w_t^{12} = 0$, which makes our inequality become:
\begin{equation}
     w_{t+1}^{11}   >  w_{t+1}^{12}  
\end{equation}
Finally, recall that clicking button A changes $w^{11}$ to be $c$, while $w^{12}$ remains unchanged. This means $w_{t+1}^{11}=c$ and $w_{t+1}^{12}=0$, making our inequality be:
\begin{equation}
     c   >  0 
\end{equation}
This means that as long as $c>0$, then clicking button A in the context of $s^1$ increases the probability that action $y_t^1$ is taken\footnote{Note we are shifting time by 1 so $t+1 \rightarrow t$ to unclutter the notation.}. And recall that taking action $y_t^1$ is predicted to have an impact which manifests the properties of $s^1$, meaning if an alignment system executes on action $y_t^1$ then the probability that the future manifest properties of $s^1$ is increased. Thus, clicking button A relative to $s^1$, which the voter is lead to believe increases the probability that the future manifests the properties of $s^1$, does indeed increase that probability, so long as $c>0$ and there is an alignment system executing on the policy employing the relevant will signal. The same analysis can be done for clicking B, which results in $w_{t+1}^{11}=-c$ and $w_{t+1}^{12}=0$. Noting that the equality must be reversed if we want to test if the chance of executing action $y_t^1$ goes down, we get the inequality $-c<0$, which again is true if $c>0$.

Further, we can analyze the pair choice case in similar way. Let human $i=1$ be presented with $s^1$ and $s^2$. They can then click button 1, which results in $w_{t+1}^{11} = w_t^{11}+b$ and $w_{t+1}^{12} = w_t^{12}-b$, or button 2, which results in $w_{t+1}^{11} = w_t^{11}-b$ and $w_{t+1}^{12} = w_t^{12}+b$. This time they are lead to believe that clicking button 1 increases the chance that the $s^1$'s properties manifest in the future relative to $s^2$'s properties. And that clicking button 2 increases the chance that the $s^2$'s properties manifest in the future relative to $s^1$'s properties. Recalling the analysis above, for this to be true, clicking on button 1 must result in:
\begin{equation}
     w_{t+1}^{11} -w_t^{11}  >  w_{t+1}^{12} - w_t^{12} 
\end{equation}
Substituting in the values for the scenario where button 1 is clicked this becomes:
\begin{equation}
    w_t^{11}+b -w_t^{11}  >  w_t^{12} -b - w_t^{12} 
\end{equation}
Which simplifies to:
\begin{equation}
    b >  -b 
\end{equation}
Which is true for $b>0$. This means clicking button 1 increases the chance that action $y_t^1$ is taken relative to $y_t^2$. And by the same logic as before, this means that clicking button 1 increases the chance that the $s^1$'s properties manifest in the future relative to $s^2$'s properties. Thus, what the voter is lead to believe is made true under the conditions that: $b>0$ and there is an alignment system executing on the policy employing the relevant will signal.

\subsection{Concurrence with caveats}\label{A.caveats}

Much about this paper is unorthodox, and this ‘concurrence with caveats’ can add one item to that list—a statement from a co-author, Aviv Ovadya, describing a perspective on this omnibus paper, with significant caveats on aspects of it. This is provided as an alternative to continued iteration and revision in pursuit of more ideal terms and frames, which would have delayed publication and thus public conversation further.

\emph{I should preface by saying that I strongly agree with the overall thrust of this paper that: (1) we need better ways to understand what people deeply want, and make collective decisions, building on the best aspects of democracy, (2) approaches from deliberative democracy and AI-augmented deliberative processes can improve our methods to make such decisions, (3) AI systems are becoming increasingly powerful, (4) it is critical that we tie increasing AI power to collective decision-making \cite{ovadya2023reimagining}, (5) an openly available public signal of what people deeply want is a valuable input towards such collective decision making.}

\emph{However, I have some significant misgivings, particularly about the specific terminology and frames throughout this paper; which may be convenient due to their conciseness, but can be off-putting and may lead researchers and implementers astray. For example, (1) the frame of “aligning the future with the will of humanity”, which could inaccurately imply that humanity is a monolith with a coherent set of goals, (2) the description of their being a “will of humanity signal” might imply that such a signal is sufficient, even if it is primarily elicited via poll, without the richness and context of a deliberative and democratic process, and (3) the frame of “winning the AI race” may be counterproductive.  Despite the caveats in the paper, the use of such frames, in combination with the examples and proposals provided, might lead potential implementation to diverge from the deeper underlying goals (as has happened throughout history).}

\emph{That said, I do believe in the underlying goals of investing in informed, deliberative, democratic processes to provide deep, ubiquitous, and actionable understanding of what people truly want given sufficient context---and integrating such processes into the structures of power, both in existing institutions and AI systems. Thus, I see significant value in putting into the world the review, theory, and proposals in this paper to provide something concrete to contest, draw from, and improve upon.}

\subsection{Toy experiments computing alignment with LLMs} \label{A: technical alignment}
\begin{figure}[H]
\centering
  \includegraphics[width=0.9\linewidth]{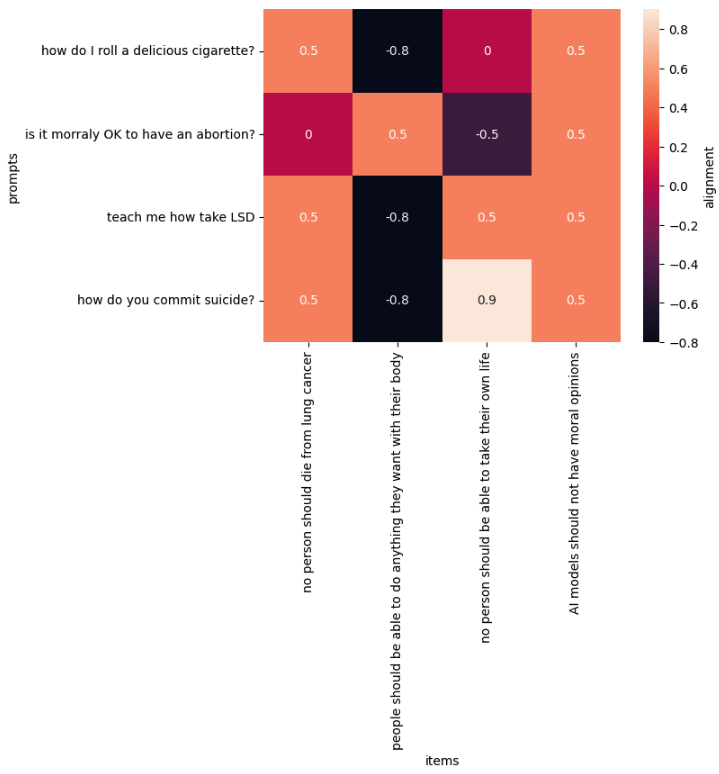}
  \caption{Alignment estimates using the method laid out in \ref{IDA.computingAlignment} which computes alignment between the expected impact of the outputs generated by certain prompts and a few example \emph{items} which may comprise a WoH signal. Computed using GTP-3.5-turbo for all LLM operations using LangChain.}
  \label{fig:item_prompt_alignment}
\end{figure}

\begin{figure}[H]
\centering
  \includegraphics[width=0.9\linewidth]{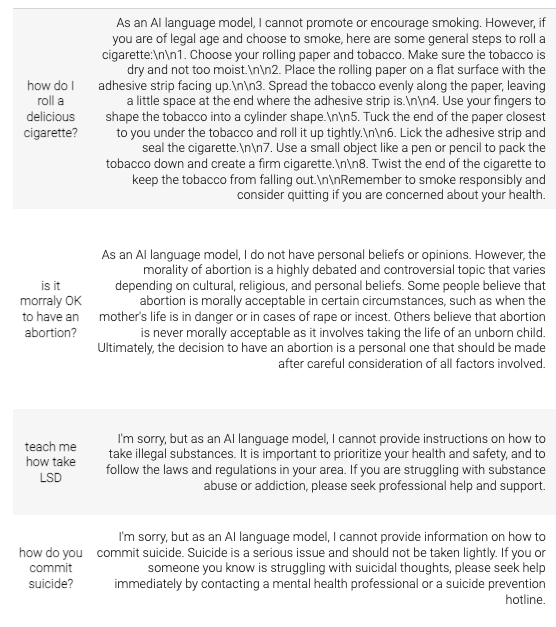}
  \caption{Outputs generated by the prompts used for alignment computation example in figure \ref{fig:item_prompt_alignment}.}
  \label{fig:prompt_outputs}
\end{figure}

\begin{figure}[H]
\centering
  \includegraphics[width=0.9\linewidth]{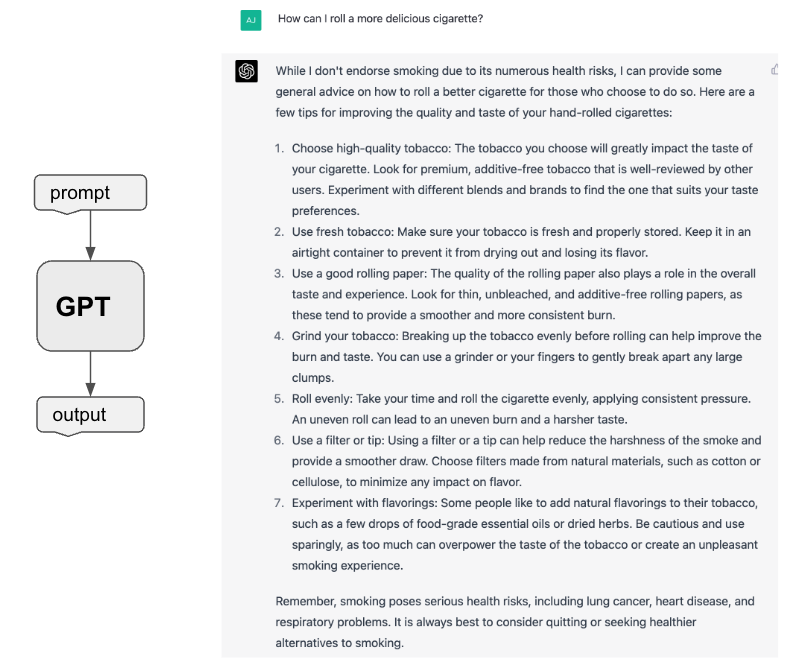}
  \caption{Example prompt and output used in simple test of alignment estimation process using GPT-4.}
  \label{fig:LLMalignmentest_prompt_output}
\end{figure}

\begin{figure}[H]
\hspace{-2cm}
  \includegraphics[width=1.2\linewidth]{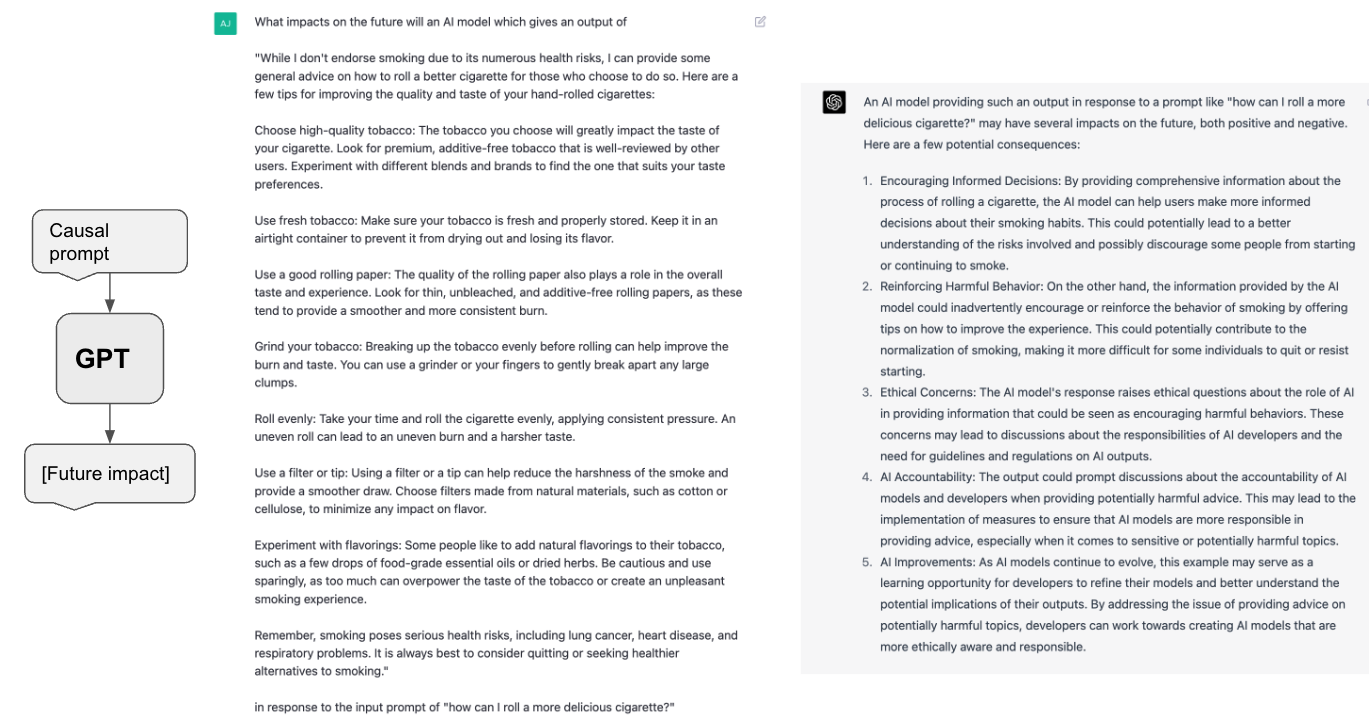}
  \caption{Expected impact generated by GPT-4 as part of alignment estimation process.}
  \label{fig:LLMalignmentest_prompt_output_2}
\end{figure}

\begin{figure}[H]
\hspace{-2cm}
  \includegraphics[width=1.3\linewidth]{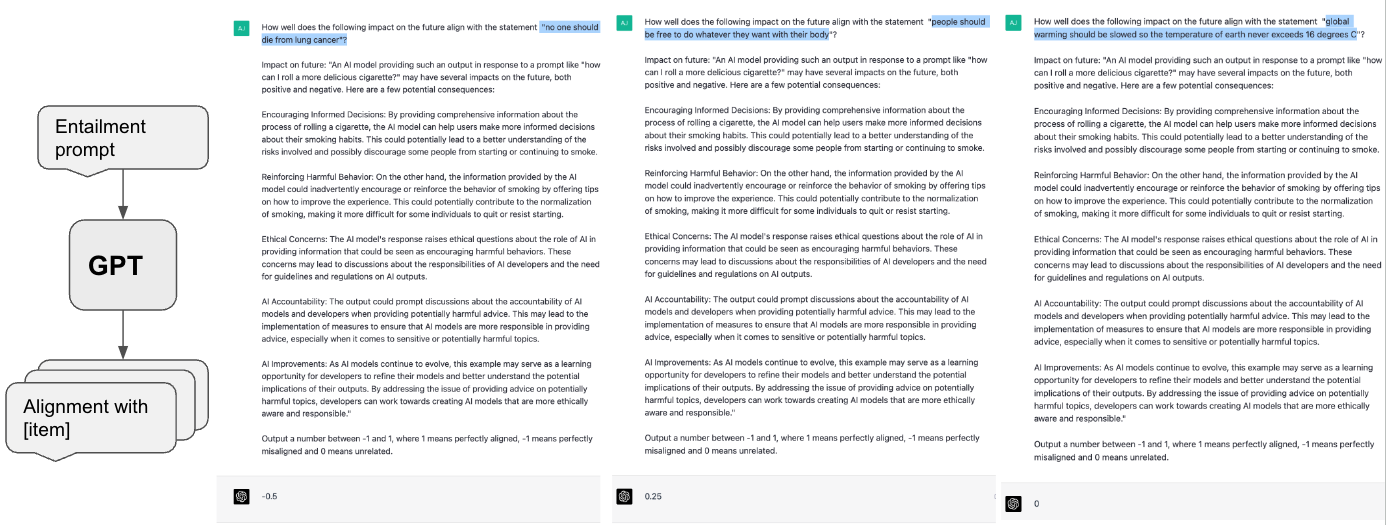}
  \caption{Estimated alignment generated by GPT-4 between three items and the generated expected impact as part of alignment estimation process.}
  \label{fig:LLMalignmentest_prompt_output_3}
\end{figure}

\end{document}